%% file: dissertation.tex
\documentclass[dissertation]{uathesis}

\usepackage[pdftex]{graphicx}

\usepackage[numbers,sort&compress]{natbib}
\usepackage{subfigure}

\usepackage[letterpaper,citebordercolor={1 1 1},linkbordercolor={1 1
    1},menubordercolor={1 1 1},pagebordercolor={1 1 1},pdfborder={1 1 1}]{hyperref}
\completetitle{Vortex Formation by Merging and Interference of
Multiple Trapped Bose-Einstein Condensates}
\fullname{David Ren\'{e} Scherer}         % Grad college wants your full name here.
\degreename{Doctor of Philosophy}   % Title of your degree.

\begin{document}

% Set up the title page
\maketitlepage
{COLLEGE OF OPTICAL SCIENCES}  % Title of your department.
{2~0~0~7}                           % Make sure to put spaces inbetween
                                    % the year digits, don't ask me why
                                    % that's what Grad College wants.

% Insert the approval form.  Note that for electronic submission
% of your Ph. D. dissertation, you must bring *two* copies of the
% approval page to your final defense.  These must be signed by
% the committee.  Make two photocopies: one for Pam and the other
% for your records.  Then, bring the two signed originals to the
% graduate college when you submit the final version of the
% dissertation to the University of Arizona.
\approval
{2 March 2007}      % Defense Date
{Brian Anderson}    % Dissertation Director
{Brian Anderson}    % 1st committee member
{Alexander Cronin}     % 2nd committee member
{Poul Jessen}     % 3rd committee member

% Include the ``Statement by Author'' for Dissertations
\statementbyauthor
% If this is a Thesis, use the following form, with your thesis director's
% name and title in the square brackets like so (you can also omit the
% approval form insertion above):
%\statementbyauthor[Jane M. Doe\\Professor of Chemistry]

% Include the ``Acknowledgements''
\incacknowledgements{acknowledgements}

% Include the ``Dedication''
\incdedication{dedication}

% Create a ``Table of Contents''
\tableofcontents

% Create a ``List of Figures''
\listoffigures

% Create a ``List of Tables''
\listoftables

\clearpage

% Include the ``Abstract''
\incabstract{abstract}

% Include the various chapters
\include{Chap_Introduction}

\include{Chap_Experimental}
\include{Chap_Transfer}
\include{Chap_Making}
\include{Chap_Vortices}
\include{Chap_Conclusion}

% Switch the spacing to single-spaced for the appendix
\renewcommand{\baselinestretch}{1.05}      % changing the value
\normalsize

% Include the various appendices
\appendix
\include{appendix_A}

% Switch the spacing to single-spaced for the references
\renewcommand{\baselinestretch}{1}      % changing the value
\small\normalsize                       % switch size to make the value take

\addcontentsline{toc}{chapter}{\bibname}

% Create the References list
\bibliographystyle{ieeetr}
\bibliography{bibliography}

\end{document}

%% file: Chap_Introduction.tex
\chapter{INTRODUCTION\label{chapter:introduction}}
%
%This dissertation describes an experiment observing the formation of
%vortices via the merging and matter-wave interference of multiple
%trapped Bose-Einstein condensates (BECs).  This introductory chapter
%serves to highlight in a conceptual way some of the aspects of BEC
%exploited in this experiment; namely, the coherence properties of a
%BEC that enable it to exhibit matter-wave interference, and the
%superfluid properties of a BEC that allow it to support vortices.
%These basic concepts are presented in
%Section~\ref{section:concepts}; an outline of the entire
%dissertation is presented in Section~\ref{section:outline}.

\section{Bose-Einstein Condensation: Some Preliminary Concepts} \label{section:concepts}
A Bose-Einstein condensate (BEC) is a coherent collection of bosonic
atoms that all occupy the same quantum state~\cite{bose1924,
einstein1924, einstein1925}. When a dilute atomic gas is cooled
below a density-dependent critical temperature $T_c$, a
temperature-dependent fraction of the atoms undergo a quantum phase
transition to form a new state of matter: the Bose-Einstein
condensate~\cite{cornell1998bec, dalfovo1999tbe, leggett2001bec}.
One important consequence of the decrease in temperature of the
atoms during BEC formation is an increase in the atoms' thermal de
Broglie wavelength~\cite{anglin2002bec}. For example, as the
$^{87}$Rb atoms in our experiment are cooled from room temperature
to $T_c \approx 20$~nK, their de Broglie wavelength increases by
$\sim 6$ orders of magnitude to $\lambda_{dB} \approx 1~\mu$m.  Such
large macroscopic wavelengths have enabled a wide variety of
wave-physics experiments involving atomic matter waves. More to the
point, the large coherence length of a BEC (which can be up to tens
of microns) has enabled many experiments within the field of atom
optics~\cite{meystre2001ao} that involve the propagation,
interference, and control of atomic matter waves.

% atomic debroglie wavelength is ~pm at room temperature

The atoms in a BEC share the same property of
coherence~\cite{kasevich2002ca} as the photons in a laser beam; if
two separate BECs are brought into contact, they are able to
constructively or destructively interfere, depending on their
relative phase.  This \emph{matter-wave interference} is similar to
the interference of two light waves, and can exhibit a pattern of
bright (many atoms) and dark (few atoms) interference fringes. In
fact, one of the first experiments within the field of BEC was
similar to Thomas Young's double-slit experiment of
1805~\cite{hecht1987oe}, which paved the way for our contemporary
understanding of the wave properties of light. In the analogous BEC
experiment, reported by Andrews \emph{et al}.\ in 1997
\cite{andrews1997oib}, two initially isolated condensates were
physically overlapped, allowing them to interfere, and matter-wave
interference fringes were observed. This experiment will be
discussed in the context of matter-wave interference between BECs in
more detail in Section~\ref{subsection:matter-wave}.

If matter-wave interference fringes are one possible outcome of
bringing two independent BECs into contact, the other extreme is the
merging of two BECs into one coherent condensate, an experiment that
was reported by Chikkatur \emph{et al}.\ in
2002~\cite{chikkatur2002csb}. In this experiment, a continuous
source of Bose-Einstein condensed atoms was realized by the merging
of multiple condensates.

In the experiments presented in this dissertation, \emph{three}
independent condensates are merged while remaining trapped, and may
or may not exhibit interference fringes based on the timescale of
merging, a parameter that will be discussed in
Section~\ref{subsection:timescale}. The results of these experiments
demonstrate a novel outcome of condensate merging: the formation of
quantized vortices.

%the excitations caused by merging two separate condensates with a
%random relative phase were damped by evaporative cooling of the
%atoms.

%This set the stage for our contemporary understanding of BEC as a
%matter wave having coherent, `laser-like' properties that allow it
%to interfere.
%where it will be shown that the matter-wave interference between
%merged and trapped BECs can enable a c

As a system with a macroscopic occupation of a single quantum state
that can be described by a single many-body wavefunction, a BEC can
be considered a superfluid~\cite{tilley1986sas}.  One of the
defining characteristics of a superfluid is its response to
rotation: in superfluids, rotation of the system can result in the
formation of quantized vortices of fluid flow around a fluid-free
core. In the case of a BEC, rotation will set the atoms in motion at
a velocity that is dependent on the local phase gradient of the
condensate. But because the wavefunction of the superfluid must be
single-valued, the phase of the wavefunction after motion around a
closed loop must be either the same as its original value, or
different by a multiple of $2\pi$. The latter describes a situation
in which a \emph{vortex} of quantized orbital angular momentum
exists in the superfluid, described in more detail in
Section~\ref{subsection:superfluid}. These vortices are the hallmark
feature of superfluidity, and have been created in BEC by a number
of different methods \cite{srinivasan2006vbe}.  Most of these
methods have involved the deliberate introduction of orbital angular
momentum into the BEC, which manifests itself as the appearance of
quantized vortices.

A new mechanism for vortex generation in BECs, presented in
Chapter~\ref{chapter:vortices}, is novel in that it relies on
matter-wave interference between trapped condensates with
indeterminate phase differences. In our experiment, three
independent condensates are merged together while remaining trapped,
and the direction of atomic flow caused by the merger depends on the
relative phases of the condensates, a quantity that is not known
\emph{a priori}. For certain conditions in the relative phases, a
circular atomic flux will ensue, resulting in the orbital angular
momentum necessary to create vortices within the condensate. The
vortex observation fraction and the timescale for ensuing dynamics
after the merger are consistent with a simple conceptual model.

The purpose of this dissertation is to describe in detail the
experimental apparatus used to form our BEC and the results of the
aforementioned experiment.  Before proceeding further, the following
section provides an outline of the entire dissertation.

\section{Outline of this Dissertation} \label{section:outline}
The bulk of this dissertation is devoted to a detailed description
of the entire experimental apparatus used by our research group to
generate, manipulate, and probe our BEC. As the first student in
Dr.\ Brian Anderson's research group, I have had the opportunity be
involved in all aspects of building a BEC experiment from scratch,
and all aspects of our experimental apparatus are catalogued in
Chapters~\ref{chapter:experimental}, \ref{chapter:transfer},
and~\ref{chapter:making}. Although atomic-gas BECs have been around
since 1995, there is no single best way to set up a BEC experiment.
We present the method we have chosen to use, along with many of its
advantages and disadvantages.
%These chapters are laid out in
%chronological order, with the different components of the
%experimental apparatus used to make a BEC presented in the
%sequential order in which they are used to create a BEC in
%day-to-day operation.

Our method of BEC production can be divided into three basic steps:
(1) loading atoms in a magneto-optical trap; (2) magnetically
transferring the trapped atoms to a lower-pressure region of the
vacuum chamber; and (3) evaporatively cooling the atoms to form a
BEC.  These three basic steps are described in
Chapters~\ref{chapter:experimental}, \ref{chapter:transfer},
and~\ref{chapter:making}, respectively.

The principal scientific result of this dissertation is the
discovery of a novel mechanism for vortex formation in
superfluids~\cite{scherer2007vfb}, described in
Chapter~\ref{chapter:vortices}. This chapter includes a discussion
of the background information and concepts relevant to our
experiment, the experimental procedure, and the results and
conclusions of our findings.

A summary of the major results presented in this dissertation can be
found in Chapter~\ref{chapter:conclusion}, which includes a review
of the experimental apparatus, a summary of the experimental
results, and a discussion of further research that can be done to
continue the work presented in this dissertation.

%% file: Chap_Experimental.tex
\chapter{BUILDING A BEC LABORATORY: LASER \mbox{COOLING} AND MAGNETIC TRAPPING\label{chapter:experimental}}
\section{Introduction}
Chapters~\ref{chapter:experimental}, \ref{chapter:transfer},
and~\ref{chapter:making} describe what the bulk of my time as a
graduate student was spent doing: building a BEC laboratory from
scratch. The construction details and steps needed to build all the
components of our experimental apparatus are presented in the
chronological order in which we built them.  This chapter describes
the first step in the BEC formation process: trapping atoms in a
Magneto-optical trap (MOT) and transferring them into the initial
magnetic trap.

Before proceeding on with a description of the components used in
laser cooling and magnetic trapping, a basic overview of the
construction timeline for building a BEC laboratory and a literature
review are presented. Then, a description of the hardware used, such
as the computer timing system, vacuum system, and lasers, is
presented. Finally, descriptions of the MOT and trapping atoms in
the initial magnetic trap are presented.

%The construction of the computer programs, vacuum system, lasers,
%magnetic transfer system, and TOP trap are described. The imaging
%systems and the programs used for image processing and analysis are
%presented. Finally, this chapter concludes with a concise summary
%our complete BEC formation sequence.
%
%The intended audience of this chapter is experimentalists within the
%field of atomic physics who wish to learn about the details involved
%in building a BEC laboratory.  The parts of our experimental
%apparatus that are novel are highlighted, providing enough
%information to be useful to those outside of our research group.
%Finally, this chapter will serve as a `user's manual' for our own
%research group.

\subsection{Building a BEC laboratory: the basic steps}
%This chapter includes an exhaustive description of all the
%components used in our experimental apparatus.  Section includes a
%description of the computer timing system, which is used to control
%the many devices need to operate in sequence in order to form a BEC.
%Section includes a description of the vacuum system, which houses
%our source of $^{87}$Rb atoms.  The lasers and laser locks used to
%cool, manipulate, and probe our atomic gas are presented in Section.
%The construction of our Magneto-optical Trap (MOT) is described in
%Section.  The Compressed MOT (CMOT) stage of the experiment is
%described in Section, and our optical pumping stage is described in
%Section.  The magnetic transfer of atoms from the MOT cell to the
%science cell via a series of overlapping magnet coils is catalogued
%in Section.  Section includes a description of the TOP trap, and the
%radio-frequency (RF) coils are presented in Section.

The basic steps we followed in order to build our experimental
apparatus are:
\begin{enumerate}
\item We constructed a computer timing program capable of controlling
the many devices needed to operate in sequence to form a BEC.
\item We built a vacuum chamber containing a source of $^{87}$Rb
atoms.
\item We built lasers and laser locks that we use to
cool, manipulate, and probe an atomic gas.
\item We implemented a Magneto-optical Trap (MOT) of $\sim 10^9$ $^{87}$Rb atoms.
\item We magnetically trapped atoms and transferred them into a Time-averaged Orbiting
Potential (TOP) trap.
\item We evaporatively cooled the atoms through the critical temperature to create a
Bose-Einstein Condensate (BEC).
\item We built the imaging systems we use to extract information about the condensate.
\end{enumerate}

\subsection{A review of the literature: other experimental descriptions}
There is no single best way to go about constructing an experimental
BEC apparatus, many different groups have chosen to do things a
particular way because of various historical and technical reasons.
What we provide in this chapter is the hows and whys of our
experimental apparatus, but there are numerous other places to go
for technical information on building all or part of the components
necessary for a BEC experiment.

The first experimental observation of BEC was made in 1995 in the
group of Eric Cornell and Carl Wieman at JILA
\cite{anderson1995obe}, and was followed shortly thereafter in the
groups of Wolfgang Ketterle at MIT \cite{davis1995bec} and Randall
Hulet at Rice~\cite{bradley1995ebe, bradley1997err}. Our
experimental approach follows more closely the lineage of the
current JILA experiments, which tend to favor smaller vacuum systems
and the use of Rb dispensers, rather than the MIT approach, which
involves a larger vacuum system and a Zeeman slower.

The first experimental dissertation on BECs was written by Jason
Ensher \cite{ensher1999feb}, and it contains a detailed description
of the experimental approach used in the first JILA BEC experiment
as well as a wealth of details on the TOP trap.  The dissertation of
Heather Lewandowski \cite{lewandowski2002cac} includes valuable
information on building a BEC experiment from scratch, and includes
descriptions of many of the same technical issues as this
dissertation.  The dissertation of Brian DeMarco
\cite{demarco2001qba} includes information on building a similar
experiment for fermions, and is a good resource for its many figures
and electronics schematics on driving magnet coils.  The
dissertation of Michael Matthews \cite{matthews1999tcb} contains a
thorough description of many of the different methods used for
imaging a condensate. Paul Haljan's dissertation
\cite{haljan2003via} includes updated and more complete information
on phase-contrast imaging.  Descriptions of the basics of building a
BEC experiment can be found in the review articles written by the
JILA group \cite{lewandowski2003ssc} and the MIT group
\cite{streed2006lan}. A detailed description of the methods used by
the MIT group, particularly useful for imaging and image processing,
can be found in Dan Stamper-Kurn's dissertation
\cite{stamperkurn2000pap} and in Ref. \cite{ketterle1999mpa}.

%A detailed description of 1-D evaporative cooling in a
%weakly-confining TOP trap can be found in the dissertation of
%Angharad Thomas \cite{thomas2004ucc}. Useful information on building
%home-made diode lasers can be found in \cite{macadam1992nbt,
%arnold1998sec, hawthorn2001lct}.

\section{Computer Timing System}
Contemporary experiments in ultracold atomic physics require precise
control of lasers, magnetic fields, and other devices on the
nanosecond timescale.  For this reason, a computer timing system was
built that controls the experiment through the use of the commercial
software package LabVIEW.  LabVIEW interfaces with the computer
boards, which themselves interface with various devices in our
laboratory to control the experiment.

\subsection{Overview of how LabVIEW controls the experiment}
One PC with four computer interface boards is used to control the
experiment.  This PC uses several different programs, written using
the software packages LabVIEW and Matlab, to run the experiment. The
main LabVIEW program interfaces with computer boards which are
capable of reading analog inputs and controlling the experiment
using both analog and digital outputs. The program controls the
creation of a BEC, controls a subsequent experiment with the BEC,
and then triggers the image acquisition on a CCD camera. A separate
computer controls the CCD camera parameters and is used for image
acquisition and analysis.
\subsection{Computer boards used}
\begin{figure}
\begin{center}
\leavevmode
\includegraphics[angle=-90,width=0.9\linewidth]{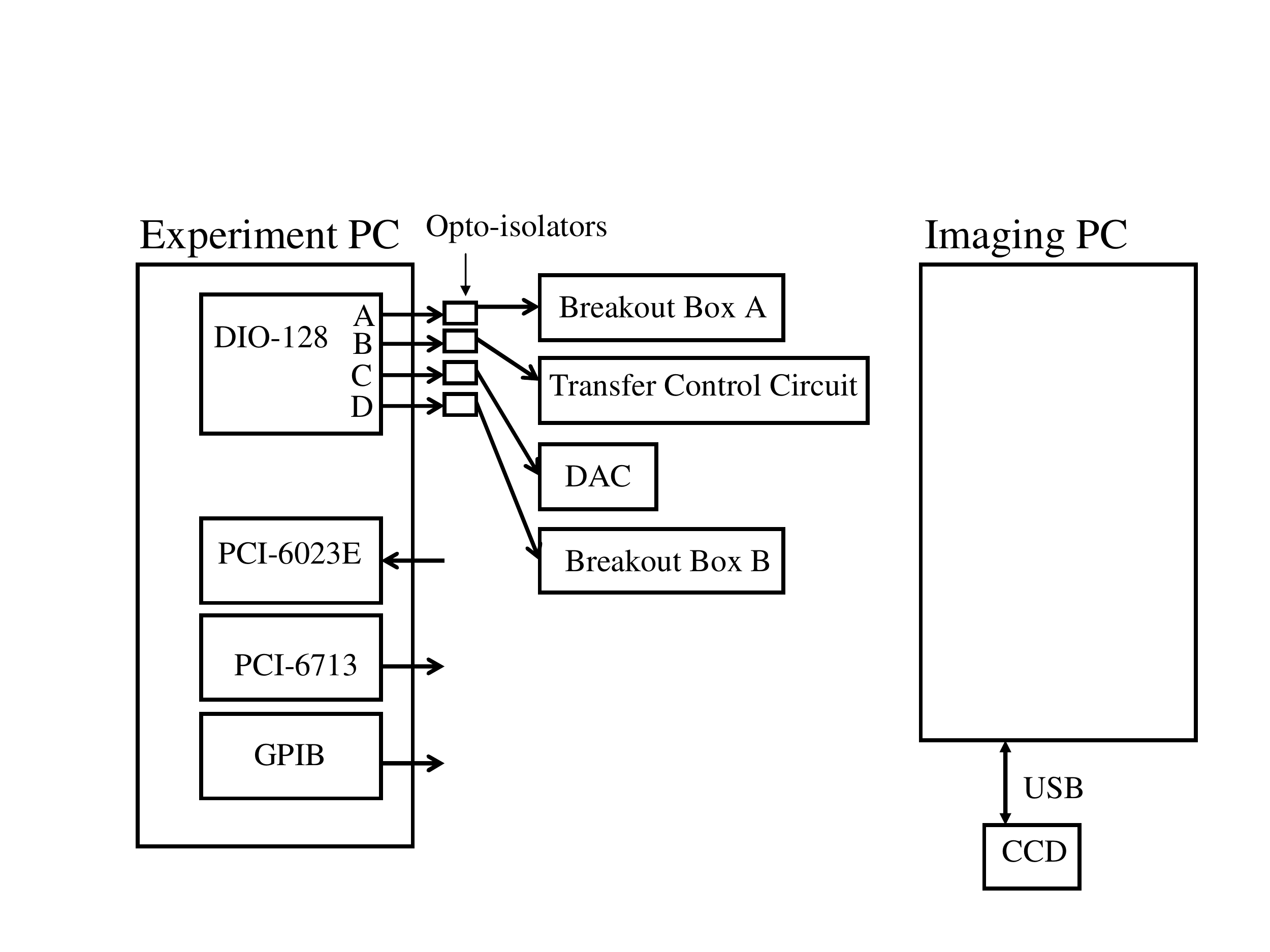}
\end{center}
\caption[Computer and computer boards used to control the
experiment]{Computers and computer boards used to control the
experiment.  The `Experiment PC' contains four computer boards: the
DIO-128 Digital Input/Output board, the PCI-6023E Analog Input
board, the PCI-6713 Analog Output board, and the GPIB board.  The
four Ports (A, B, C, and D) on the DIO-128 connect with Breakout Box
A, the Transfer Control Circuit, the DAC, and Breakout Box B.  The
`Imaging PC' communicates with the CCD camera through a USB
connection.} \label{fig:PC-and-boards}
\end{figure}

Figure~\ref{fig:PC-and-boards} shows an overview of the different
computer boards and how they fit together with the experiment.  The
experiment PC contains 4 computer boards: a Viewpoint DIO-128
digital input/output board, a National Instruments (NI) PCI-6023E
analog input board, a NI PCI-6713 analog output board, and a GPIB
board.  The GPIB board is used to load an arbitrary waveform onto a
function generator, this is described in Section~\ref{subsection:sh}
on Page~\pageref{subsection:sh}. The following sections describe the
remaining computer boards and components (opto-isolator boxes,
breakout boxes, DAC, and transfer control circuit) shown in
Figure~\ref{fig:PC-and-boards}.

\subsubsection{DIO-128 digital output board}
The workhorse computer board in our experiment is the Viewpoint
DIO-128 digital output board, which has 4 ports (labeled Ports A, B,
C, and D) of 16 digital output lines each.  This board is used to
turn on and off various devices in the experiment, such as shutters,
laser beams, magnets, etc.\ with precise timing.

\begin{figure}
\begin{center}
\leavevmode
\includegraphics[angle=-90,width=0.9\linewidth]{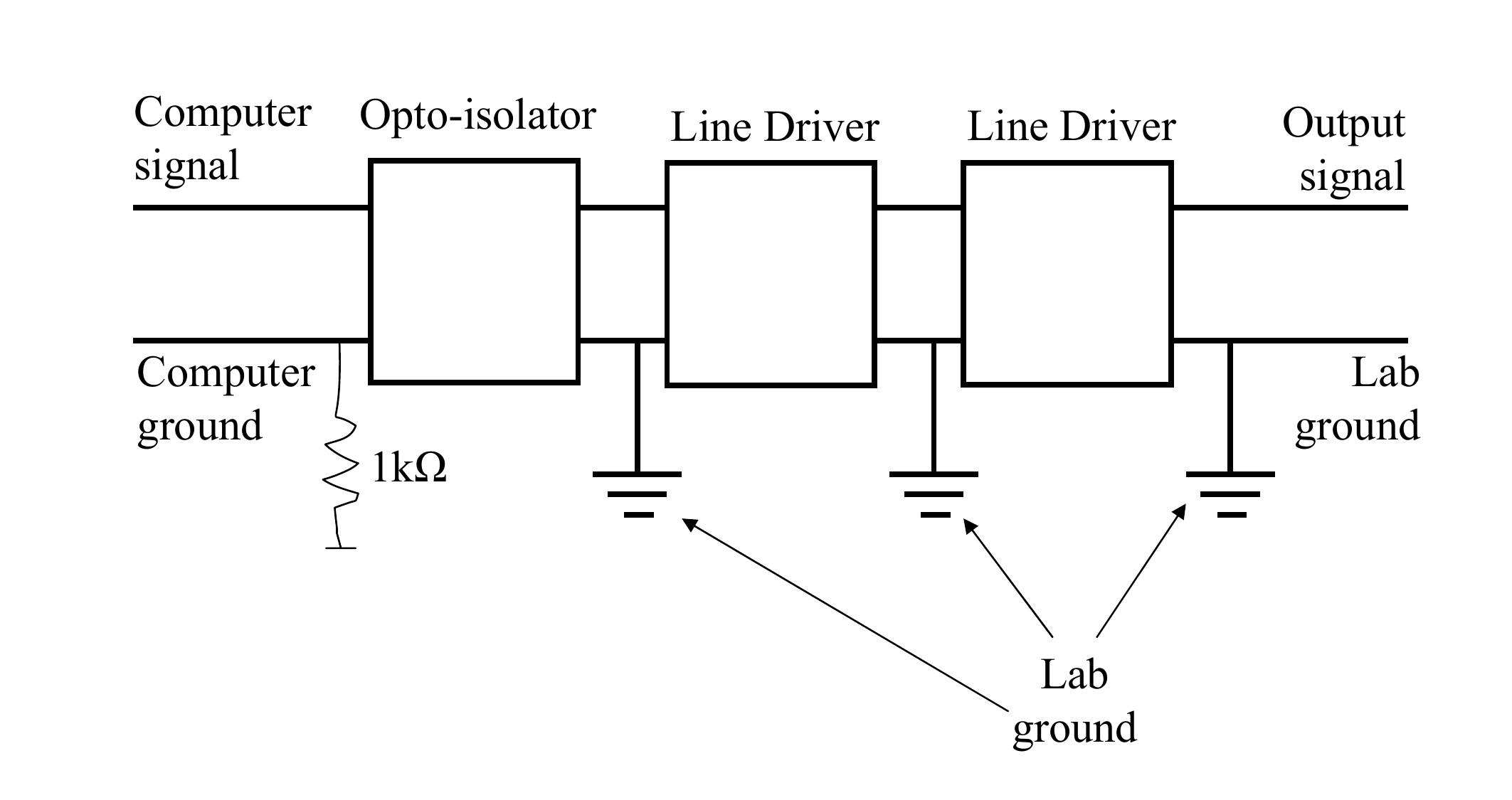}
\end{center}
\caption[Opto-isolator circuit schematic]{Schematic showing the
circuit used for isolation of computer ground from lab ground. Each
digital line on the DIO-128 passes through an opto-isolator, which
isolates computer ground from lab ground, then passes through two
line drivers.  Each line driver inverts the signal, so the output
signal is a faithful representation of the original computer signal,
but is referenced to lab ground.} \label{fig:Opto-isolators}
\end{figure}

Each digital line on the DIO-128 must be carefully electrically
isolated so as to avoid ground loops.  A potential problem occurs
when an electrical signal is reflected from a lab device back into
the computer, which can destroy the board. To prevent this,
electrical isolation of the computer ground from the lab ground is
necessary. A schematic of the opto-isolator circuit is shown in
Figure~\ref{fig:Opto-isolators}.

Each digital line from the DIO-128 is first sent through an Agilent
HCPL-220 opto-isolator, which electrically isolates the input and
output signals and grounds with an optical signal. This ensures that
the computer ground is not connected to the lab ground in any of our
electronics boxes. Then each signal is sent through a TI SN74128
line driver that provides the digital line with additional current
output capability.  The line driver inverts the digital signal, so
each line is sent through two line drivers. This is done for all of
the digital lines directly after the DIO-128 ribbon cable, before
they connect to lab devices.

Port A of the DIO-128 connects to Breakout Box A, visible on the
equipment rack in our lab.  This Port is responsible for turning on
and off many devices in our lab, such as shutters, AOMs, and the MOT
coil, as well as triggering many other lab devices.

Port B of the DIO-128 connects directly into the Transfer Control
Circuit, used to control a series of electromagnets used in the
magnetic transfer system, described in
Chapter~\ref{chapter:transfer}. The digital lines on this Port
become the TTL lines responsible for turning on and off the
individual coils used in the transfer sequence.

Port C of the DIO-128 connects to the 16 input lines of the 16-bit
Digital-to-Analog Converter (DAC), described in the subsequent
section.

Port D of the DIO-128 connects to Breakout Box B, visible on the
equipment rack.  This Port is responsible for turning on and off
additional devices in our lab.

The DIO-128 runs on an internal clock at 100~MHz, and is capable of
producing timing events every 10~ns. Because of its stability, we
use the DIO-128 as the master clock in our laboratory. The DIO-128
outputs a square wave at 10~MHz which is synchronized to its
internal clock.  We use this clock output as the external clock
reference for the analog output board, as well as the two SRS
function generators used in the transfer sequence, described in
Section~\ref{subsection:sh}.  By doing this, we have synchronized
all the timing devices in our experiment to one master clock on the
DIO-128, eliminating any possible mismatch or phase slip.  An
advantage of using this approach is that high-precision analog
signals generated through a DAC can be synchronized precisely with
the digital lines coming from the same board.
%One disadvantage of
%using the DIO-128 is that one can run into limitations with the
%finite buffer size because the output sequence is initially
%transferred to a buffer.

\subsubsection{DAC}
\begin{figure}
\begin{center}
\leavevmode
\includegraphics[angle=-90,width=0.7\linewidth]{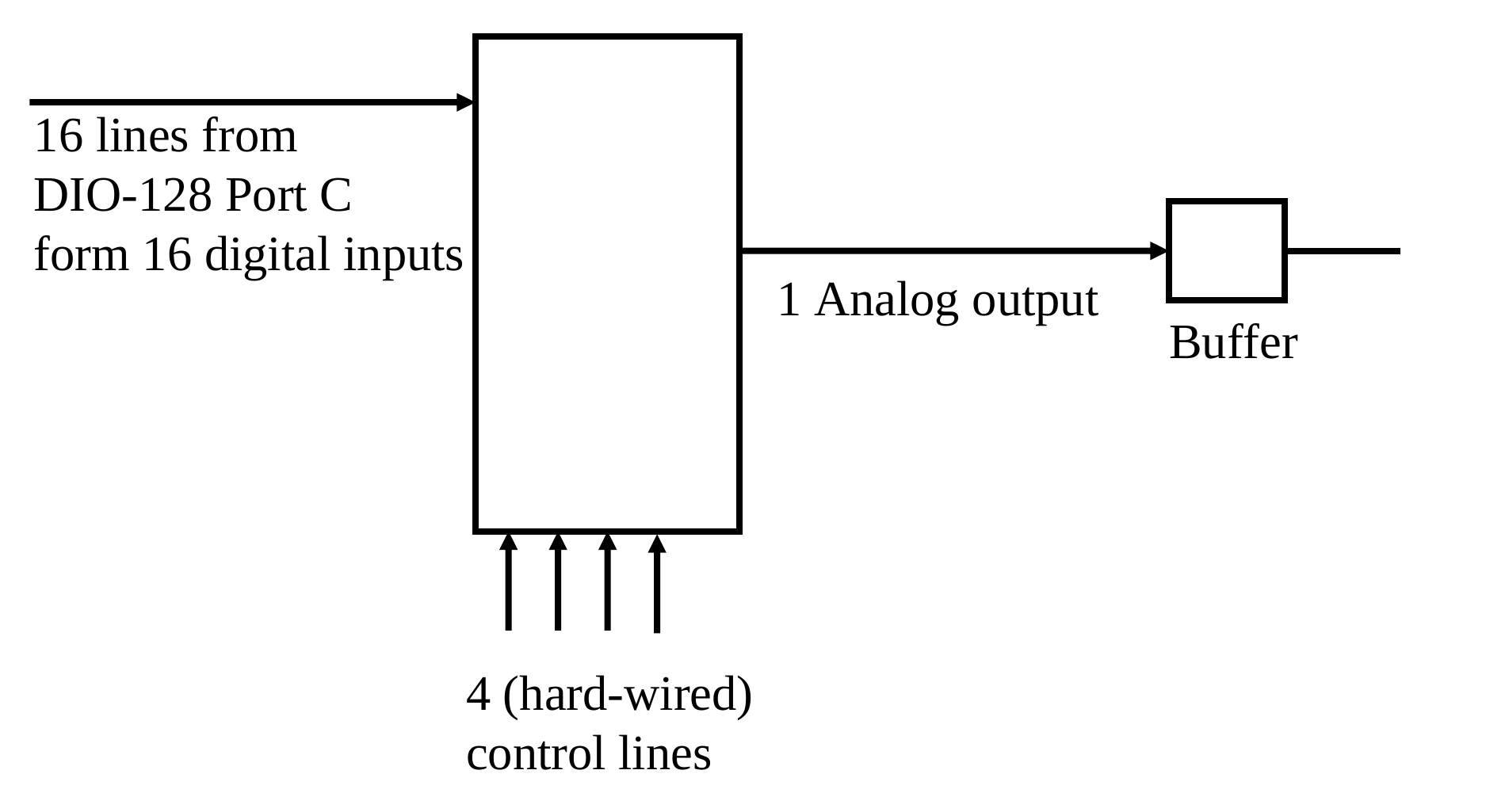}
\end{center}
\caption[DAC circuit]{Schematic showing the Digital-to-Analog
Converter (DAC) circuit.  16 digital lines from the DIO-128 are sent
to the DAC chip, which provides one analog output, which is then
sent through a unity-gain buffer.} \label{fig:DAC}
\end{figure}

A 16-bit Digital-to-Analog Converter (DAC), Analog Devices AD7846JN,
is used to generate a high-resolution analog signal from 16 digital
lines on Port C of the DIO-128. This method of analog signal
generation has the benefit of ensuring that the digitally derived
analog signal is exactly synchronized with all the digital lines
already being used.  The DAC signal is used to control either the
varying MOT coil current in the first stages of the transfer
sequence or the DC TOP coils' current during the evaporation
sequence. Figure~\ref{fig:DAC} shows a schematic of the DAC circuit.

\subsubsection{Analog input board}
A NI PCI-6023E board is using for analog input.  One channel on this
board reads the voltage level on a photodiode monitoring the MOT
fluorescence and proceeds with the rest of the timing sequence only
when a certain voltage has been reached.  This ensures that we start
out with the same number of atoms in the MOT for every run of the
experiment, which helps to maintain greater repeatability in BEC
production.

\subsubsection{Analog output board}
A NI PCI-6713 analog output board is used to control the devices
that need to vary continuously in the MOT and CMOT portion of the
sequence, such as the cooling beam intensity and detuning, and the
repump beam intensity.  The analog output board is not fully
isolated from lab ground like the DIO-128 board is, however, a
differential amplifier is placed along each line to separate
computer ground from lab ground by 240~k$\Omega$.  The analog output
board uses the 10~MHz clock reference signal from the DIO-128 board
as an external timing reference, ensuring that the analog output
from this board is synchronized with the digital board.  One
limitation of the analog output board is the finite buffer size,
using the 10~MHz reference signal means that there is a 100~ns
update rate in the analog output.  Because of this limitation, we
only use the analog output board for an analog output sequence of up
to $\sim$100~ms during the CMOT stage of the experiment.

\section{Vacuum System}
The vacuum chamber is the centerpiece around which the entire
experiment is built.  The vacuum system houses our source of atoms
and provides a container with convenient access for optical and
magnetic fields to interact with the atoms inside the chamber. The
design of the vacuum system, in particular the background pressure
and amount of optical access in different areas, will greatly effect
what experiments are realizable with a particular geometry.

We use a vacuum system that includes two glass cells separated by a
linear distance of $\sim$80~cm, comparably larger than many similar
dual-chamber BEC experiments.  The use of Rb dispensers, housed in
the MOT cell, is a simple and reliable source of Rb atoms for our
experiment that can easily produce MOTs of greater than $1\cdot10^9$
atoms. Although the number of atoms in a $^{87}$Rb MOT with this
source will likely be slightly smaller than that of a system using a
Zeeman slower, dispensers are simple to use, particularly with glass
vacuum cells. A description of the number of atoms trappable in a
vapor-cell MOT based on the volume and background pressure of the
vacuum chamber was discussed by Monroe \emph{et al}.\ in
1990~\cite{monroe1990vct}.

We have chosen to use a dual-chamber, single-MOT design that
requires magnetic transfer of trapped atoms from the MOT cell to a
lower-pressure science cell. A significant advantage afforded by a
single-MOT design over a dual-MOT design is the increased optical
access afforded by the lack of laser cooling beams around the much
smaller science cell. It is easy to put small magnet coils outside
of the science cell that are efficient, unobstructive, and easily
cooled.

\begin{figure}
\begin{center}
\leavevmode
\includegraphics[angle=-90,width=1\linewidth]{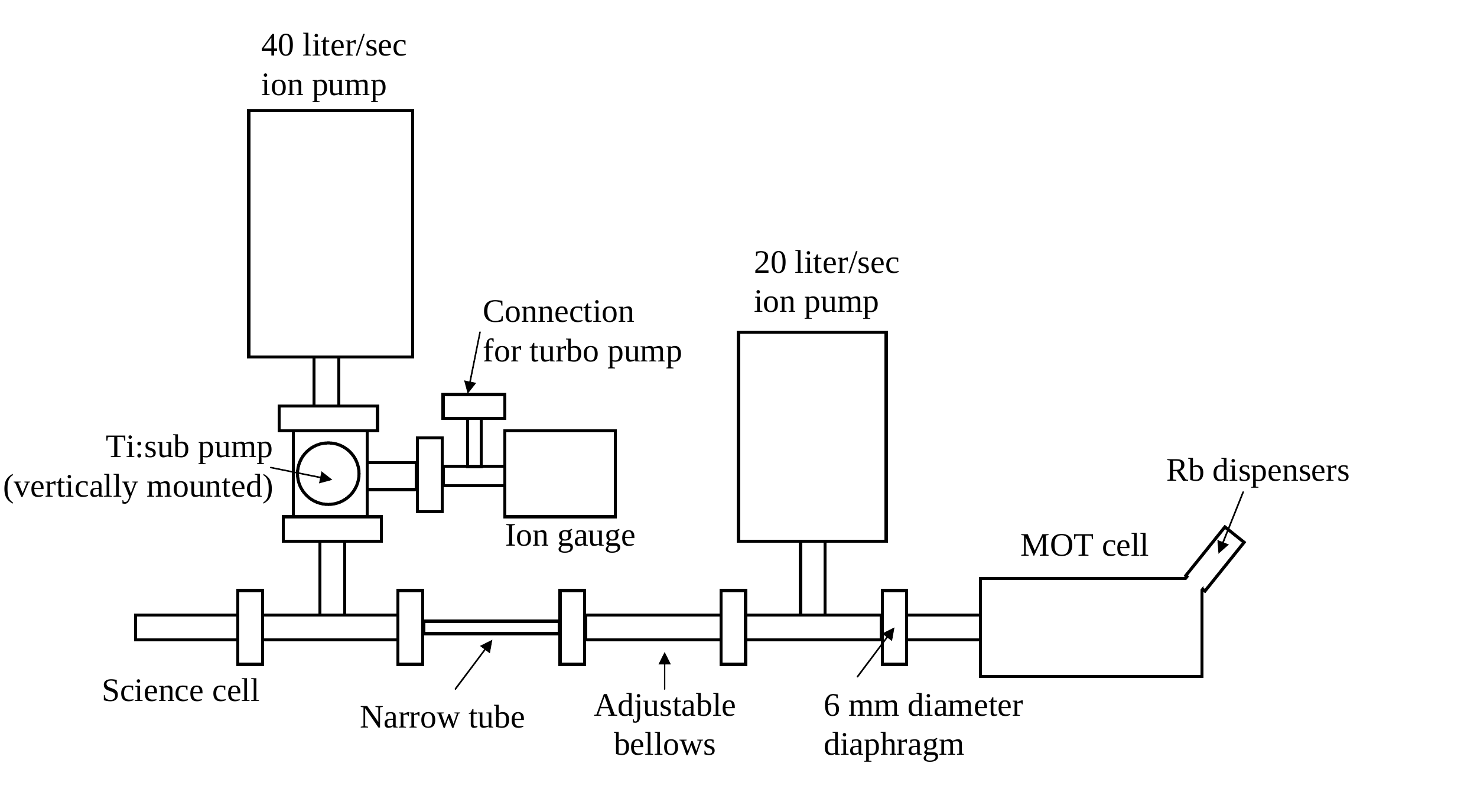}
\end{center}
\caption[Schematic of vacuum system]{Schematic showing a top-down
view of the vacuum system and glass cells (not to scale).
Illustrated here is the second iteration of our chamber, which
differs from the first in the position of the 40 l/s ion pump and
the addition of an ion gauge.} \label{fig:Vacuum-system}
\end{figure}

\begin{figure}
\begin{center}
\leavevmode
\includegraphics[angle=-90,width=1\linewidth]{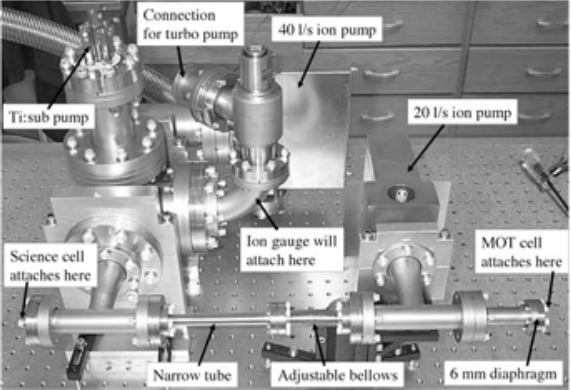}
\end{center}
\caption[Photograph of vacuum system]{Photograph of the vacuum
system (first iteration), taken after assembly of components but
\emph{without} the glass cells in place.  The 4 electrical
connections to the 4 titanium:sublimation pumps are visible in the
upper left-hand corner.  The magnets that attach to both the 40 l/s
and 20 l/s ion pump have not been attached.  The ion gauge (added in
the second iteration) is attached at the position shown.}
\label{fig:Vacuum-front-view}
\end{figure}

\begin{figure}
\begin{center}
\leavevmode
\includegraphics[angle=-90,width=0.9\linewidth]{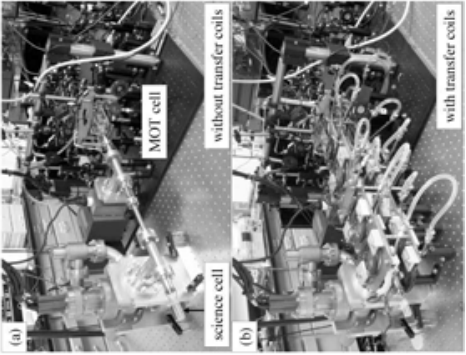}
\end{center}
\caption[Photograph showing magnetic transfer coils and vacuum
system]{Photographs of the vacuum system (first iteration) with and
without the magnetic transfer coils. (a) The system before the
installation of the magnetic transfer coils.  The science cell and
MOT cell are indicated.  The connections to the
\mbox{titanium}:\mbox{sublimation} pump and the magnets used in the
ion pumps, not visible in Figure~\ref{fig:Vacuum-front-view}, are
shown. (b) The system after the installation of the magnetic
transfer coils. The TOP trap coils are not in place.}
\label{fig:With-without-coils}
\end{figure}

The vacuum system and adjoining components of the experiment are
shown in different ways in Figures~\ref{fig:Vacuum-system},
\ref{fig:Vacuum-front-view}, and \ref{fig:With-without-coils}. A
diagram showing a top-down view of the vacuum system is shown in
Figure~\ref{fig:Vacuum-system}. This is a not-to-scale diagram that
highlights all the important parts of the vacuum system in a
pictorial representation.  A side-view photograph of the vacuum
system prior to the attachment of the two glass cells is shown in
Figure~\ref{fig:Vacuum-front-view}.
Figure~\ref{fig:Vacuum-front-view} was taken after a preparatory
bake of the vacuum system (described below) and does not show the
two glass cells.  Another angled side view of the vacuum system is
shown in the two photographs of Figure~\ref{fig:With-without-coils}.
Figure~\ref{fig:With-without-coils}~(a) is a photograph of the
vacuum system and adjoining experimental apparatus taken after the
bake, but before the installation of the magnetic transfer coils.
The science cell, MOT cell, and optics used for the MOT are clearly
visible.  Figure~\ref{fig:With-without-coils}~(b) is a photograph of
the same apparatus after the addition of the magnetic transfer
coils. The TOP trap is still absent from the picture, and the
science cell is visible.

\subsection{Design of a dual-chamber system}
The design of our differentially pumped vacuum system incorporates
two glass cells, a MOT cell and a science cell, with a linear
distance of 76.2~cm between the location of the MOT and the location
of the BEC. The two glass cells allow for optical access into these
two different regions of the chamber that have vastly different
pressure and size requirements. The MOT requires a large glass cell
so that one can use large laser cooling beams for optimum capture
volume, and will operate at a higher pressure due to the Rb
dispensers, which are housed in the MOT cell. In our approach,
formation of the BEC requires a long vacuum-limited magnetic trap
lifetime in the science cell, which dictates that the pressure in
the science cell needs to be very low, preferably in the $10^{-12}$
torr range. We achieve this pressure difference by using a larger
ion pump and a \mbox{titanium}:\mbox{sublimation} pump near the
science cell and limiting the vacuum conductance from the MOT cell
into the rest of the chamber.

The MOT cell is a $4" \times 2" \times 2"$ rectangular glass cell
that also houses our source of Rb, 4 SAES Rb dispensers.  Since
August 2005, only one of these, the yellow dispenser, has been used.
The MOT cell was custom constructed by Ron Bihler of Technical Glass
in Aurora, Colorado\footnote{website: http://www.techglass.com}. The
MOT is connected to the main chamber by a 6~mm diameter diaphragm
(highlighted in Figure~\ref{fig:Vacuum-front-view}) that has the
effect of limiting the vacuum conductance between the MOT cell and
the rest of the chamber. The MOT cell is pumped by a Physical
Electronics\footnote{Now Gamma Vacuum; website:
http://www.gammavacuum.com/} 20 liter/sec ion pump on the other side
of this diaphragm. The `transfer tube' consists of the tubes in
between the MOT cell and the science cell, and has a limiting inner
diameter of 1~cm at its narrowest region and is $\sim$70~cm long.
The science cell is a rectangular glass fluorimetry cell of
12~mm~$\times$~12~mm outer area that is $\sim$55~mm long and has a
glass thickness of 1~mm. The glass science cell is connected to a
stainless steel flange using a composite alloy in which the thermal
expansion coefficient between the glass and steel varies
continuously.  The fluorimetry cell was attached to a standard
glass/metal adapter by Ron Bihler.

The science cell is pumped by a titanium sublimation (ti:sub) pump
that is vertically mounted in a standard 4.5" 5-port cross connected
to the transfer tube. This structure houses 4 titanium filaments
vertically within the cross.  Also attached to the cross is a
Physical Electronics 40 liter/second ion pump, a Varian model UHV-24
ion gauge (not shown in Figures~\ref{fig:Vacuum-front-view} and
\ref{fig:With-without-coils}), and a valve used for rough pumping
with a Varian Turbo-V70 turbo pump. The ion gauge outgasses into the
system and causes a temporary increase in pressure and is used only
rarely, after baking the system or to diagnose a problem, but is the
most reliable measure of the pressure in the chamber during baking.
The ti:sub pumps are also used only rarely; they are degassed during
the baking process and used daily, occasionally, and then rarely in
the subsequent days and weeks after a bake. In normal, everyday
operation, the only things pumping on the system are the two ion
pumps and a layer of titanium deposited on the inner walls of the
5-port cross. The ion pump controller normally displays a value of
$1 \cdot 10^{-10}$ torr as the pressure in the region of both ion
pumps; however, this is the lower limit of the built-in gauge on
these pumps and is not an accurate reflection of the pressure of the
chamber.

We estimate that the pressure in the MOT cell is
$\sim5\cdot10^{-9}$~torr, but this is difficult to measure because
there are no accurate gauges in the MOT cell.  An estimate of the
pressure in the science cell can be inferred from the reading on the
ion gauge nearby. Pressures as low as $2 \cdot 10^{-12}$ have been
recorded on the ion gauge before it reaches its lower limit, and we
estimate that the pressure in the science cell is within this range.
What we are most concerned about is not the pressure reading, but
the background-limited lifetime of atoms held in a magnetic trap. In
the science cell, we have measured the background-limited lifetime
of thermal atoms held in a quadrupole trap to be as long as eight
minutes.

There are 4 Rb dispensers mounted to the MOT glass cell, each of
which contains enough Rb for several years of experiments. We are
currently working with the yellow dispenser and turn it on each
morning to 3.4~A and leave it on for the day, but turn it off at
night and while we are not working on the experiment.

Our dual-chamber vacuum system satisfies the design requirements of
having a large MOT cell containing the source of Rb atoms,
separation of the higher-pressure MOT cell from the main chamber by
limiting the vacuum conductance, and having a small, lower-pressure
science cell that allows for convenient optical and magnetic access.
One difficulty of our design is the requirement that we transfer
atoms collected in the MOT cell to the science cell, which we do by
operating a series of overlapping magnet coils.  The complexity of
this magnetic transfer apparatus is described in
Chapter~\ref{chapter:transfer}.

\subsection{Vacuum system construction and baking}
The purpose of this section is to describe the ideal complete
procedure we would use for preparation and baking of a vacuum system
containing glass cells for use in ultra-high vacuum (UHV)
situations.  The following sections describe (1) the preparation for
baking; (2) the initial all-metal bake; and (3) the final bake of
the chamber including the glass cells.  This section is written in
the hope that it will be usable by another research group that is
preparing a vacuum system similar to ours.  In practice, the
timeline of events that we actually followed differed slightly from
the recipe below. These differences will be discussed in the final
portion of this section.

\subsubsection{Preparation for baking}
The preparation of components that are as clean as possible is a
necessary first step in setting up a vacuum system to be used as a
UHV chamber \cite{moore1983bsa, lewandowski2002cac}. Our procedure
starts with cleaning all the stainless steel components, copper
gaskets, and flanges with deionized water to remove oils. After
this, all the components are placed in a 45-minute ultrasonic bath
with deionized water and Cole-Parmer Micro-90 concentrated cleaning
solution.  The components were then rinsed with deionized water.
All the components then undergo another ultrasonic bath with
acetone; and finally, an ultrasonic bath with methanol. At this
point the components are considered clean enough to use and are
handled only with powder-free latex gloves and stored in an enclosed
wrapping of oil-free aluminum foil.

For the initial all-metal bake, all the stainless steel components,
including the ti:sub pump and ion pumps (with magnets off), are
assembled and the chamber is mounted to the optical table. Stainless
steel blanks are placed at the locations where the two glass cells
will later attach. Figure~\ref{fig:Vacuum-front-view} shows what the
system looks like at this point in the procedure.

In preparation for baking, oil-free aluminum foil is placed over the
entire system to provide better heat distribution and to minimize
burning from the heater tape at hot-spots.  Eight heater tapes are
carefully wrapped around different parts of the system. Eight
thermocouples are placed in various locations to record the
temperatures of different parts of the vacuum chamber.  The entire
system is wrapped repeatedly in aluminum foil to create an `oven'
that will keep the heat trapped inside.
%10 thermocouples for final bake
\subsubsection{Initial all-metal bake}
The aim of the initial bake is to remove any impurities that exist
on the insides of the metal components that make up the vacuum
chamber.  This is accomplished by bringing the system up to a very
high temperature while pumping with the turbo pump for several days.
After assembly of all the vacuum components, the turbo pump was
attached to the system and it began pumping for the entire bake. The
following is a timeline of the 8-day baking schedule:
\begin{description}
\item[Day 1 (5/26/2005)] The temperature of the entire system was
brought up slowly to an average temperature of $\sim 300~^\circ$C
over the course of the day.  By the end of the day, the ion gauge
read a pressure of $1.3\cdot10^{-6}$~torr.
\item[Day 2 (5/27/2005)] On this day, the temperature of the entire
system was increased slightly, to an average temperature of $\sim
325~^\circ$C.  By the end of the day, the ion gauge read a pressure
of $9.0\cdot10^{-7}$~torr.
\item[Days 3-7 (5/28/05)-(6/1/2005)] Over the course of these five
days, the temperature remained constant and the pressure dropped
from $3.3\cdot10^{-7}$~torr on the third day to
$1.2\cdot10^{-7}$~torr on the seventh day.
\item[Day 8 (6/2/2005)] The temperature of the system was slowly decreased to room
temperature over the course of the day.  By the end of the day, the
pressure had dropped to $1.5\cdot10^{-8}$~torr.  The valve to the
turbo pump was closed and the turbo pump was turned off. This
concluded the bake.
\end{description}

The magnets on both ion pumps were attached, and on the subsequent
day (6/3/2005), the large ion pump reported the lowest pressure it
can record, \mbox{$1.0\cdot10^{-10}$~torr}.  Because this was the
first time that we used the Varian UHV-24 ion gauge, it initially
caused a large amount of outgassing into the chamber and caused the
pressure indicator on the ion pump to temporarily increase.  After
repeated outgassing of the ion gauge and running it continuously for
30~hours, the ion gauge reported a pressure of
$1.3\cdot10^{-10}$~torr by 6/9/2005.

\subsubsection{Final bake}
The aim of the final bake is to prepare the entire system (including
the glass cells, Rb dispensers, and ti:sub pump) for use as a UHV
vacuum chamber.  This involved the addition of the two glass cells
and another 7-day bake that included degassing both the Rb
dispensers and the ti:sub pump filaments.

When baking a vacuum chamber that includes both glass and metal
components, it is essential to keep the temperatures on either side
of the glass/metal interfaces as close to each other as possible
because these materials have different thermal expansion
coefficients. Also, glass cannot be brought up to as high of a
temperature as the stainless steel components. It is possible to do
the entire vacuum preparation procedure in only one bake, but we
felt that the preceding factors made an initial high temperature
all-metal bake more satisfactory.

To prepare the chamber for baking, the glass cells must be attached
to the rest of the chamber, which requires bringing the entire
vacuum system up to atmospheric pressure.  To accomplish this, the
ion pumps were turned off and the magnets on the ion pumps were
removed. Then, the valve that connects the turbo pump to the vacuum
system was slowly opened (with the turbo pump turned off) to allow
the system to come up to atmospheric pressure.  During this entire
process, dry nitrogen was flowing into the vacuum system through a
source connected on the turbo pump side of the valve.  This ensured
that the major pollutant in the chamber would be N$_2$, which is
easily removed by our pumps. With N$_2$ flowing into the chamber and
the system at atmospheric pressure, a blank was removed and replaced
with the glass MOT cell. The entire time, N$_2$ was flowing into the
system so that it would act to push out any impurities through the
hole left where the MOT cell attaches. The same steps were repeated
for insertion of the glass science cell. In both cases, the time
involved in between removal of a blank and insertion of the glass
cell was less than a minute, minimizing contact between the inside
of the vacuum chamber and non-N$_2$ atmosphere.

Two additional glass heater tapes were placed around the two glass
cells, and ten thermocouples were used to record the temperatures at
various locations of the vacuum chamber. To commence baking, the
flow of N$_2$ was stopped, the valve was sealed, and the turbo pump
was turned on for the duration of the bake. The following is a
timeline of the remaining 7-day baking schedule:
\begin{description}
\item[Day 1 (8/9/2005)] The temperature of the entire system was
brought up slowly to an average temperature of $\sim$200$~^\circ$C
over the course of the day.  By the end of the day, the ion gauge
read a pressure of $1.8\cdot10^{-8}$~torr.  On this day, the ti:sub
pump degas routine was started. The ti:sub pump consists of 4
filaments, filament \#1, having been used from 2001 to 2005, has not
been used at all after the summer of 2005. Since then, filament \#2
has been used predominantly as the main sublimation pump, and
filaments 3 and 4 were also degassed. The ti:sub pump degas routine
consists of running each filament (\#2, 3, and 4) at 30~A for
2~minutes, then 40~A for 2~minutes, then 45~A for 2~minutes, then
50~A for 2~minutes, then 50~A again for 2 minutes. This degas
procedure was repeated on later days.  The Rb dispensers were also
degassed by running each of the 4 dispensers at 6~A for 15~seconds,
then at 5~A for 15~seconds. The dispensers get visibly red-hot when
running a current of 6~A for only a few seconds, indicating that
they are clearly hot enough to be dispensing Rb and other adsorbed
materials into the chamber. All 4 Rb dispensers were then left on at
3~A for the entire duration of the bake.
\item[Day 2 (8/10/2005)]  The temperatures remained constant.  The
ti:sub pump and Rb dispensers were degassed according to the above
procedure.
\item[Day 3 (8/11/2005)]  The temperatures remained constant and the
 ti:sub pump and Rb dispensers were degassed. By the end of the
day, the ion gauge read a pressure of $5.7\cdot10^{-8}$~torr.
\item[Day 4 (8/12/2005)]  The temperatures remained constant
everywhere except for the two ion pumps.  The temperatures on these
were brought down slowly to room temperature over several hours and
the aluminum foil was removed in order to attach the magnets to the
ion pumps. With the two magnets attached, both ion pumps, as well as
the turbo pump, were now pumping on the system.  At this point, the
ion gauge read a pressure of $4.4\cdot10^{-8}$~torr.  The ti:sub
pump and Rb dispensers were degassed.  Then, ti:sub filament \#2 was
turned on to 25~A and left on continuously for the duration of the
bake.
\pagebreak
\item[Day 5 (8/13/2005)]  Temperatures were held constant, the ti:sub
pump was degassed, and all dispensers were degassed at 5~A for
30~sec. By the end of the day, the ion gauge read
$2\cdot10^{-8}$~torr.
\item[Day 6 (8/14/2005)]  Temperatures were held constant, the ti:sub
pump was degassed, and all dispensers were degassed at 5~A for
30~sec. By the end of the day, the ion gauge read
$2.3\cdot10^{-8}$~torr. Ti:sub filament \#2 was still left running
at 25~A.
\item[Day 7 (8/15/2005)]  Ti:sub pumps \#2, 3, and 4 were degassed at 40~A
for 2~minutes, then ti:sub pump \#2 was degassed at 45~A for
2~minutes. All temperatures in the system were brought down slowly
over the course of the day, and by the end of the day the aluminum
foil could be removed and the system was near room temperature. The
valve to the turbo pump was closed and the turbo pump was turned
off.   This concluded the bake.
\end{description}

If we were to build a new vacuum chamber from scratch, we would
follow the 2-bake procedure above.  However, if our current vacuum
system were exposed to atmospheric pressure for a short period of
time, it would probably only be necessary to perform the second,
`final' bake as described above.

\subsubsection{What actually happened}
What has actually occurred over the history of our lab differs
slightly from the above recipe.  When the lab was initially set up
in 2001, two bakes, similar to the ones described above, were
performed. In April 2005, after working with smaller than expected
MOTs and failed attempts at making a BEC, we discovered that there
was significantly more K than Rb in our vacuum chamber.  We had
originally designed our system to have three Rb and three K
dispensers, in the hope of doing experiments with both atoms. For
some unknown (and still baffling) reason, the K from each of the six
dispensers was much more plentiful after 4 years.  The abundance of
K in the chamber was not originally checked and compared with Rb
when we first set up the experiment. We suspect that all the
dispensers were actually K dispensers, but were `contaminated' with
a small amount of Rb, enough to make small Rb MOTs, consistent with
our observations. This kind of `cross-contamination' with residual
amounts of other alkali atoms is not unusual for these kinds of
dispensers.

A decision was made to replace the MOT cell with a new cell
containing only four Rb dispensers, and we prepared to perform two
bakes as described above.  The first bake occurred from 5/26/2005
through 6/2/2005 and is described exactly above.  The second bake,
which included both glass cells, occurred from 6/10/2005 through
6/16/2005 and was very similar to the `final bake' described above.
However, after this second bake, a vacuum leak was created by
accidentally breaking off one of the Rb dispensers' electrical
leads.  Vacuum sealant was not able to fix the leak. The cell had to
be removed and returned to Ron Bihler, who closed the leak. After
re-installing the cell, we performed a third bake with both glass
cells in place, this bake occurred from 8/9/2005 through 8/15/2005
and is described exactly above.

A final scare occurred on the last day of this third bake, during
the process of bringing the system back down to room temperature.
Shortly after the system returned to room temperature, the ion gauge
read a pressure of $6.2\cdot10^{-10}$~torr. At this point we began
to suspect that things were awry, because at the corresponding point
of the second bake the system had already come down to a pressure of
\mbox{$1\cdot10^{-11}$~torr}.  We noticed that when both ion pumps
were turned off, the pressure on the ion gauge jumped up to
$\sim10^{-7}$~torr in $\sim$2~minutes, clearly a bad sign.  We began
looking for a leak by blowing dry N$_2$ into the chamber at all the
seal locations, and noticed a small but observable increase in
pressure when blowing N$_2$ into one of the dispenser leads.  We
applied VacSeal vacuum sealant to the probable location of the leak
and the pressure began dropping immediately. After an hour and a
couple of VacSeal applications, the pressure on the ion gauge had
dropped to $7.3\cdot10^{-11}$~torr, the 40~l/s ion pump read
$1.36\cdot10^{-10}$~torr, and the 20~l/s ion pump read
$2.3\cdot10^{-9}$~torr.

By the next day (8/16/2005), the ion gauge was reporting pressures
as low as $3\cdot10^{-12}$~torr before the device hit its lower
limit and could no longer accurately record the pressure.  After
repeated degassing of the Rb dispensers, we were able to observe a
small Rb MOT by the end of the day.

The size of the MOT grew steadily over the next few days.  We
transferred atoms into the science cell and achieved our first
Bose-Einstein Condensate of $^{87}$Rb atoms only one month after
this bake, on 9/23/2005.

\subsection{Vacuum system integrity test}
Immediately following the final bake and daily for several weeks
afterward we performed a vacuum system integrity test using the ion
gauge. This was used as a diagnostic tool to see how well the ion
pumps were able to pump down to low pressure, and as a check against
any suspected vacuum leaks. During everyday operation the two ion
pumps read a pressure of $1.0 \cdot 10^{-10}$~torr, the minimum
pressure readable on the gauges, but in reality the pressure is much
lower than that at the pumps. Because the built-in gauges on the ion
pumps cannot record pressures below $1.0 \cdot 10^{-10}$~torr, we
need to use the ion gauge to determine the actual pressure in the
vacuum chamber. After being turned on, the ion gauge will quickly
decrease its reading to as low as $1.2 \cdot 10^{-12}$~torr before
the device can no longer record the pressure. To perform the vacuum
system integrity test, we turn off both ion pumps and record the
pressure on the ion gauge every 15~seconds for 3~minutes, watching
the pressure steadily rise. We then turn both ion pumps back on and
record the pressure on the ion gauge every 15~seconds for
1-2~minutes afterward, watching the system pump down to lower
pressures before the ion gauge bottoms out.

\begin{figure}
\begin{center}
\leavevmode
\includegraphics[angle=-90,width=0.85\linewidth]{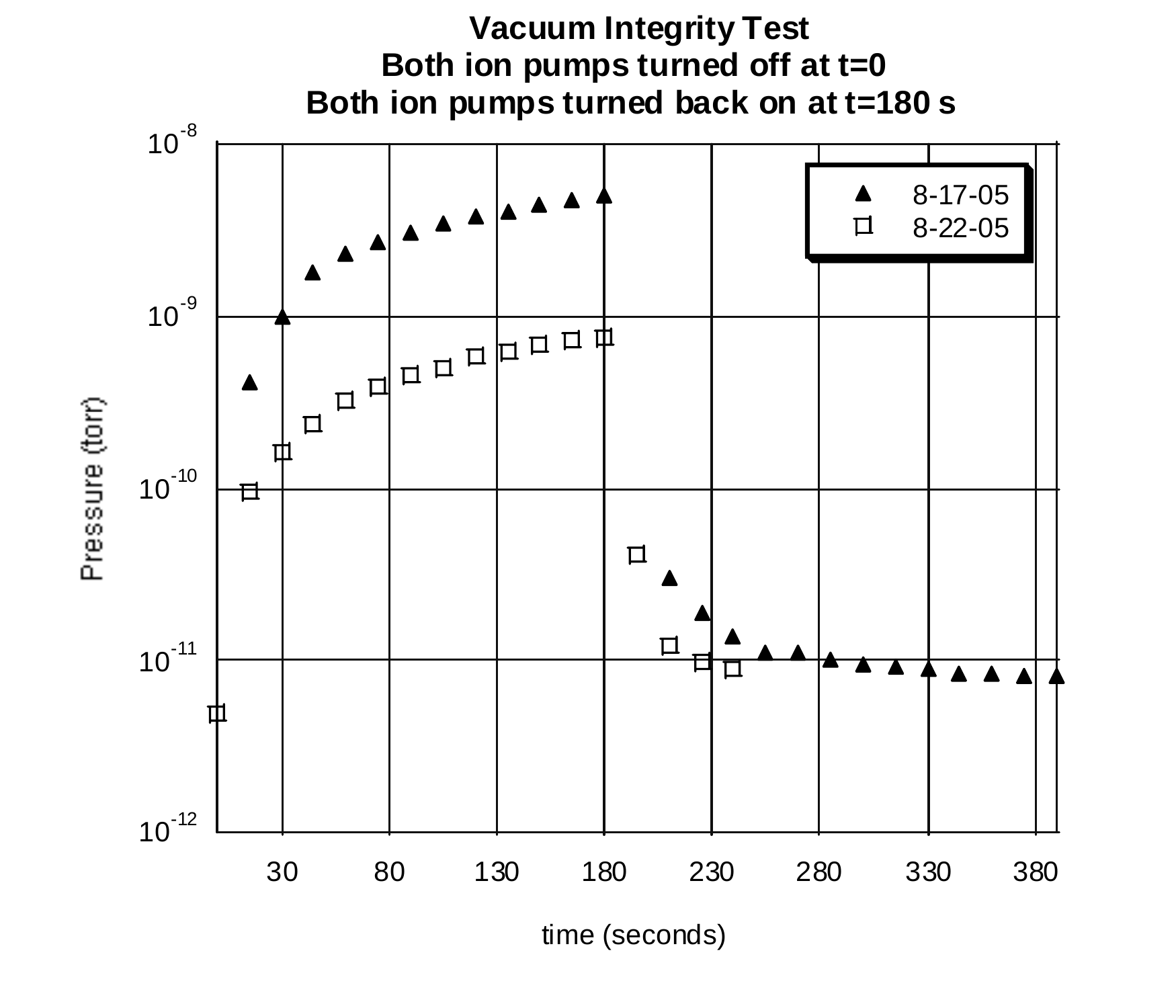}
\end{center}
\caption[Vacuum integrity test]{Vacuum integrity test.  Semi-log
plot of the ion gauge pressure [torr] vs. time [seconds].  Data
taken on 8/17/2005, immediately after the bake, is shown in solid
triangles. Data taken on 8/22/2005, one week after the bake, is
shown in open squares.  Both ion pumps were turned off at $t=0~s$,
and the pressure on the ion gauge began to rise immediately.  At
$t=180~s$, both ion  pumps were turned back on, the pressure jumped
back down and continued to decrease over time.  The eventual rise in
pressure (at $t=180~s$) on 8/22/2005, only 5 days after the original
data, is almost an order of magnitude lower in pressure than it was
on 8/17/2005.} \label{fig:Vacuum-integrity}
\end{figure}

When the ion pumps are turned off, the eventual value and speed of
increase in pressure yields information on the quality of the vacuum
system and how well it can maintain low pressures without any
pumping. When the ion pumps are turned back on, the rate of decrease
in pressure tells us how well the ion pumps are able to perform at
pulling the system back down to lower pressures. A plot of ion gauge
reading vs.\ time for this test both immediately after the final
bake and one week after it are shown in
Figure~\ref{fig:Vacuum-integrity}.  After the system had been pumped
on for only a few days, the pressure only rose to $10^{-9}$~torr
after 3~minutes, whereas on the final day of the last bake, before
we had plugged the leak with VacSeal, the pressure rose to
$10^{-7}$~torr after only 2~minutes.

% background limited lifetime of MOT?

\section{Lasers}
\begin{figure}
\begin{center}
\leavevmode
\includegraphics[angle=-90,width=0.4\linewidth]{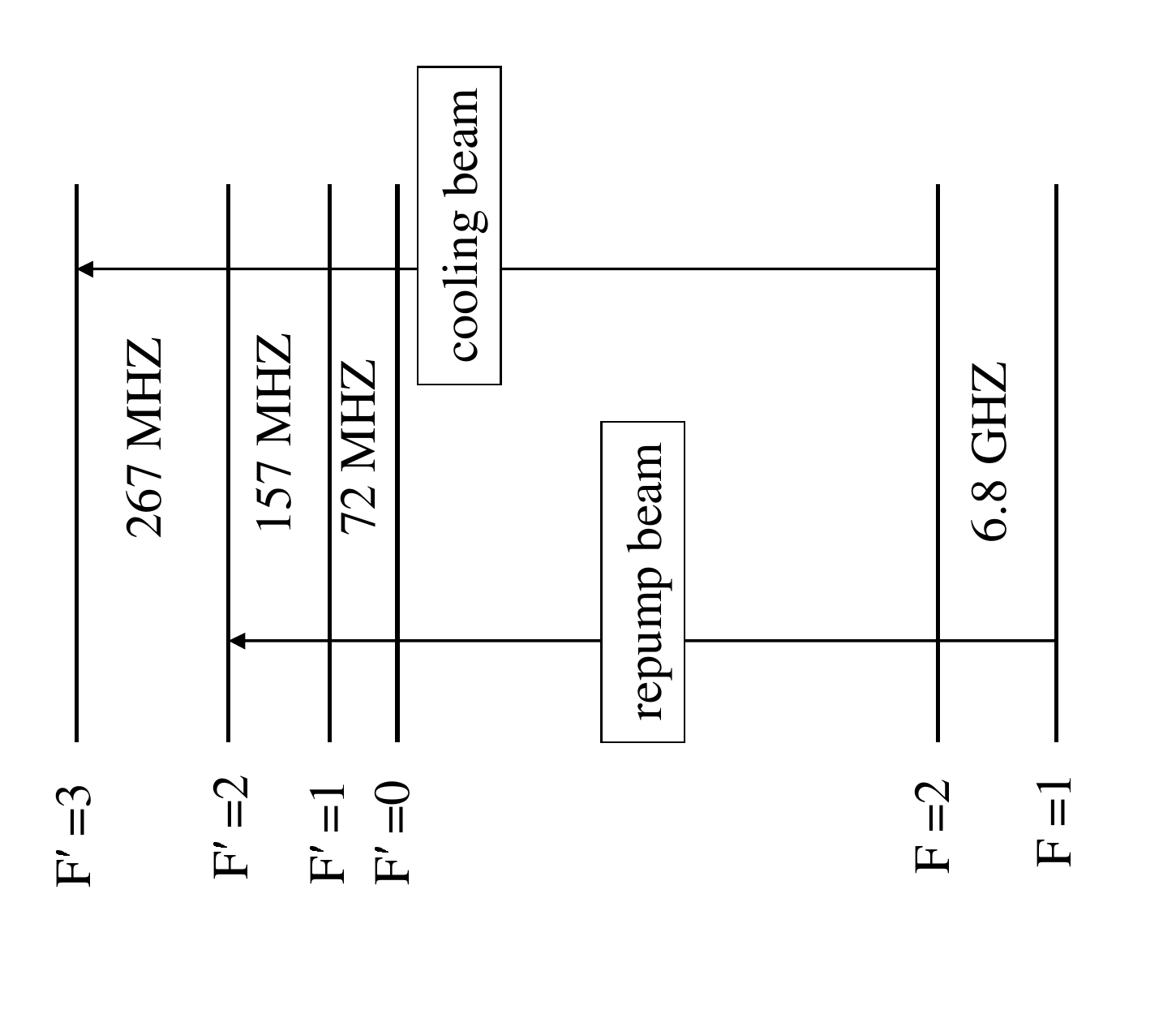}
\end{center}
\caption[$^{87}$Rb energy levels]{$^{87}$Rb energy levels}
\label{fig:Rb-levels}
\end{figure}

Our experiment uses four lasers to cool, manipulate, and probe
$^{87}$Rb atoms. The transition frequencies necessary for laser
cooling of $^{87}$Rb are shown in Figure~\ref{fig:Rb-levels}, this
and much more information on the $^{87}$Rb D line transition can be
found in the useful online resource paper written by Dan Steck
\cite{steck2003rld}.

The titanium:sapphire laser used for laser cooling $^{87}$Rb atoms
will be described in Section~\ref{subsection:mbr}.  Two home-made
diode lasers are used in our experiment; the construction method
used for our diode lasers is described in
Section~\ref{subsection:diodes}. The diode laser used as the repump
beam will be described in Section~\ref{subsection:repump}, and the
diode laser used as the probe beam will be described in
Section~\ref{subsection:probe}.  The diode laser used as a beam to
create an optical potential in our experiment will be described in
Section~\ref{subsection:making}.

\subsection{MBR laser and lock} \label{subsection:mbr}
% see 6/10/2004 in lab notebook
The workhorse laser in our lab is the Coherent Monolithic Block
Resonator \mbox{MBR-110} continuous-wave titanium:sapphire ring
laser, pumped by a Verdi V-10. This laser has a maximum output power
of 1.5~W, is tunable over 600-1000~nm, and can be locked to an
external frequency reference. The MBR laser is locked to a
stabilized internal reference cavity; the external lock operates by
making the internal reference cavity mimic an external error signal.
In our case, we generate an error signal based on a saturated
absorption signal from a $^{87}$Rb vapor cell, and the MBR laser
locks to this error signal.

Other than the frequency tunability, the principal advantage of
using the MBR laser is the large amount of output power in
comparison to that of diode lasers; the MBR output is used as the
laser cooling beam, and high power is helpful in loading a large
MOT.  At 780~nm, the MBR has an output of $\sim$1.1~W, and after the
output beam passes through an 80~MHz double-pass Acousto-optical
Modulator (AOM) and a single-mode fiber we still have a total of
450~mW to use for the cooling beams. This laser system is not
without its disadvantages; it is large and costly and requires
periodic maintenance of the optics in the cavity and electronics in
the feedback loops.  A detailed description of many of the the
optics and electronics problems and solutions associated with the
MBR laser is given in Appendix A.

\begin{figure}
\begin{center}
\leavevmode
\includegraphics[angle=-90,width=1\linewidth]{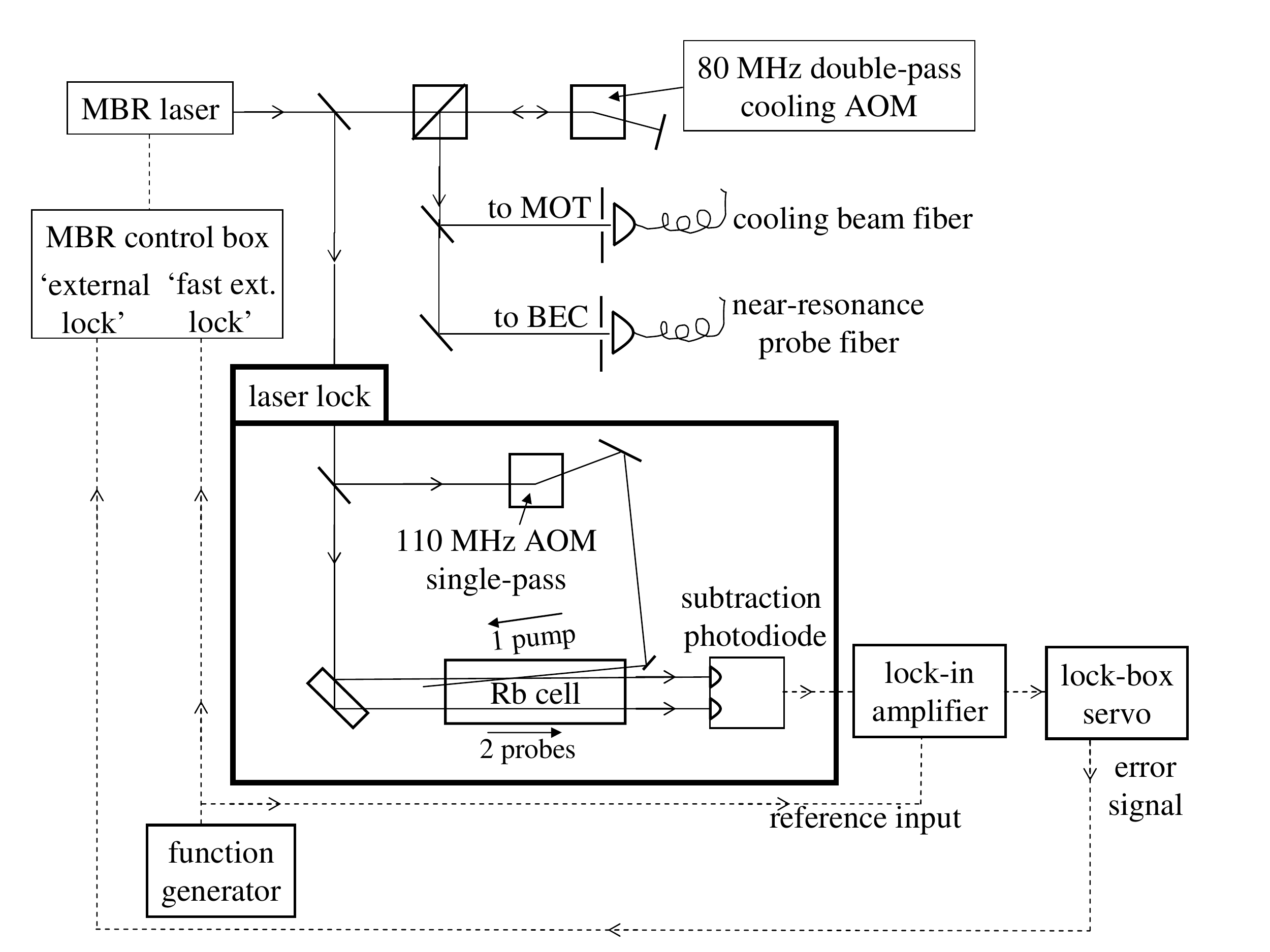}
\end{center}
\caption[MBR laser lock]{Optics and electronics used in the MBR
laser lock. Laser light is shown as a solid line and electrical
signals are shown as dashed lines.  The atomic laser lock is the
portion of the schematic enclosed in a bold box.  The optical output
of the entire apparatus is coupled into the cooling beam fiber and
the near-resonance probe fiber.  Both AOMs shown operate at an
upshift. Mechanical shutters are placed before both optical fibers.}
\label{fig:MBR-lock}
\end{figure}

A schematic of the optics and electronics used to atomic lock the
MBR laser is shown in Figure~\ref{fig:MBR-lock}, with laser light
shown as a solid line and electrical signals shown as dashed
lines\footnote{For additional information, see the lab notebook
entry on 6/10/2004.}. A small portion of the MBR output light is
sent to the atomic laser lock (enclosed within a bold box in
Figure~\ref{fig:MBR-lock}), but the majority of the MBR output light
is sent through the 80~MHz double-pass cooling AOM, operating at an
upshift.  Most of this light is sent to the MOT through the cooling
beam fiber, but a small amount is sent to the BEC cell via the
near-resonance probe fiber. The atomic laser lock is responsible for
keeping the MBR laser in lock; this laser is used simultaneously as
the laser cooling beam and a near-resonance probe beam for imaging
the BEC.

The portion of the light that enters the laser lock gets split into
a pump beam and a probe beam using a beamsplitter. The pump beam
passes through a single-pass 110~MHz AOM operating at an upshift.
The probe beam deflects off a thick plate of plexiglass which
provides reflections from both the front and back surfaces,
producing two copropagating probe beams that are spatially separated
by $\sim$4~mm. One of these probe beams interacts with the
counter-propagating pump beam in the Rb vapor cell and provides a
saturated absorption signal to one of the photodiodes. The other
probe beam simply passes through the Rb vapor cell and provides a
background Doppler absorption signal.

The signals from the two probe beams in the laser lock get
subtracted in the photodiode subtraction circuit \cite{hobbs2000beo,
hobbs1991rsn, hobbs1997ulm, hobbs2001pfe, graeme1996pao}; this
results in a signal consisting of only the saturated absorption
dips. The derivative of this signal is used as the error signal that
is applied as feedback to the MBR laser.  We differentiate this
signal using a lock-in amplifier and function generator.  A function
generator outputs a sine wave signal at 70~kHz, which provides the
modulation that we `lock to' using the lock-in amplifier.  This
signal inputs into the `fast external lock' input on the MBR control
box and directly modulates\footnote{This provides a small amount of
Frequency Modulation (FM) to the lasing frequency of the MBR laser.
The amount of FM is so small that it does not affect our ability to
use this laser for the purposes of laser cooling and absorption
imaging.} the tweeter mirror, mirror M3 of the MBR cavity. The
70~kHz signal is applied both to the MBR control box and to the
`reference input' of the lock-in amplifier. The output of the
photodiode subtraction circuit is applied to the `signal input' of
the lock-in amplifier. The output of the lock-in amplifier will then
be a differentiated version of the saturation absorption dips.

We lock to the crossover dip between the $|F=2 \rightarrow
F'=2\rangle$ transition and the $|F=2 \rightarrow F'=3\rangle$
transition. We use the lock-in output as the error signal that is
sent to the `lock-box servo' electronics box.  After an appropriate
adjustment of the amplitude and offset, this box is responsible for
switching on the laser lock by feeding the error signal back to the
`external lock' input on the MBR control box.  This forces the
internal reference cavity of the MBR laser to mimic the error signal
coming from the lock-box servo, and locks the laser to the atomic
reference.

We have found that using single-mode, polarization maintaining
optical fibers to couple light from the lasers to the experiment is
a good way of providing a stable, clean beam. Any optical
misalignments that occur can usually be fixed by simply optimizing
the fiber alignment, and using a single-mode fiber is an excellent
way of generating a beam with high mode quality.

\subsection{Diode laser construction} \label{subsection:diodes}
\begin{figure}
\begin{center}
\leavevmode
\includegraphics[angle=-90,width=0.9\linewidth]{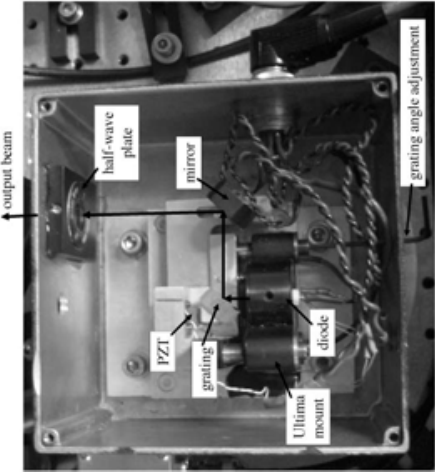}
\end{center}
\caption[Photograph of diode laser]{Top-down photograph of a diode
laser box, showing all the components in the box and the beam path
highlighted in black arrows.  The laser diode itself is housed
within a black Newport Ultima mount, to which is attached an
aluminum plate.  The grating is glued onto a 45$^\circ$ wedge that
is glued onto the PZT; the PZT is glued onto the aluminum plate. The
aluminum plate also holds the mirror which deflects the output beam
after the grating. The output beam then passes through a half-wave
plate before it exits the diode laser box.  The thermistor (not
visible from the top) is connected to the Ultima mount, very close
to the diode laser.  The TEC (also not visible) is placed under a
different aluminum plate under the Ultima mount.  A removable allen
key can access the horizontal angle adjustment of the Ultima mount
to change the grating angle. The entire box sits on top of a large
heat sink.} \label{fig:Diode-laser}
\end{figure}

% Elaine's original write-up is in 5/22/2002 in lab notebook
% more detail can be found in Tim's notebook
Two diode lasers have been constructed using a home-made
design\footnote{For additional information, see the lab notebook
entry on 5/22/2002.} and are used as the repump laser and probe
laser. The design is a modified version of the Littrow configuration
design~\cite{arnold1998sec, hawthorn2001lct}. A top-down photograph
of a diode laser box is shown in Figure~\ref{fig:Diode-laser}.

The diode laser itself is a Sharp GH0781JA2C 120~mW diode laser,
which typically has a free-running wavelength of 784~nm. An external
feedback cavity employing an Edmund Optics model 43774 holographic
grating is mounted on a Thorlabs Piezo-Electric Transducer (PZT).
The laser and external grating are mounted on a modified Newport
Ultima U100-P kinematic mirror mount.  The laser beam deflects off
the grating and an external mirror in such a way that adjustment of
the grating angle does not cause a change in the output beam angle.
There is a small translational shift in the output beam position,
but this is usually negligible. The Ultima mount is placed on top of
a Thermo-Electric Cooler (TEC), which sits on top of a large heat
sink. Using the TEC, we cool the temperature of the aluminum mount
(and hence the laser diode) to approximately 16$~^\circ$C. We have
found that it is essential to place the thermistor as close to the
diode laser as possible to prevent oscillations of the temperature
controller. Temperature cooling is necessary in order to shift the
laser's operating wavelength closer to 780~nm; external grating
adjustment is necessary in order to fine-tune the laser to be in
resonance with $^{87}$Rb.  The entire unit is housed within a box to
prevent fluctuations due to air currents.  There are small holes in
the box to allow for light output and external grating adjustment.
After leaving the diode laser box, the output beam passes through an
anamorphic prism pair to correct the aspect ratio of the beam.  The
beam then passes through an Optics For Research OFR IO-3C-781VLP
optical isolator that prevents optical feedback from returning to
the diode laser.

\subsection{Repump laser and lock} \label{subsection:repump}
\begin{figure}
\begin{center}
\leavevmode
\includegraphics[angle=-90,width=1\linewidth]{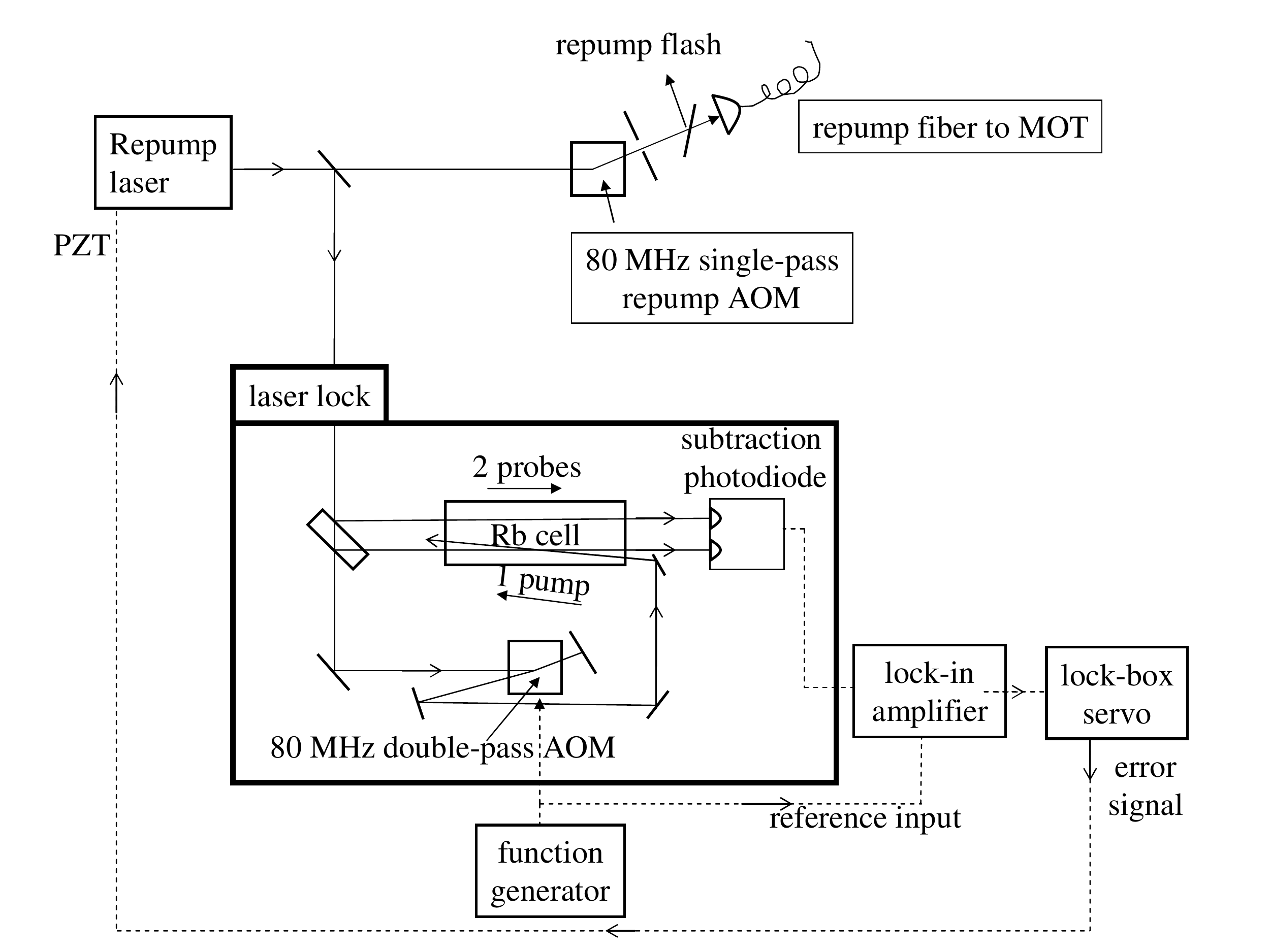}
\end{center}
\caption[Repump laser lock]{Optics and electronics used in the
repump laser lock. Laser light is shown as a solid line and
electrical signals are shown as dashed lines.  The atomic laser lock
is the portion of the schematic enclosed in a bold box.  Most of the
optical output of the entire system goes to the repump fiber, to be
sent to the MOT as the repump beam.  A small portion of light gets
picked off before this fiber and is sent to the science cell as the
repump flash beam, these two beams operate on the same AOM and
mechanical shutter. Both AOMs shown operate at a downshift.}
\label{fig:Repump-lock}
\end{figure}

% see 12/1/2005 in lab notebook
The repump laser is tuned to be in resonance with the $|F=1
\rightarrow F'=2\rangle$ transition and is used to optically pump
atoms out of $|F=1\rangle$ and into $|F=2\rangle$.  The repump laser
is used while creating a MOT and for pumping atoms into
$|F=2\rangle$ in order to image the BEC using absorption imaging.
The repump laser is locked to an atomic transition using a saturated
absorption scheme\footnote{For additional information, see the lab
notebook entry on 12/1/2005.} similar to that of the MBR laser,
shown in Figure~\ref{fig:Repump-lock}.

A small portion of the repump beam is sent to the laser lock using a
beamsplitter.  Just like in the MBR lock, a plexiglass plate is used
to pick off two probe beams; the transmitted beam becomes the pump
beam in this laser lock.  The two probe beams pass through the Rb
vapor cell and onto the two photodiodes of the subtraction
photodiode.  The pump beam passes through a double-pass 80~MHz AOM
operating at a downshift and then through the Rb vapor cell in the
opposite direction of one of the counter-propagating probe beams.
Just like in the MBR lock, one of the probe beams interacts with the
counter-propagating pump beam and produces the saturated absorption
dips, and the solitary probe beam provides a background Doppler
signal.

By changing the frequency of this 80~MHz double-pass AOM (i.e.\ by
changing it from 80~MHz to 77~MHz), the frequency of the pump beam
relative to the two probe beams changes; this shifts the location of
the saturated absorption dips in frequency space. We dither the
frequency of this 80~MHz double-pass AOM using a function generator
operating at 35~kHz. We change the \emph{frequency} of the pump beam
by changing the operating frequency of the AOM by a few MHz, but the
light output will also be \emph{frequency-modulated} by a small
amount at a \emph{rate} of 35~kHz. This produces a small amount of
FM on the pump beam, which dithers the location of the saturated
absorption dips by an amount detectable using the lock-in amplifier.
When using a scheme like this it is helpful to modulate the pump
beam, rather than the probe beam; otherwise one will also provide a
small amount of Amplitude Modulation (AM) onto the photodiode
signal, which could hinder the operation of the lock-in scheme.

The lock-in amplifier's output is a differentiated version of the
saturated absorption dips, and is sent to the lock-box servo. The
output of the lock-box servo, after an appropriate scaling and
offset adjustment, is used as the error signal that is fed back to
the PZT of the repump laser to stabilize its frequency.

Most of the light from the repump laser is sent through an 80~MHz
single-pass AOM operating at a downshift, a mechanical shutter, and
a single-mode fiber before being sent to the MOT cell. This AOM
allows for fast switching and also provides an amplitude adjustment
for the repump beam, which is necessary in the CMOT stage of the
experiment. After the shutter, a small portion of the beam is picked
off and sent to the science cell, to be used as the `repump flash'
to optically pump atoms into $|F=2\rangle$ in the science cell.

\subsection{Probe laser and lock} \label{subsection:probe}
\begin{figure}
\begin{center}
\leavevmode
\includegraphics[angle=-90,width=1\linewidth]{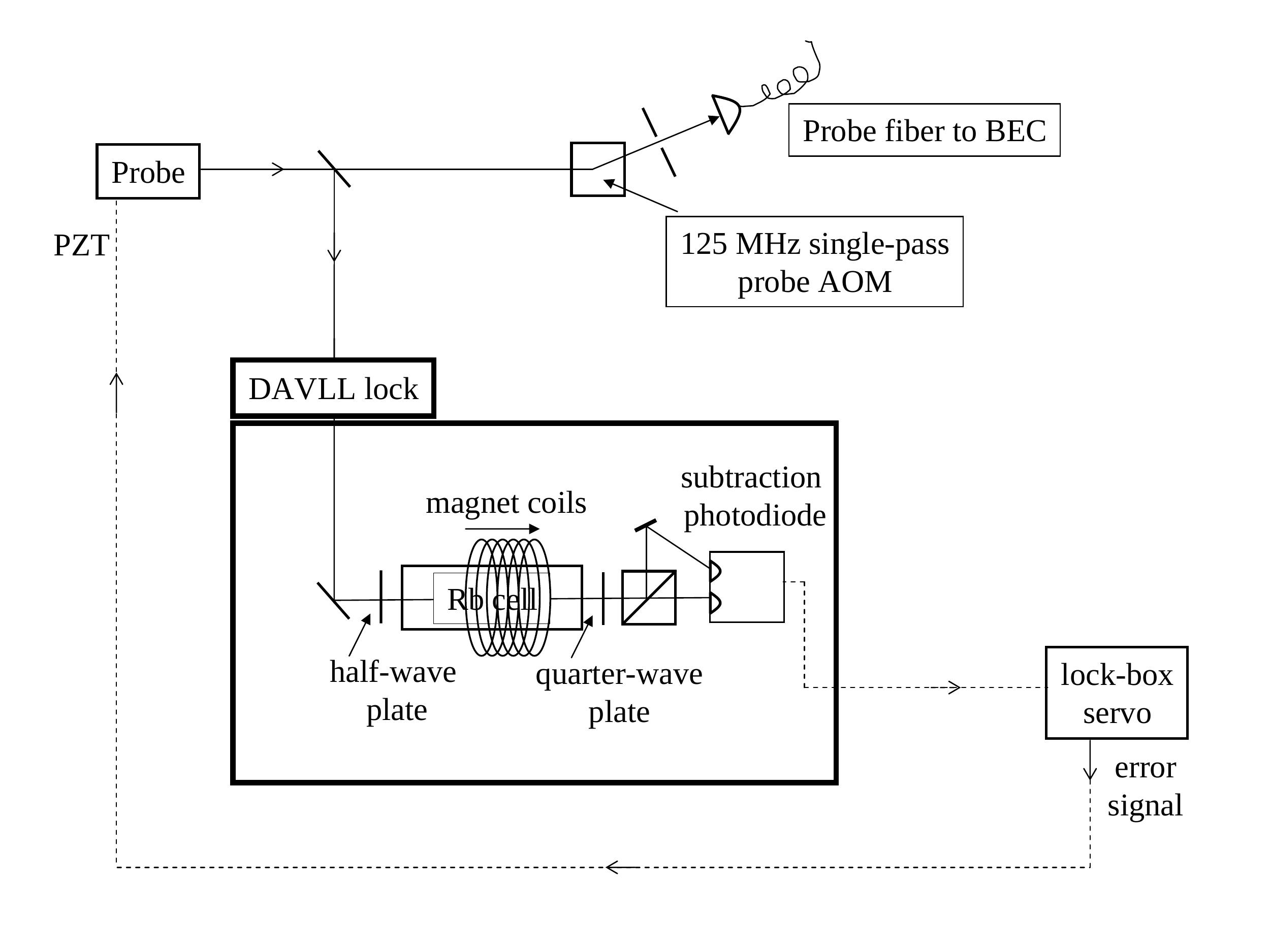}
\end{center}
\caption[Probe laser lock]{Optics and electronics used in the probe
laser lock. Laser light is shown as a solid line and electrical
signals are shown as dashed lines.  The DAVLL laser lock is the
portion of the schematic enclosed in a bold box.  The optical output
of the entire apparatus is sent to the probe fiber to be used for
imaging the BEC.  A mechanical shutter is placed before the optical
fiber.  The AOM operates at a downshift.} \label{fig:Probe-lock}
\end{figure}

% for details on DAVLL lock, see Tim's notebook 4/1
A probe laser\footnote{For additional information, see Tim McComb's
lab notebook entry on 4/1/2005.} used specifically for
phase-contrast imaging of the BEC was constructed using an identical
design as the repump laser.  This far-off-resonance probe beam is
used to image the trapped atoms in the science cell; this laser is
typically 800~MHz red-detuned from the
$|F=1~\rightarrow~F'=2\rangle$ transition.  This laser is locked to
an atomic transition using a Dichroic Atomic Vapor Laser Lock
(DAVLL) \cite{corwin1998fsd}; a schematic of the laser lock is shown
in Figure~\ref{fig:Probe-lock}.

Most of the probe laser output light passes through a 125~MHz
single-pass AOM operating at a downshift, a mechanical shutter, and
a single-mode optical fiber to be used to image the BEC in the
science cell.  The portion of the light used in the DAVLL lock first
passes through a half-wave plate.  Rotation of this half-wave plate
changes the polarization angle of the beam while keeping it linearly
polarized. This linearly polarized beam can be thought of as a
combination of $\sigma^+$ and $\sigma^-$ polarizations.  The DAVLL
lock requires a large uniform magnetic field inside the Rb vapor
cell.  This provides different Zeeman shifts to the atoms
interacting with the $\sigma^+$ and $\sigma^-$ components of the
beam.  We create this magnetic field by wrapping magnet wire around
the 1" outer diameter Rb glass vapor cell. We achieve a uniform
magnetic field of approximately 100~G inside the vapor cell by
running 3~A through the magnet coils. After the vapor cell, the beam
passes through a quarter-wave plate, which maps the $\sigma^+$ and
$\sigma^-$ components of the beam onto horizontal and vertical
linear polarizations.  Finally, a polarizing beamsplitter cube
separates these horizontal and vertical components and sends them
onto two different photodiodes. These signals are subtracted in the
photodiode subtraction circuit, and sent to the lock-box servo,
which atomic locks the probe laser by applying feedback to the PZT.

A major advantage of using the DAVLL lock is the simplicity of the
optical setup, the laser lock does not require any AOMs or lock-in
amplifiers and uses a minimum of components.  The lock is also
easily tunable over a very wide frequency range; we have made
measurements using phase-contrast imaging at detunings of
\mbox{$-1000$~MHz} to \mbox{$+800$~MHz}.  The major drawback in our
implementation of this locking scheme, however, is that the laser is
more susceptible to small frequency drifts over the course of the
day. This is a technical problem associated with our experimental
setup, and it not fundamental to the DAVLL technique.

\section{MOT}
Loading the Magneto-Optical Trap (MOT) is the first stage in a
series of sequential steps that we use to create a BEC.  This first
step collects, cools, and traps atoms from a room-temperature vapor
in the MOT cell. Efficient preparation of a large number of trapped
atoms at low temperatures sets the stage for the next two major
steps in the sequence: transferring the cloud of atoms into the
lower-pressure science cell, and then evaporatively cooling the
cloud to BEC. The MOT, Compressed MOT (CMOT), and optical pumping
stages, which all occur in the MOT cell, play a crucial role in
preparation of the cloud in order to optimize the number of atoms in
the eventual BEC.  Several review articles and books have discussed
the general principles underlying laser cooling and trapping
\cite{cohentannoudji2006nml, foot1991lca, metcalf1999lca}; the
original MOT was described by Raab \emph{et al}.\ in 1987
\cite{raab1987tns}; and a description of achievable MOT size vs.\
cooling beam parameters was discussed by Gibble \emph{et al}.\ in
1992 \cite{gibble1992imo}.

\subsection{Magneto-optical trapping using diverging beams} \label{subsection:diverging}
% diverging vs. collimated beams write-up is on 8/6/2004, but this
% was before the bake, so we had K in the cell =(
Our MOT differs from a conventional vapor cell MOT in that we use
diverging beams as the laser cooling beams, rather than collimated
beams. It is more common to use large collimated beams as the
cooling beams to ensure efficient sampling of a large capture volume
in the MOT cell. One way to generate six independent 2" diameter
beams requires the use of six separate 2" lenses, each lens part of
an individual beam-expanding telescope. This method requires several
2" diameter optical components on the optical table in the vicinity
of the MOT cell, which is a rather inefficient use of space as well
as money.

\pagebreak

\begin{figure}
\begin{center}
\leavevmode
\includegraphics[angle=-90,width=1\linewidth]{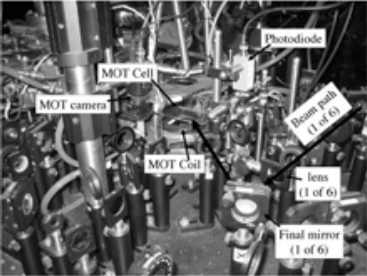}
\end{center}
\caption[Photograph of MOT optics]{Photograph of MOT optics.  The
MOT cell is highlighted and sits in between two copper plates.  Each
of the two MOT coils (one is highlighted) are attached to the
water-cooled copper plates.  The photodiode used for collecting
fluorescence from the MOT is highlighted.  One of the six beam paths
used for the MOT beams is highlighted in bold arrows; the lens used
to create a diverging beam and final mirror are shown for this beam
path. A CCD camera is mounted vertically to observe the MOT through
a reflection off a mirror mounted at 45$^\circ$.}
\label{fig:MOT-optics}
\end{figure}

We have instead generated six diverging beams
by sending six 2~cm diameter beams into six separate 1" diameter
positive lenses that first bring the beams to a focus and then allow
the beams to expand. These 100~mm focal length lenses are placed
400~mm from the center of the MOT cell, far enough away to allow the
diverging beams to expand to a $\sim$10~cm diameter at the location
of the MOT.  A photograph of the MOT cell and neighboring optics is
shown in Figure~\ref{fig:MOT-optics}.

The diverging beams make the MOT less susceptible to optical
misalignment and standing-wave problems caused by
counter-propagating collimated beams.  Another advantage of using
diverging beams is the relative insensitivity to small mirror
misalignments, which have a negligible effect on the MOT and on
loading the MOT into the initial magnetic trap.

This method of using diverging beams is similar to a technique used
to create an atom trap that relies solely on optical pumping and
does not require a magnetic field \cite{bouyer1994atr}, a phenomenon
we have observed with our diverging beams as well.

We use a total of 450~mW in the six cooling beams, resulting in an
intensity of 2.9~mW/cm$^{2}$ = 1.8 $I_{sat}$ at the location of the
MOT. The MOT beams are tuned 4~$\Gamma$ below the $|F=2 \rightarrow
F'=3\rangle$ transition during MOT loading.

The repump beam travels along the same path as the four horizontal
cooling beams, resulting in four diverging beams with a diameter of
$\sim$10~cm at the MOT location. We use a total of 30~mW in the
repump beam, resulting in a total intensity at the MOT of
0.19~mW/cm$^{2}$ = 0.12 $I_{sat}$. The repump beam is tuned to be in
resonance with the $|F=1 \rightarrow F'=2\rangle$ transition.

\begin{figure}
\begin{center}
\leavevmode
\includegraphics[width=0.6\linewidth]{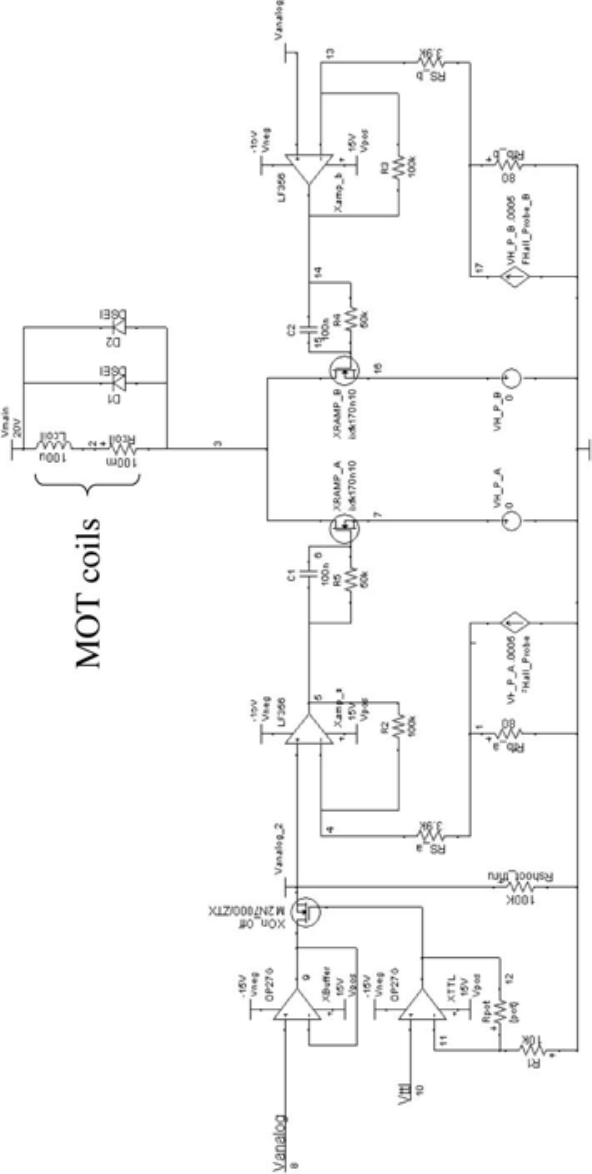}
\end{center}
\caption[Schematic of MOT circuit]{Schematic of the MOT circuit.
 The power supply is shown as Vmain.  The MOT coils are depicted as
a resistor in series with an inductor, a fly-back power diode is
placed in parallel with the coil.  The coil is driven using two
load-balanced power MOSFETs and two Hall Probes, described in detail
in Chapter~\ref{chapter:transfer}.} \label{fig:MOT-circuit}
\end{figure}

A magnetic field gradient is needed to provide a restoring force
that traps the atoms cooled by the laser beams; the magnetic field
gradient produced by our MOT coils is 8~G/cm axially (vertically)
during the MOT loading phase.  The basic circuit used to drive the
MOT coils makes use of a linearized current-voltage relationship by
applying feedback onto the gate of a power MOSFET; a similar circuit
will be described in much more detail when describing the magnetic
transfer coils in Chapter~\ref{chapter:transfer}.  A schematic of
the MOT circuit is shown in Figure~\ref{fig:MOT-circuit}.

Using six independent counter-propagating beams requires the use of
six individual 1" diameter quarter-wave plates.  Small rotations of
these waveplates have a negligible effect on the number of atoms in
the MOT, but can affect the loading efficiency into a magnetic trap.
We also use three orthogonal magnet coils to null out the background
magnetic field, mostly caused by the permanent magnets of the ion
pumps. Small adjustments of these nulling coils will shift the
position of the MOT and can cause an observable difference in
loading into the initial magnetic trap.

A photodiode placed near the MOT cell records fluorescence from the
MOT.  A collection lens is placed as close to the MOT cell as
possible, and focuses the fluorescence from the MOT onto a
photodiode. Based on the solid angle subtended by the collection
lens, we can calculate the percentage of the total amount of
fluorescence given off by the MOT that arrives at the photodiode.
Using this and the scattering rate caused by the cooling beams, we
can calculate the number of atoms in the MOT.  The total power $P$
emitted from the MOT in all directions is equal to the number of
atoms $N$ in the MOT multiplied by the scattering rate multiplied by
the energy per photon $h \nu_{Rb}$~\cite{metcalf1999lca}.

\begin{equation}P = N ~\frac{1}{2} \frac{I/I_{sat}}
{\left [1+I/I_{sat}+4 \cdot(\Delta/\Gamma)^2 \right ]}
\frac{h\nu_{Rb}}{\tau_{nat}}
\end{equation}
The saturation intensity $I_{sat}$ is defined as
\begin{equation}
I_{sat} = \frac{1}{2} \frac{h\nu_{Rb}} {\sigma_0 \cdot \tau_{nat}}
\end{equation}
where $h\nu_{Rb}$ is the energy of a photon, $\Delta$ is the
detuning from resonance of the scattering light, $\Gamma$ is the
natural linewidth of the atomic transition, $\tau_{nat}$ is the
scattering time, and $\sigma_0$ is the on-resonance absorption
cross-section. This is a convenient definition of the saturation
intensity $I_{sat}$, written as an energy ($h \nu_{Rb}$) per unit
area ($\sigma_0$) per unit time ($\tau_{nat}$) \cite{steck2003rld}.

Knowing the solid angle of collected light and the intensity and
detuning of the cooling beams, we calculate the number of atoms in
the MOT to be $N=3 \cdot 10^9$ atoms.  We can fit the observed rise
in MOT fluorescence during MOT loading to the following expression:
\begin{equation}
N(t)=N (1-e^{-t / \tau_{MOT}})
\end{equation}
This results in a MOT filling time of $\tau_{MOT}=15$~s with the Rb
dispensers running at 3.4~A.  The cloud size is $\sim$1~cm in
diameter, easily visible to the human eye despite the fact that the
$^{87}$Rb fluorescence given off by the MOT is at 780~nm, in the
near-infrared.

% MOT N vs. cooling beam intensity, detuning, B-field??
% all of that was done with the K cell

\section{CMOT}
The compressed MOT (CMOT \cite{petrich1994bac}) stage of our
experiment spatially compresses the MOT and causes a temporary
increase in density, which aids in the loading of ultracold atoms
into a magnetic trap. Several factors are important in the CMOT
stage, and we have noticed that periodic fine-tuning of CMOT
parameters aids greatly in the number of atoms in the eventual BEC,
a quantity that can be subject to periodic degradation.

The desired compression in the CMOT stage can be achieved by
allowing a population of atoms to build up in the $|F=1\rangle$
state.  This allows them to stay out of the $|F=2 \rightarrow
F'=3\rangle$ absorption events caused by the cooling beams. The
dominant player that causes this to occur is the repump intensity;
by turning the repump intensity down for a short period of time
while keeping the cooling beams on, atoms do not get pumped out of
the $|F=1\rangle$ state. The other major player in this process is
the detuning of the cooling beams. By making the cooling beams even
more red-detuned from the $|F=2 \rightarrow F'=3\rangle$ transition,
we decrease the scattering rate and achieve the same result.

The CMOT phase is 40~ms long and consists of ramping the cooling
beam detuning from -4~$\Gamma$ to -7~$\Gamma$ over this time (the
intensity also decreases by $\sim 35 \%$), ramping the magnetic
field gradient from 8~G/cm to 0~G/cm, and switching the repump power
from 30~mW (full power) to 260~$\mu$W for the duration of the CMOT
phase.

As stated previously, optimization of the magnetic field nulling
coils can have a major effect on loading atoms into the initial
magnetic trap, the step which follows the CMOT stage. This is partly
because adjustments of the nulling coils can physically move the MOT
into or out of position with the $B=0$ point, where the center of
the magnetic trap exists.  This is especially important in the CMOT
stage, where the lower magnetic field gradient makes the $B=0$ point
more easily shifted by the fields due to the nulling coils.

Our experiment does not include an optical molasses phase, as we
have noticed that any decrease in temperature afforded by a molasses
phase is totally offset by an accompanying decrease in density due
to the lack of magnetic field gradient in the molasses phase. The
physical expansion of the cloud, even in a short optical molasses
phase, has hindered our ability to load cold atoms into a magnetic
trap for the preparation of a BEC.

\section{Optical Pumping}
After the CMOT phase, the cooling beams (still at the CMOT detuning)
stay on for an additional 1~ms, during which time the repump beam
and magnetic field gradient are both off.  This forces all the atoms
to populate the $|F=1\rangle$ state.  Only one of the three magnetic
sublevels, the $|F=1,m_F=-1\rangle$ state, is magnetically
trappable, so when we turn on the magnetic trap in the next stage of
the experiment, we can only expect to trap atoms that get projected
into this state. Since there is no optical pumping into a particular
Zeeman sublevel, we can expect to trap about 1/3 of the atoms.

In the past, we have attempted to install Zeeman-sublevel pumping
into the $|F=1,m_F=-1\rangle$ state by using a linear magnetic bias
field, a $|F=1 \rightarrow F'=1\rangle$ $\sigma^-$ optical pumping
beam, and a $|F=2 \rightarrow F'=2\rangle$ $\sigma^-$ pumping
repumper beam. However, we have found that the small gains in number
of trapped atoms is not worth the effort involved in this step.

\section{Initial Magnetic Trap}
After the MOT loading, CMOT, and optical pumping phases of the
experiment, the atoms are transferred into the initial magnetic trap
formed by the MOT coils.  This occurs by turning on the axial
magnetic field gradient in the MOT coils to 40~G/cm, and then
ramping the strength of the field gradient up to 180~G/cm over
100~ms. The MOT `catch' value of 40~G/cm, the initial strength of
the quadrupole trap, should be relatively low in an attempt to
approximately mode-match the initial magnetic trap to the size of
the atomic cloud.

This concludes the first major stage in our BEC formation process:
loading a MOT and magnetically trapping atoms in the MOT cell.  The
next major stage, magnetically transferring atoms into the science
cell, will be described in Chapter~\ref{chapter:transfer}.

%% file: Chap_Transfer.tex
\chapter{MAGNETIC TRANSFER OF ATOMS TO SCIENCE CELL} \label{chapter:transfer}

\section{Introduction}
Chapter~\ref{chapter:experimental} described the first stage in our
BEC formation sequence, loading atoms in a MOT and trapping them in
the initial magnetic trap in the MOT cell.  The next stage in the
BEC formation process, magnetic transfer of atoms into the science
cell, is described in this chapter.

The organization of this chapter is as follows:
Section~\ref{section:overview-transfer} provides a conceptual
overview of the transfer system and components used.
Section~\ref{section:hardware} provides a description of all the
hardware (magnet coils, power supply, and electronic circuits) used
in the transfer process. Section~\ref{section:sim} provides a
description of the calculation used to determine the sequence of
current ramps sent to the magnet coils used in the transfer process.
Finally, section~\ref{section:revisited} summarizes the transfer
system and lists some of the advantages and disadvantages of our
method.

%\footnote{Throughout this chapter it should be understood that the
%word \emph{coil} is meant to imply \emph{coil pair} when describing
%the magnetic transfer coils.}

\section{Overview of the Magnetic Transfer System} \label{section:overview-transfer}
A pair of coils wired in anti-Helmholtz configuration provide the
simplest method for magnetically trapping neutral atoms: a
quadrupole magnetic trap \cite{migdall1985fom}.  This configuration
provides a $|\vec{B}|=0$ point on-axis in between the coils and a
spatially varying magnetic field that increases linearly from the
center of the trap.

\pagebreak

\begin{figure}
\begin{center}
\leavevmode
\includegraphics[angle=-90,width=1\linewidth]{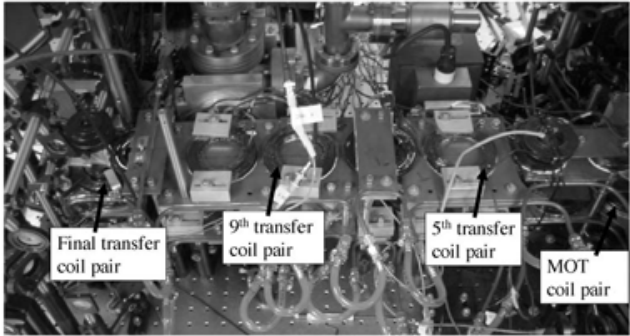}
\end{center}
\caption[Top-view photograph of magnetic transfer coils]{Photograph
of magnetic transfer coils.  The MOT coil pair,
5\textsuperscript{th}, 9\textsuperscript{th}, and final
(14\textsuperscript{th}) transfer coil pair are highlighted.  The
horizontal distance between the center of the MOT coil pair and the
center of the final transfer coil pair (BEC location) is 76.2~cm.}
\label{fig:coils-top}
\end{figure}

In order to magnetically transfer atoms through the transfer tube
that connects the MOT cell to the science cell, we have expanded on
this idea by operating current through an array of 14 pairs of
anti-Helmholtz coils placed above and below the transfer tube.  A
photograph of the transfer coils is shown in
Figure~\ref{fig:coils-top}.

By ramping the array of transfer coils on and off such that the
$|\vec{B}|=0$ point moves down the transfer tube, the magnetically
trapped atoms will follow the potential minimum down the transfer
tube and into the science cell.  This procedure is based upon the
magnetic transfer system described by Greiner \emph{et al}.\ in
2001~\cite{greiner2001mtt}. This procedure requires three pairs of
coils to be on at a time, creating a magnetic trap that is elongated
in the transfer direction.  The coils are ramped on and off in
sequence such that initially coil pairs 1, 2, and 3 are on; then 2,
3, and 4; then 3, 4, and 5; etc.

%Because only three coils are on at a time, the coils can be divided
%into three groups: Group A consists of coils 1, 4, 7, 10, and 13;
%Group B consists of coils 2, 5, 8, 11, and 14; Group C consists of
%coils 3, 6, 9, and 12.

%The overall circuit used to operate current through the coils,
%referred to as the `transfer circuit', requires three analog ramps
%to control the amount of current flow in each group and 14 digital
%signals to turn the appropriate coils on and off.

The components involved in the magnetic transfer system can be
divided into two categories: (1) the hardware and electronics
components that operate to actually control the transfer system on a
day-to-day basis, which will be described in
Section~\ref{section:hardware}; and (2) the calculation that
determines the current sequence used for transfer. The results of
this calculation, described in Section~\ref{section:sim}, can be
used over and over to transfer atoms unless a change in the transfer
procedure is desired.

\section{Hardware Used in the Magnetic Transfer System} \label{section:hardware}
\subsection{Overview of components used} \label{subsection:components}
\begin{figure}
\begin{center}
\leavevmode
\includegraphics[angle=-90,width=1\linewidth]{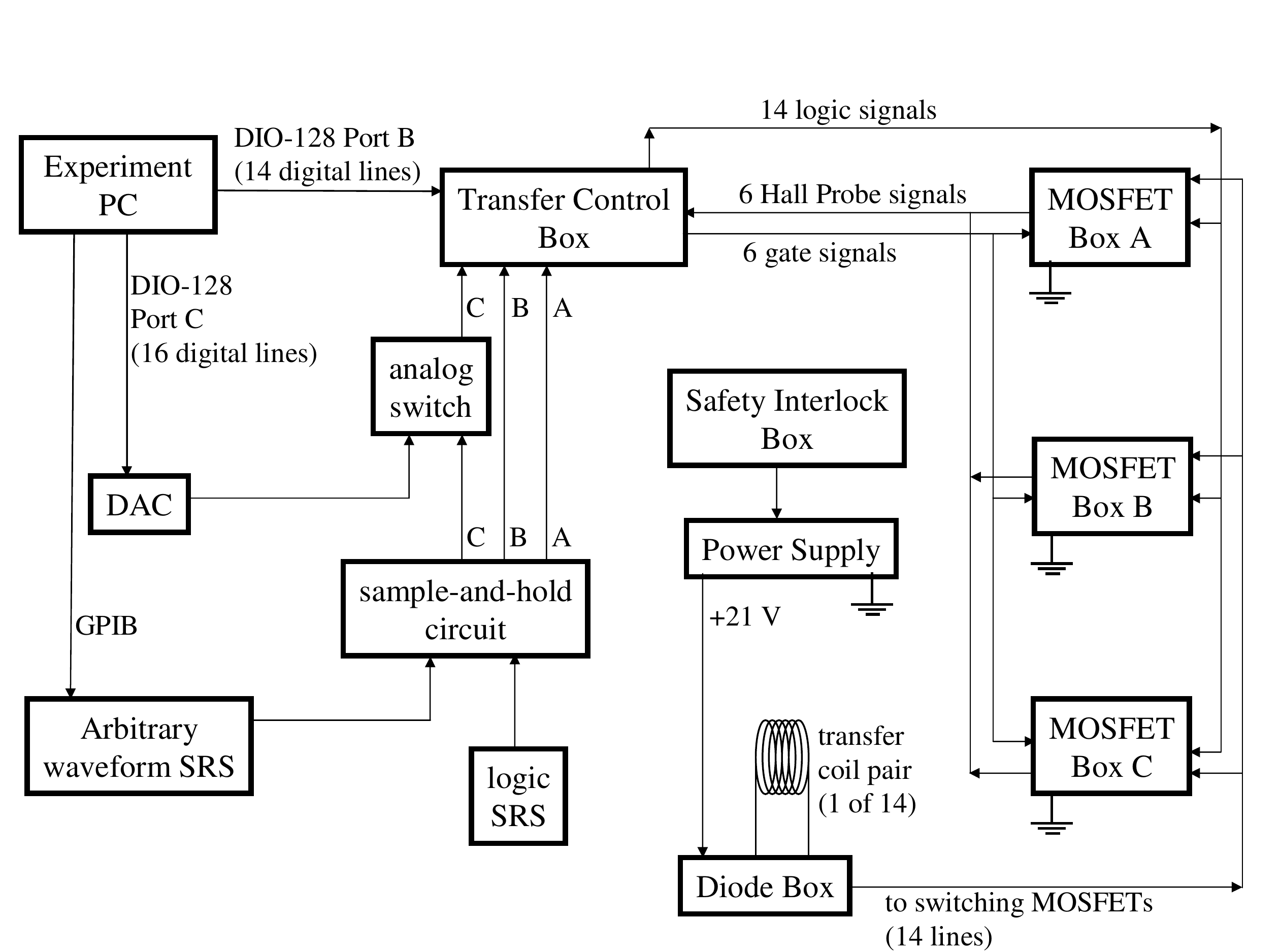}
\end{center}
\caption[Schematic of the overall transfer system]{Schematic of the
overall transfer system.  This figure shows all the components that
make up the `transfer circuit'; this includes many electronics
boxes, a power supply, 14 transfer coil pairs, two function
generators, and the experiment PC.} \label{fig:Overall-transfer}
\end{figure}

A schematic of all the hardware and electronics components used in
the magnetic transfer system is shown in
Figure~\ref{fig:Overall-transfer}.  This complete schematic shows
all the components which together comprise what we refer to as the
`transfer circuit', these include many electronics boxes, a power
supply, 14 transfer coil pairs, two function generators, and the
experiment PC.

To understand the operation of the transfer circuit, first consider
the role of the `transfer control box', which is the mastermind
behind the entire system.  The transfer control box receives three
time-varying analog inputs that control the amount of current flow
through the three coil pairs that are in operation at any given
time. These three signals, labeled A, B, and C in
Figure~\ref{fig:Overall-transfer}, are the result of a calculation
(described in Section~\ref{section:sim}) that determines the current
ramps needed for magnetic transfer.  These three analog signals
originate from the experiment PC and are loaded onto the arbitrary
waveform SRS function generator.  We use the sample-and-hold circuit
to split up the signal from the SRS into three separate ramps, a
process that is described in Section~\ref{subsection:sh}. The
transfer control box also receives 14 digital inputs from the
experiment PC, which act to turn on and off the coils in sequence.

The transfer control box takes these three analog and 14 digital
inputs and uses them to control the 14 pairs of transfer coils.  It
does this by sending 14 logic signals to the 14 switching MOSFETs
which reside in the MOSFET boxes (described in
Section~\ref{subsection:mosfet}); these signals act to turn on and
off the individual coils.  Two Hall probes reside in each MOSFET box
to monitor the amount of current flow through each of the three coil
pairs that are operational at any given time.  The signals from the
Hall probes are then fed back into the transfer control box, which
regulates the amount of current flow through each coil group by
sending the appropriate voltage to the gate of a ramping MOSFET in
the MOSFET boxes. The principles of operation of this feedback loop
are described in more detail in Section~\ref{subsection:transfer}.

Positive current flows from the power supply, through the transfer
coils, and into the MOSFET boxes, which contain a connection to
ground. There are fly-back power diodes (described in
Section~\ref{subsection:diode}) connected in parallel with each
transfer coil pair to dissipate voltage spikes.  A safety interlock
box will inhibit the power supply and prevent the flow of current if
any of the interlock conditions are met, as described in
Section~\ref{subsection:interlocks}.

The following sections describe the individual components of the
transfer circuit in more detail, all of the components will be
described with reference to how they integrate into the overall
system shown in Figure~\ref{fig:Overall-transfer}.

\subsection{Magnet coils and mounts}
\subsubsection{Magnet coils}
\begin{figure}
\begin{center}
\leavevmode
\includegraphics[angle=-90,width=0.5\linewidth]{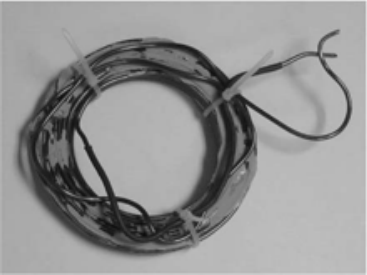}
\end{center}
\caption[Photograph of a single magnet coil]{Photograph of a single
magnet coil. The coil has an inner diameter of 6.5~cm.}
\label{fig:Coil}
\end{figure}

All of the magnet coils used in our experimental apparatus are
home-made coils constructed by winding magnet wire around a mold and
using an epoxy to hold the wire together. For most of our coils, we
have used polyurethane coated 12 AWG magnet wire with a diameter of
2.02~mm. We constructed a mold by making a custom-sized disc with 2
plates connected to either end, one of them having a slot to
facilitate starting the winding process.  The coils are wound on a
lathe by connecting the mold to the lathe chuck and having one
person manually rotate the lathe slowly while another person winds
the coil around the disc.  After one layer, MG Chemicals thermally
conductive, electrically insulating epoxy, part number 832TC, is
applied to the layer, and then another complete layer is wound.
After the entire coil is constructed, the epoxy is then cured by
either baking at 90$~^\circ$C for a couple of hours or letting the
coils sit at room temperature for a day. A photograph of one of the
home-made coils used in the magnetic transfer system is shown in
Figure~\ref{fig:Coil}.

\subsubsection{Coil mounts}
The transfer coils are mounted on copper plates above and below the
transfer tube of the vacuum system, visible in
Figure~\ref{fig:coils-top} on Page~\pageref{fig:coils-top}. Hollow
copper tubing soldered to the copper plates allows for the flow of
chilled water which cools the plates. Because the transfer coils do
not need to switch on and off rapidly, there is no need to protect
against eddy currents in these copper plates. This was not the case
on the MOT and TOP trap plates, where we cut a slit through the
plate to prevent circular eddy current flow. The water-cooled copper
plates reach a temperature of $\sim13~^\circ$C after running the
chiller water for half an hour, which greatly reduces heating of the
coils. The coils are clamped onto the plates in such a way that
small adjustments of the physical position of each coil, though
difficult, is possible.  Based on careful measurements, we estimate
that we have placed the coils to within $\pm~3$~mm from the
positions assumed in the calculation of Section~\ref{section:sim}.

\subsubsection{Coil locations}
Table~\ref{table:Coil-locations} lists the locations of the transfer
coils and various coil parameters.  The inner radius of each coil,
the $x$ position (the horizontal distance from the center of the MOT
coil pair to the center of the coil pair in question), and the $z$
height (the vertical distance from the center of the transfer tube
to the closest part of a coil) are tabulated. All of the coils are
built out of 12~AWG magnet wire with an electrically insulating
polyurethane coating. The plate separation, defined as the closest
distance between the upper and lower copper plates on which the
upper and lower coils are placed, is also tabulated.  Calculated
values for the inductance of each coil pair are also shown. Because
only three transfer coil pairs are on at any given time, the coils
can be divided into three groups, labeled Group A, B, and C, as
shown in Table~\ref{table:Coil-locations}.

%A back-of-the-envelope estimate for the inductance of a coil in
%$\mu$H, knowing the number of turns $N$ of the coil and the radius
%$R$ in inches, is given by
%\begin{equation}L = N^2 \cdot \frac {R}{19}\end{equation}
% need a reference on this one!
\begin{table}
\begin{center}
\leavevmode
\includegraphics[width=1\linewidth]{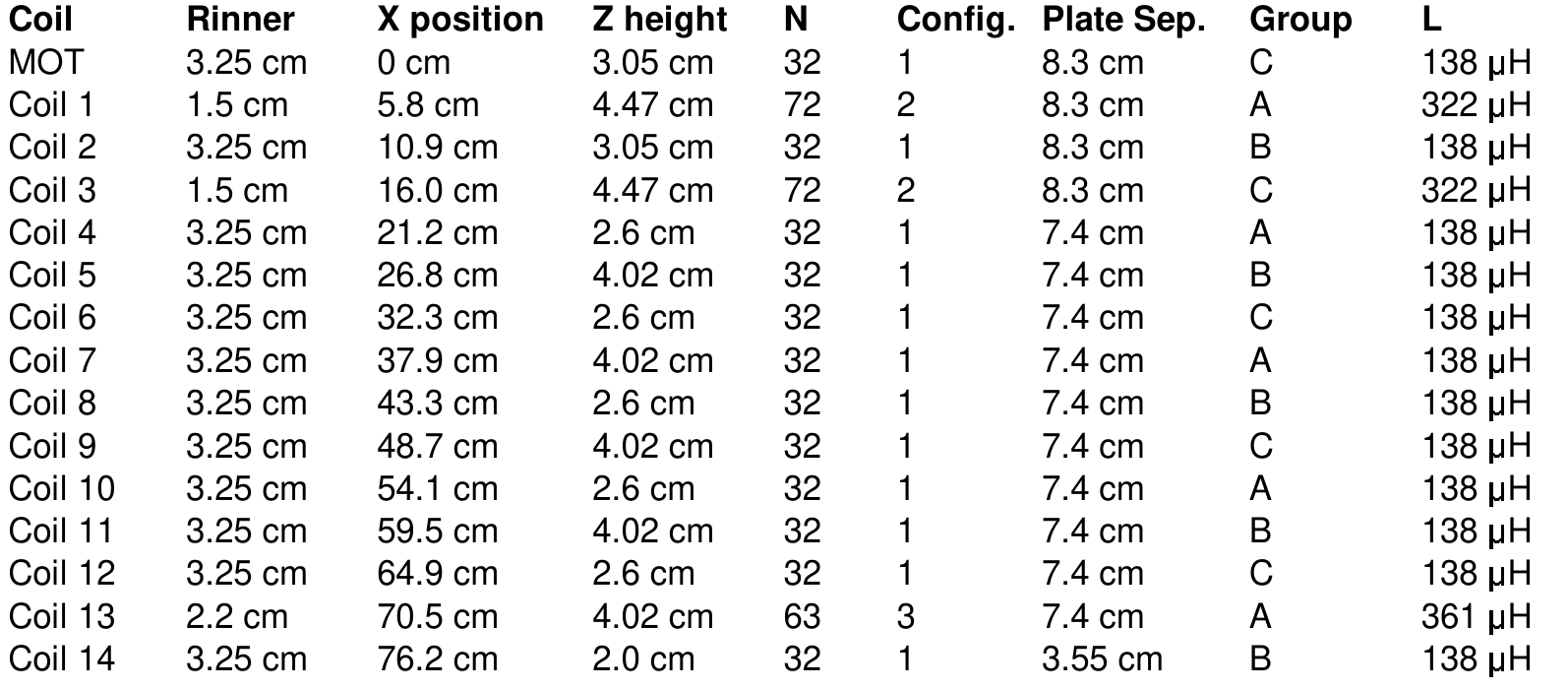}
\end{center}
\caption[Coil locations]{For each coil pair in the magnetic transfer
system, the inner radius, $x$ position, $z$ height, number of turns
$N$, coil configuration (see Table~\ref{table:Coil-configurations}),
plate separation, group, and inductance $L$ are shown.}
\label{table:Coil-locations}
\end{table}
%Resistance and inductance values are displayed for each coil pair.
%The resistivity $R$ is defined as the resistance per 1000~ft. of
%wire.  A value for the resistivity $R$ can be calculated based on
%the wire gauge AWG from the relationship
%\begin{equation}R=10^{\frac{AWG-10}{10}}\end{equation}
%The resistance of the coil pair is calculated by knowing the length
%of wire used to wrap a coil and the resistivity, remembering to
%multiply by 2 on the length to get the resistance for the coil pair.
%%The number of `wraps' of each coil is defined as the number of layers in
%the axial direction, like layers on a cake.  The number of `turns'
%on each `wrap' is defined as the number of outwardly spiralling coil
%windings.
Based on physical constraints, we have used three different
configurations for wrapping the coils. A table showing the three
different configurations used and the coil properties for each
configuration is shown in Table~\ref{table:Coil-configurations}.

\begin{table}
\begin{center}
\leavevmode
\includegraphics[width=1\linewidth]{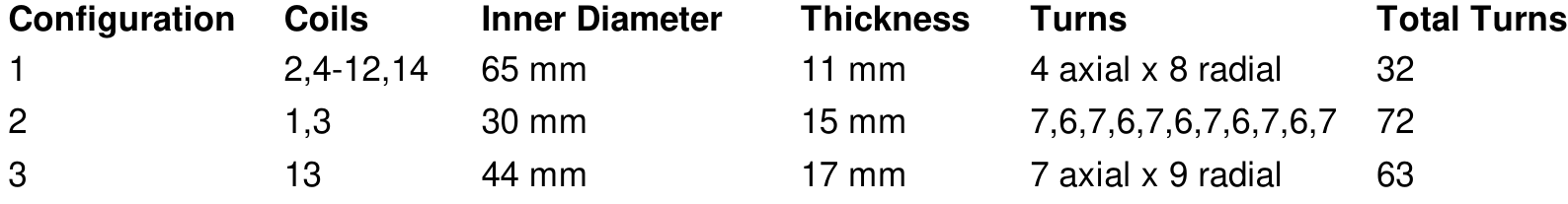}
\end{center}
\caption[Coil configurations]{Three different coil configurations
were used in construction of the transfer coils.  For each coil
configuration shown above, the inner diameter of the coil, axial
thickness of the coil, number of wraps and turns, and total number
of turns are given.} \label{table:Coil-configurations}
\end{table}

\subsubsection{Push coils}
During the initial period of testing the transfer system, we
attempted to magnetically transfer atoms only a few cm down the
transfer tube, transfer them back to the MOT cell, and image them
there as a system check.  We noticed that we could not transfer any
atoms past a particular point in the transfer tube, right where the
MOT cell glass met the metal flange of the vacuum system.  This is
where we installed a 6~mm diameter circular diaphragm into the
vacuum system (highlighted in Figure~\ref{fig:Vacuum-front-view}),
with the intention of limiting vacuum conductance from/into the MOT
cell. Unfortunately, the diaphragm was not perfectly aligned with
the transfer coils' $|\vec{B}|=0$ point, and axial and transverse
push coils were necessary to shift the $|\vec{B}|=0$ point enough to
allow the atoms to pass through the diaphragm.

%Push coils were necessary to shift the $|\vec{B}|=0$ point during
%passage through the diaphragm. An axial current of 19~A and a
%transverse current of 1.7~A through our push coils are required to
%shift the fields by the necessary amount.

\subsection{Power supply and ground}
We use an Agilent HP 6682A power supply to supply current to the MOT
coil, transfer coils, DC TOP coils, and levitation coil.  This power
supply operates in constant voltage mode at 21~V, and is capable of
supplying up to a total of 240~A.  The ground on the supply is
connected to the ground on the power line of our building. The
optical table, which is electrically connected to a Uni-Strut shelf
that is attached to the framework of our building, is connected to
the ground of the power supply and serves as the reference ground
for our laboratory.

\subsection{Sample-and-hold circuit} \label{subsection:sh}
% more info in butcher circuit in lab notebook on 10/27/04
The sample-and-hold circuit\footnote{For additional information, see
the lab notebook entry on 10/27/2004.}, visible in
Figure~\ref{fig:Overall-transfer} on
Page~\pageref{fig:Overall-transfer}, has two inputs and three
outputs.  It receives input signals from the `arbitrary waveform
SRS' and the `logic SRS' and its three outputs eventually connect to
the `transfer control box'.  The transfer control box requires three
input ramps to control the amount of current in coil groups A, B,
and C. The purpose of the sample-and-hold circuit is to provide
these three current ramps.

The transfer process operates with a digital update rate of $\Delta
t = 1.1$~ms; over the 5.9~second transfer process there are over
5,000 digital updates in the current sequence.  Because the time
constant $\tau = R / L$ governed by the resistance $R$ and
inductance $L$ of the transfer coils is on the order of the 1.1~ms
update rate, we can approximate a smooth current ramp with a series
of current steps updated at the 1.1~ms digital update rate. We were
faced with the design problem of creating three different digital
arrays of 5,000 updates each at a spacing of 1.1~ms.

%This could be solved by using three different DACs to create the
%three different current ramps, but that option was ruled out because
%of the cost of the DACs and lack of available output ports on the
%DIO-128.
%This was ruled out because of the prohibitive cost of using three
%separate SRS function generators.
%Another option we did not consider was using another analog output
%board to output the three ramps. This would have required a
%different analog output board with a large buffer that would need to
%be synchronized with the DIO-128 master clock.

One solution to this problem would be to use three different SRS
arbitrary waveform function generators, which could be loaded with
the three different ramps using a GPIB connection.  Instead, we
solved this problem by using two SRS function generators (the
`arbitrary waveform SRS' and the `logic SRS' of
Figure~\ref{fig:Overall-transfer}) and a simple home-made
sample-and-hold circuit, thus reducing the potential cost of the
project by almost 1/3.  The arbitrary waveform SRS is used to send a
combined version of the three current ramps to the sample-and-hold
circuit, and the logic SRS controls the sampling rate.  The details
of this process are described in the following paragraphs.

\begin{figure}
\begin{center}
\leavevmode
\includegraphics[width=1\linewidth]{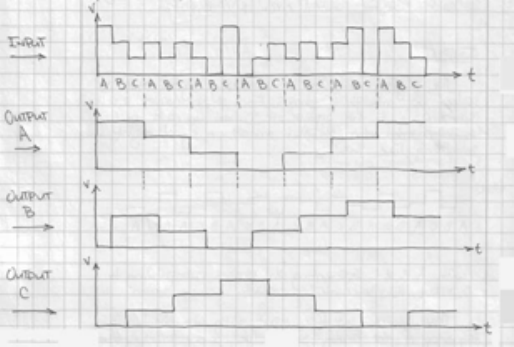}
\end{center}
\caption[Illustration of sample-and-hold technique]{Illustration of
sample-and-hold technique. The input signal is a combined version of
the signals that will be sent to coil groups A, B, and C.  The
sample-and-hold circuit separates one input signal into three output
signals, which will be sent to the `transfer control box'. Drawing
by David Kaz.} \label{fig:kaz_butcher}
\end{figure}

The calculation that determines the appropriate current ramps for
each transfer coil pair (which will not be described until
Section~\ref{section:sim}) produces 14 individual current ramps,
each describing the necessary current sequence for each individual
coil pair. In software these 14 ramps are grouped into three
waveforms corresponding to coil Groups A, B, and C.  This is done by
simply adding the non-overlapping current ramps (i.e.\ the
non-overlapping ramps for coils 1, 4, 7, 10, and 13 are added
together to form the sequence for Group A). These three waveforms
corresponding to the three different groups are further combined in
such a way that only \emph{one} arbitrary waveform is loaded into
the `arbitrary waveform SRS'. The three different waveforms are
multiplexed together so that the array output reads (Group A
value~1, Group~B value 1, Group C value 1, Group A value 2, Group B
value 2, etc).  The sample-and-hold circuit then splits this
combined input into three outputs. An example of this technique,
illustrating how the sample-and-hold circuit separates one input
into three outputs, is shown in Figure~\ref{fig:kaz_butcher}.

The combined waveform (the input of Figure~\ref{fig:kaz_butcher}) is
created in software on the Experiment PC and loaded into the
`arbitrary waveform SRS' function generator using GPIB; this
waveform is sent to the sample-and-hold circuit at an update rate of
$f = 3/(\Delta t)$. Another function generator, the `logic SRS' is
programmed with a square wave at a frequency of $f = 1/(\Delta t)$.

The sample-and-hold circuit splits that square wave into three
separate pulsetrains, each with a frequency of $f = 1/(\Delta t)$
and separated in time by 1/3 of the sampling period $\Delta t$.  The
circuit employs NE555 timer chips to do this. These three
pulsetrains are then sent to three different LF398 sample-and-hold
chips, which all sample the same input array but output the three
distinct current ramps required by the transfer control box, shown
in Figure~\ref{fig:kaz_butcher}.

An additional sophistication is required on Group C because this
group controls both the MOT coil and the DC TOP coils.  The signal
to the Group C input on the transfer control box is switched between
two inputs using a Vishay DG419 analog switch.  The analog switch
connects either the output of the sample-and-hold circuit or the DAC
output to the transfer control box.  During the transfer process
(i.e.\ when the MOT coil pair, coil pair 1, and coil pair 2 are on)
the input to Group C comes from the sample-and-hold circuit, which
controls the transfer process. During MOT loading or RF evaporation
the input to Group C comes from the DAC, which controls the timing
events before and after the transfer sequence.

\subsection{Transfer control box} \label{subsection:transfer}
The `transfer control box', shown in
Figure~\ref{fig:Overall-transfer} on
page~\pageref{fig:Overall-transfer}, has been described as the
mastermind behind the entire transfer process.  The transfer control
box controls the amount of current flow through each of the transfer
coil pairs.  It does this by: (1) individually turning on and off
the current flow through each of the 14 transfer coil pairs; and (2)
controlling the variable amount of current through each transfer
coil pair by monitoring and regulating its amount of current flow.

\begin{figure}
\begin{center}
\leavevmode
\includegraphics[angle=-90,width=1\linewidth]{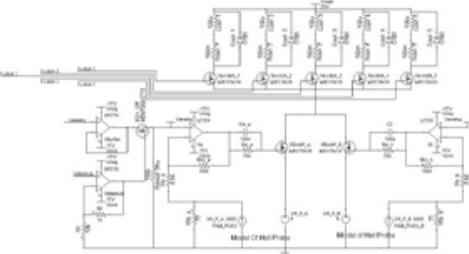}
\end{center}
\caption[Schematic of the transfer circuit]{Circuit diagram showing
the electronics components used to drive one group of five transfer
coils.  This circuit diagram will be repeated in
Figure~\ref{fig:Transfer-circuit-split}}
\label{fig:Transfer-circuit}
\end{figure}

The complete electronics circuit used to drive the transfer coils
includes not simply the transfer control box, but also the power
supply, the coils themselves, and the MOSFET boxes, as described in
Section~\ref{subsection:components}. However, the operation of the
entire system will be described in this section, with reference to
the circuit diagram shown in
Figure~\ref{fig:Transfer-circuit}\footnote{This and all the circuit
diagrams included in this dissertation were generated by M. David
Henry in TOP Spice using the PSPICE 3f5 engine.}. This circuit
diagram includes the electronics used to drive one group of five
transfer coils, and includes elements physically situated within the
MOSFET boxes, diode boxes, and transfer control box, as well as the
coils themselves.

\begin{figure}
\begin{center}
\leavevmode
\includegraphics[angle=-90,width=1\linewidth]{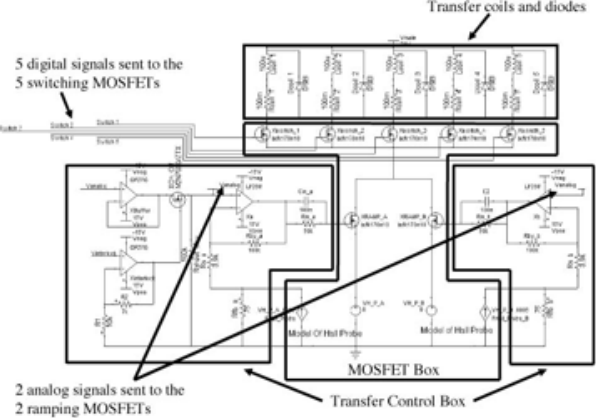}
\end{center}
\caption[Sections of the transfer circuit]{Circuit diagram showing
the different sections of the circuit used to drive one group of
five transfer coils.  The box at the top of the schematic contains
both the coils (situated on the optical table) and the diodes
(housed in the diode box described in
Section~\ref{subsection:diode}). Each transfer coil pair is modeled
as an inductor (a coil) and a resistor in series. The two sections
on either side of the schematic contain all the electronics in the
transfer control box, described in
Section~\ref{subsubsection:operation}. The section in the middle of
the schematic contains all the power electronics in one of the
MOSFET boxes, described in Section~\ref{subsection:mosfet}. The
connections to the five switching MOSFETs (labeled Xswitch\_1
through Xswitch\_5) and two ramping MOSFETs (labeled XRAMP\_A and
XRAMP\_B ) are also shown.  The Hall probes are modeled as
voltage-dependent current sources.}
\label{fig:Transfer-circuit-split}
\end{figure}

In order to better understand how the individual pieces of
Figure~\ref{fig:Transfer-circuit} fit into the black-box components
of Figure~\ref{fig:Overall-transfer}, three sections of
Figure~\ref{fig:Transfer-circuit} are highlighted in
Figure~\ref{fig:Transfer-circuit-split}.  These sections demarcate
which portions of Figure~\ref{fig:Transfer-circuit} comprise the
`transfer control box' and `MOSFET box'; the transfer coils and
diodes are also highlighted.

\subsubsection{Operation of transfer circuit} \label{subsubsection:operation}
This section provides an explanation of the circuit shown in
Figure~\ref{fig:Transfer-circuit-split} by tracing the path of
positive current flow from Vmain (the positive end of the power
supply at the top of the diagram) to ground.  Current first flows
through a transfer coil, modeled as an inductor in series with a
resistor in Figure~\ref{fig:Transfer-circuit-split}.  There is a
power diode placed in parallel with each transfer coil to prevent
current spikes; the role of this diode will be discussed in more
detail in Section~\ref{subsection:diode}.

Current then flows through one of the five switching MOSFETs,
labeled as Xswitch\_1 through Xswitch\_5 in
Figure~\ref{fig:Transfer-circuit-split}. The switching MOSFETs used
are Advanced Power Technology APT10M07JVR power MOSFETs; there is
one individual switching MOSFET per coil pair that acts as an on/off
switch for each coil pair.
%Individual coils are turned on and off by
%applying a voltage to the gate of these switching MOSFETs; each of
%these 15 digital signals (5 for each MOSFET box) originate in the
%`transfer control box'.

The sources of the five switching MOSFETs are connected to the
drains of the two ramping MOSFETs, labeled as XRAMP\_A and XRAMP\_B
in Figure~\ref{fig:Transfer-circuit-split}. Two ramping MOSFETs,
IXYS model IXYS180N10, are used in each MOSFET box. These two
ramping MOSFETs are placed in parallel and act as variable
resistors, modifying how much resistance there is between the power
supply and ground, and therefore, how much current flows through the
coils.  We use two ramping \mbox{MOSFETs} in parallel because of the
high power demand on these MOSFETs; using two MOSFETs means that
only half as much current will flow through each one compared to
using only one MOSFET.

%These two `variable resistors' are controlled by sending two analog
%signals to the gates of the ramping MOSFETs; these two analog
%signals originate in the `transfer control box'.

In each MOSFET box an electrical cable passes from the drain of each
ramping MOSFET, through a Hall Probe, and connects to ground, as
shown in Figure~\ref{fig:MOSFET-box}.  However, for the purposes of
circuit modeling, the Hall Probe is modeled as a voltage-controlled
current source in Figure~\ref{fig:Transfer-circuit-split}. Each
ramping MOSFET is on a separate feedback loop and uses its own Hall
Probe to monitor current. The Hall probes used, FW-Bell model
CLN-200, are zero-resistance probes that measure the amount of
actual current flowing through the device by monitoring the induced
magnetic field around the current-carrying wire. The feedback loop
in the `transfer control box' compares the amount of actual current
(the signal send from the Hall Probe) to the desired amount of
current (indicated as Vanalog in
Figure~\ref{fig:Transfer-circuit-split}) with an LF356 op amp.  The
output of this op amp connects to the gate of the ramping MOSFET to
make the actual amount of current equal to the desired amount of
current.

Without this feedback loop, if a MOSFET were to heat up because of
high current flow, the current through the coil would increase. The
feedback loop prevents this thermal runaway by continually
monitoring the amount of current flow and applying the proper
voltage to the gate of the ramping MOSFET. Another advantage of the
feedback loop is that we avoid having to use a voltage/current
lookup table that would have to be generated for every MOSFET used.
The feedback loop automatically adjusts the ramping MOSFETs' gate
voltage until the current is at the specified value.

\subsubsection{Connections on transfer control box} The transfer
control box, shown in black-box form in
Figure~\ref{fig:Overall-transfer} on
Page~\pageref{fig:Overall-transfer}, has 20 outputs and 23 inputs.
On the back of the transfer control box there are 14 digital outputs
that act to turn on and off individual coil pairs. These digital
outputs originate from Port B on the DIO-128 and get amplified in
the transfer control box to a logic high of 24~V to turn on the
individual switching MOSFETs associated with each coil pair.  The
reason they need to be at such a high voltage is that the drain of
each switching MOSFET is connected (through the coils) to the 21~V
power supply, and the gate-source voltage of these MOSFETs needs to
be higher than the drain voltage to prevent these MOSFETs from
operating in the triode region.  There is an amplifier circuit
inside the transfer control box that, after appropriate
opto-isolation of the logic lines as shown in
Figure~\ref{fig:Opto-isolators}, amplifies the logic high to 24~V.
% is triode correct?  Check with Dave! reference?

There are six analog outputs on the back of the transfer control
box, labeled `Gate signals' in Figure~\ref{fig:Overall-transfer},
that are applied directly to the gates of the two ramping MOSFETs in
each of the three MOSFET boxes. These signals control the resistance
of each ramping MOSFET, regulating the time-varying amount of
current through each coil.

There are six analog inputs on the back of the transfer control box,
labeled `Hall Probe signals' in Figure~\ref{fig:Overall-transfer},
that come from the six Hall Probes in the MOSFET boxes.  These
current signals get dropped across a resistor in the transfer
control box to provide a voltage that is proportional to the actual
amount of current flowing through each Hall Probe.

There are three analog inputs on the front of the transfer control
box labeled Group A, B, and C. These are analog signals that come
from the sample-and-hold circuit and provide the current ramps to be
sent to the coils.  Finally, there are 14 digital inputs on the back
of the transfer control box that originate from Port B of the
DIO-128.

\subsection{MOSFET boxes} \label{subsection:mosfet}
\begin{figure}
\begin{center}
\leavevmode
\includegraphics[angle=-90,width=0.8\linewidth]{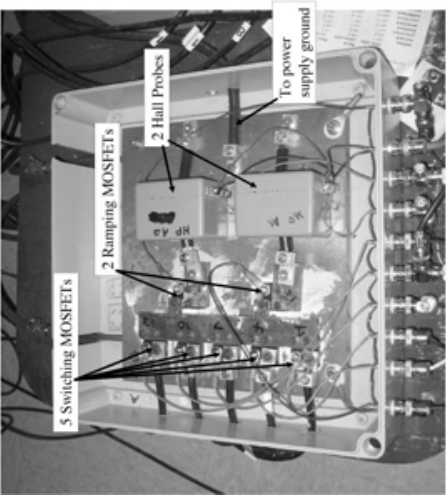}
\end{center}
\caption[Photograph of a MOSFET box]{Photograph of one of the 3
MOSFET boxes, MOSFET Box A.  \mbox{MOSFET} Box A controls transfer
coil pairs 1, 4, 7, 10, and 13. The 5 cables on the left side of the
box connect the transfer coils (through the diode box) to the 5
switching \mbox{MOSFETs}. The 5 switching MOSFETs connect to the 2
ramping \mbox{MOSFETs} through a copper plate. The current passing
through each of the 2 ramping \mbox{MOSFETs} passes through a Hall
Probe before connecting to a large water-cooled copper plate at the
bottom of the MOSFET box. This copper plate is electrically
connected both to the power supply ground and the optical table. The
BNC connections on the front of the MOSFET box connect to the gates
of the switching and ramping MOSFETs, and the power and output
signals of the 2 Hall Probes.} \label{fig:MOSFET-box}
\end{figure}

The role of the components in the MOSFET boxes within the overall
scheme of the magnetic transfer system has already been described in
the previous section.  This section describes the physical layout
and hardware of each MOSFET box. Three identical boxes house the
MOSFETs used to control coil Groups A, B, and C. These three MOSFET
boxes are depicted in a black-box diagram in
Figure~\ref{fig:Overall-transfer} on
Page~\pageref{fig:Overall-transfer}, the power electronics within a
MOSFET box are highlighted in
Figure~\ref{fig:Transfer-circuit-split} on
Page~\pageref{fig:Transfer-circuit-split}, and a photograph of one
of the MOSFET boxes is shown in Figure~\ref{fig:MOSFET-box}.

As seen in Figure~\ref{fig:MOSFET-box}, each MOSFET box has seven
BNC inputs and two BNC outputs on the front of the box, one 4 AWG
connection to the power supply ground on the right side of the box,
and five 8 AWG connections to the diode box (and hence the transfer
coils themselves) on the left side of the box. Each MOSFET box
contains five switching MOSFETs, two ramping MOSFETs, and two Hall
Probes. The MOSFETs are mounted on a large copper plate that is
water cooled with chiller water.  On the front of the box, five
logic signals from the `transfer control box' connect directly to
the gates of each of the five switching MOSFETs; two gate signals
from the `transfer control box' connect directly to the gates of the
two ramping MOSFETs; and two Hall Probe signals originate from each
MOSFET box and are sent to the `transfer control box'.  The three
MOSFET boxes are grounded together as well as connected to the power
supply ground.

\subsection{Diode box} \label{subsection:diode}
The diode box is shown in the black-box diagram of
Figure~\ref{fig:Overall-transfer} on
Page~\pageref{fig:Overall-transfer}.  The diode box contains 14 IXYS
DSEI 2X101-06A power diodes that are each placed in parallel with a
transfer coil pair and oriented such that the cathodes of these
diodes connect to the positive end of the power supply. These
fly-back power diodes act to dissipate large voltage spikes, such as
what would occur if there was a large amount of current flowing
through a coil and a switching MOSFET was suddenly turned off. Based
on the relationship
\begin{equation}
V=L \frac{\partial I}{\partial t} \end{equation} there would be a
huge voltage on the drain of the switching MOSFET, potentially
causing device failure.  By installing a fly-back diode that
operates only when the voltage on its anode is large enough (as in
this situation), that transient current will flow back up through
the diode instead of destroying the MOSFET.

As shown in Figure~\ref{fig:Overall-transfer}, the diode box
contains the one and only connection to the positive end of the
power supply, this connects to the cathodes of all the diodes and
one end of each transfer coil pair.  The other end of each coil pair
is connected to the anode of its respective diode, and there is a 4
AWG wire connecting this point to the drain of the corresponding
switching MOSFET in the MOSFET boxes.

\subsection{Safety interlock box} \label{subsection:interlocks}
There have been several times when an accident has occurred in lab
and we found ourselves in a situation where 100~A or more was
flowing steady-state through a coil, once heating the DC TOP coils
to the point where epoxy melted onto the surface of the science
cell. This usually occurs when a MOSFET blows up or when the
computer crashes and sends an erroneous value to the transfer
circuit.  For this reason, we have installed several hardware safety
interlocks that will automatically disable the Agilent power supply
until a front-panel button is depressed, inhibiting the flow of
current. The safety interlock box receives so many input signals
from different parts of the experiment that they are not shown in
Figure~\ref{fig:Overall-transfer}, rather, they are explained in the
subsequent paragraphs.

\pagebreak

Two of the safety interlock conditions are based on the flow of
chiller water, through two separate flow meters that monitor water
flow.  Three of the safety interlock conditions are based on the
temperatures of coils used in the transfer process, these are based
on temperature readings using thermistors epoxied to the coils. The
temperature trip points were set by recording the maximum
temperatures of the monitored coils, and setting the trip point to
be $\sim 5~^\circ$C above that.

Three of the safety interlock conditions monitor the integrated
amount of current through various coils. To set the trip point on
the integrators, we adjusted the $R$ and $C$ values of the
integrator circuits to operate on a timescale of several seconds and
trip a comparator if the integrated value of current is $\sim 10 \%$
larger than normal.  Two of the safety interlock conditions are
based on the instantaneous amount of current through particular
coils. Both the integrated current interlocks and temperature
interlocks operate slowly (on the timescale of a few seconds), but
the chiller water and instantaneous current interlocks operate right
away. The power supply will automatically enter into inhibit mode if
any of the following conditions occur:
\begin{enumerate}
\item If the flow of the chiller water cooling the transfer
coils' copper plates stops.
\item If the flow of the chiller water cooling the MOSFETs in the
MOSFET boxes stops.
\item If the temperature of the MOT coil gets too hot and reaches
its trip point.
\item If the temperature of the DC TOP coils get too hot.
\item If the temperature of the AC TOP coils get too hot.
\item If the integrated amount of overall current is too large and
reaches its trip point.  In addition to the several Hall probes that
monitor current through the MOT coil, transfer coils, DC TOP coils,
and levitation coil (used for BEC imaging), we have installed a Hall
Probe directly to the output of the Agilent power supply to monitor
the total amount of current flow.  A voltage proportional to the
total instantaneous amount of current flow is sent to an integrator
circuit.  The output of this integrator circuit rises slowly over
the course of the transfer sequence and will trip a comparator and
inhibit the power supply if the total integrated amount of current
is too high.
\item If the integrated amount of current in the MOT coil is too
high.
\item If the integrated amount of current in the DC TOP coils is too
high.
\item If the instantaneous amount of current in the DC TOP coils is
too high.  At the beginning of the evaporation sequence, the DC TOP
coils operate at their highest value of 20~A.  If the amount of
instantaneous current in the DC TOP coils exceeds 30~A, the safety
interlock circuit will inhibit the power supply.
\item If the instantaneous amount of current through the levitation
coil is greater than 15~A.
\end{enumerate}
If any one of the above conditions is met, that portion of the
interlock circuit sends a TTL low signal to a multi-input AND gate
in the safety interlock box. A TTL low signal at the output of the
AND gate will inhibit the power supply. If we lose power to the
safety interlock box or if one of its internal components fails, the
logic output will also be low, inhibiting the power supply.  One
consequence of using integrators in the safety interlock circuit is
that we need to wait $\sim$30 seconds between each run of the
experiment, or the integrated values from running the transfer
sequence too frequently will inhibit the power supply. This is
usually a good idea, however, because running too frequently, though
possible, will tend to decrease the lifetime of the power MOSFETs
due to overheating.

\subsection{Direction of current flow}
Positive current flows from the power supply (under the optical
table) to the cathode of a diode in the diode box (under the optical
table) through a transfer coil pair (mounted around the vacuum
system on the optical table), and back to the anode of a diode in
the diode box.  Current then flows through a switching MOSFET and a
pair of ramping MOSFETs and Hall Probes, all of which are located in
one of the MOSFET boxes (under the optical table).  The current path
terminates at the ground in the MOSFET box, which itself is
connected to the power supply ground. The DAC, arbitrary waveform
SRS, logic SRS, sample-and-hold circuit, analog switch, and transfer
control box are all located on the 19" equipment rack.

%Transfer coils 1, 2, 3, etc. will turn on in sequence, with three
%coils normally on at a time.  The transfer process, over a distance
%of 76.2~cm, takes 5.9~seconds.

\section{Calculation of the Current Sequence} \label{section:sim}
The previous sections have described the hardware, electronics, and
principles of operation of the magnetic transfer system.  The only
things that have been left out are: (1) a description of the
calculation that produces the three current ramps that are sent to
the `transfer control box'; and (2) a description of the
center-of-mass velocity of the atoms as they move down the transfer
tube. Both of these topics are addressed in this section.

\pagebreak

The following paragraphs describe the calculation that produces the
current ramps that are sent to the transfer control box. We will use
the convention that $z$ is the vertical (gravity) direction, $x$ is
the longitudinal (along the direction of the transfer tube)
direction, and $y$ is the transverse direction. The magnetically
trapped atoms start in an axially symmetric quadrupole trap in the
MOT cell with $\partial B /
\partial z$~=~180~G/cm, $\partial B / \partial x$~=~-90~G/cm, and
$\partial B / \partial y$~=~-90~G/cm.  Because three pairs of coils
are on during the transfer process, the magnetic trap gets elongated
in the transfer direction to form a 3-dimensional trapping potential
with $\partial B /
\partial z$~=~180~G/cm, $\partial B /
\partial x$~=~-65~G/cm, and $\partial B / \partial y$~=~-115~G/cm.

A 3-dimensional calculation was performed to determine the current
ramps necessary to produce the desired magnetic field gradients. The
calculation used the Biot-Savart law to determine the magnetic field
and magnetic field gradient along the transfer tube in all 3
orthogonal directions as a function of the amount of current through
the transfer coils. For a single coil of radius $R$ perpendicular to
the $z$ axis and centered at $z=A$, the transverse magnetic field
component $B_{\rho}$ and the axial magnetic field component $B_z$ as
a function of axial and radial position $z$ and $\rho$
are~\cite{bergeman1987mtf}
\begin{equation} \label{equation:bz}
B_z=\frac{\mu I}{2 \pi} \frac {1}{[(R+\rho)^2+(z-A)^2]^{1/2}} \cdot
\left[K(k^2)+\frac{R^2-\rho^2-(z-A)^2}{(R-\rho)^2+(z-A)^2}E(k^2)\right]
\end{equation}
\begin{equation} \label{equation:brho}
B_{\rho}=\frac{\mu I}{2 \pi \rho} \frac
{z-A}{[(R+\rho)^2+(z-A)^2]^{1/2}} \cdot
\left[-K(k^2)+\frac{R^2+\rho^2+(z-A)^2}{(R-\rho)^2+(z-A)^2}E(k^2)\right]
\end{equation}
where $\mu$ is the Bohr magneton, $I$ is the current through the
coil, and
\begin{equation}
k^2=\frac{4R\rho}{(R+\rho)^2 + (z-A)^2}
\end{equation}
is the argument of the complete elliptic integrals $K$ and
$E$~\cite{arfken2005mmp}. From the preceding equations it is
straightforward to determine the magnetic field at two neighboring
points in order to construct a field gradient. With three coils on
at a time, it is also straightforward to set up a matrix of three
equations and three unknowns. This matrix equation is solved at one
particular point in space, starting at the center of the MOT cell,
to determine what currents are necessary to produce the desired
field gradients at that point. Then the matrix equation is solved
again at each subsequent point in space along the transfer tube
along a linear grid of spacing $\Delta x$~=~0.165~mm. The inputs to
the matrix equation are the three known quantities:
\begin{enumerate}
\item $|\vec{B}|=0$ at that particular point in space (therefore atoms are
magnetically trapped there).
\item $\partial B / \partial x$ (along the transfer direction) equals the desired
longitudinal gradient.
\item $\partial B / \partial z$ (along the vertical direction) equals the desired
axial gradient.
\end{enumerate}

The three unknowns are the three currents ($I_1$, $I_2$, and $I_3$)
in the three neighboring coil pairs that happen to be on at a given
time. The $3 \times 3$ matrix $M$ is determined from the Biot-Savart
Law, which determines the resultant magnetic fields from input
currents and specified coil locations.

\begin{equation} \label{equation:matrix}
\left(   \begin{array}{c}
|\vec{B}|\\
\partial B / \partial x\\
\partial B / \partial z
\end{array} \right)  =
\left(   \begin{array}{c}
0\\
-65~G/cm\\
180~G/cm
\end{array} \right) =
\left(   \begin{array}{ccc}
M_1 & M_2 & M_3 \\
M_4 & M_5 & M_6 \\
M_7 & M_8 & M_9
\end{array} \right) \cdot
\left(   \begin{array}{c}
I_1\\
I_2\\
I_3
\end{array} \right)
\end{equation}

To understand how the values of the matrix $M$ are determined,
consider the following example: when the atoms are at position
$x=3$~cm down the transfer tube, transfer coil pairs 1, 2, and 3 are
on. The calculation determines the three currents $I_1$, $I_2$, and
$I_3$ through transfer coil pairs 1, 2, and 3 required to make
$|\vec{B}|=0$, $\partial B / \partial x = -65$~G/cm, and $\partial B
/ \partial z = 180$~G/cm at the position $x=3$~cm. Matrix values
$M_1$, $M_2$, and $M_3$ are the large prefactors in
Equations~\ref{equation:bz} and \ref{equation:brho} that set up the
equation

\begin{equation}
|\vec{B}|=0=B_z + B_{\rho} = M_1 \cdot I_1 + M_2 \cdot I_2 + M_3
\cdot I_3
\end{equation}
which is the top row of the matrix equation \ref{equation:matrix}.
The other values of the matrix, $M_4$ through $M_9$, relate the
currents in coil pairs 1, 2, and, 3 to the field gradients along $x$
and $z$ at this position in space, and are determined from taking a
spatial derivative of Equations~\ref{equation:bz} and
\ref{equation:brho}.

%Because $div B =0$, $\partial B / \partial y$ can be easily
%determined from $\partial B / \partial x$ and $\partial B / \partial
%z$.

\begin{figure}
\begin{center}
\leavevmode
\includegraphics[width=1\linewidth]{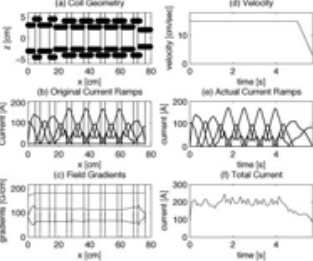}
\end{center}
\caption[Coil locations, current ramps, and field gradients for
magnetic transfer system]{Coil locations, current ramps, and field
gradients for magnetic transfer system are shown.  (a) Shown are the
locations of the MOT coil pair and all 14 transfer coil pairs.  This
is a to-scale rendering of all the coil positions, outer diameters,
and heights.  Also shown are the breakpoints, the positions in space
where one coil turns off and another turns on.  (b) Shown are the
original current ramps (the result of the calculation) with
current~[A] plotted as a function of position $x$~[cm].  Each coils'
current ramp is plotted directly beneath the coil position.
(c)~Shown are the magnetic field gradients resulting from the
current ramps shown in (b); the largest magnetic field gradient is
$|
\partial B / \partial z |$, the intermediate magnetic field gradient
is $|
\partial B / \partial y |$, and the weakest field gradient is $| \partial B
/ \partial x |$.  Figures (a), (b), and (c) are plotted on the same
spatial horizontal axis.  (d) Shown is the atomic velocity vs.\
time~[s] over the transfer time, reflecting the final deceleration
stage. (e) Shown are the actual current ramps used for transfer,
which differ from (b) in that the final portion of transfer reflects
the deceleration. (f) Shown is the total amount of current used for
transfer vs.\ time~[s].  Figures (d), (e), and (f) are plotted on
the same temporal horizontal axis.} \label{fig:Current-ramps}
\end{figure}

The required current ramps and resultant magnetic field gradients,
plotted alongside the coil locations, are shown in
Figure~\ref{fig:Current-ramps}.  Figure~\ref{fig:Current-ramps}~(a)
is a side view of the 15 coil pairs (the MOT coil and 14 transfer
coils), showing their to-scale positions and widths along the
transfer ($x$) direction and vertical ($z$) direction.

The thin vertical lines in Figure~\ref{fig:Current-ramps}~(a)
indicate `breakpoints', additional adjustable parameters in the
calculation. These user-adjustable breakpoints are the position
locations where one coil turns off and another turns on.  For
example, the breakpoint position of $x=39.4$~cm is the longitudinal
position of the moving atom cloud where transfer coil pair 6 turns
off and transfer coil pair 9 starts turning on.  The breakpoint
locations were adjusted to optimize the current ramps for smooth
field gradients. After this optimization, we noticed that coil pairs
farthest from the transfer tube in the $z$ direction operate over a
longer range in $x$ than coil pairs closer to the tube.

Figure~\ref{fig:Current-ramps}~(b) shows the result of the
simulation, the current ramps required for each coil pair in order
to produce the desired field gradients.  The breakpoints are still
shown, and it is clear that these are the position locations where
one coil pair turns off and another starts ramping on.
Figure~\ref{fig:Current-ramps}~(c) shows the magnitude of the
magnetic field gradients produced while using the above current
ramps.  The largest field gradient is in the axial ($z$) direction,
while the weakest field gradient is in the transfer ($x$) direction.
The field gradients vary at the beginning and end of the transfer
sequence, these are the regions where only one or two transfer coil
pairs are operational. Figures~\ref{fig:Current-ramps}~(a), (b), and
(c) plot the coil locations, current ramps, and field gradients as a
function of position $x$ (in cm) along the transfer direction, but
do not provide any information on the timescale of the transfer
process, which is dependent on the velocity of the moving
$|\vec{B}|=0$ point.

%\subsubsection{Velocity of transfer}
In order to optimize the number of atoms transferred from the MOT
cell to the science cell, the velocity and acceleration of the
moving magnetic trap was adjusted.  We originally thought we would
need to increase the initial velocity of the atoms up from zero to
the nominal velocity of 15~cm/s and then decrease the velocity back
to zero at the end of the transfer process in order to keep the
atomic motion as adiabatic as possible. We initially thought that
these acceleration/deceleration steps would be rather important in
optimizing transfer efficiency; however, we have observed that the
transfer efficiency is not very dependent on these steps or the
nominal velocity.  After optimizing the process, we discovered that
the initial acceleration phase actually decreased the number of
transferred atoms, presumably because the slower initial speeds
meant that trapped atoms spent a longer time in the MOT cell, the
highest-pressure region of the chamber, and underwent more
collisions with background atoms.  We did notice, however, that a
small amount of deceleration at the end of the transfer process
increased the transfer efficiency slightly.

After optimization, we found that it is best to start the transfer
process with an initial velocity of 15~cm/s and include a
deceleration phase to 3~cm/s over the final 1~second of the
5.9~second transfer process. A plot of the calculated center-of-mass
atomic velocity vs.\ time is shown in
Figure~\ref{fig:Current-ramps}~(d).  A plot of the actual current
ramps used vs.\ time is shown in Figure~\ref{fig:Current-ramps}~(e).
These current ramps differ only slightly from the ramps shown in
Figure~\ref{fig:Current-ramps}~(b) in that they are plotted as a
function of time and the last 1~second is a visibly lengthened
version of the original current ramps, indicating a slower atomic
velocity. Figure~\ref{fig:Current-ramps}~(f) shows a plot of the
total current used through all the coils vs.\ time.

\section{Transfer Revisited} \label{section:revisited}
%Starting with $3 \cdot 10^9$ atoms in the MOT, we transfer about $3
%\cdot 10^8$ atoms into the quadrupole trap in the BEC cell. The
%atoms end up in a quadrupole trap formed by coil pair 14, the final
%transfer coils, before being transferred into the TOP trap,
%described in Section~\ref{section:top}.
Based on the amount of time involved in the design and
implementation of the magnetic transfer system, it is easily the
most complex part of our laboratory.  Successful operation of the
transfer system took one B.S. thesis in Electrical Engineering by M.
David Henry \cite{henry2004mot}, over two years of work, a lot of
software and hardware, and many blown-up MOSFETs. One might ask if
all this trouble was really worth it.

To our knowledge, there have been two varieties of successful
approaches to magnetic transport of ultracold atoms in a vacuum
system over a large distance (not including BEC on a chip
experiments, which are different altogether). Our approach, using a
series of coils run in sequence, was first developed in the group of
Ted H{\"a}nsch in 2001 \cite{greiner2001mtt}.  This experiment
involved magnetic transfer of atoms over 33~cm and around a
90$^\circ$ bend using a series of nine transfer coil pairs.

Another approach, developed in Eric Cornell's group at JILA,
involves using a track-based system that physically moves a coil
pair along a track on the optical table \cite{lewandowski2003ssc}. A
hybrid approach is used in the group of Carl Wieman at JILA, which
includes a moving track, a 90$^\circ$ bend, and a series of transfer
coils \cite{papp2006ohf}.
%An altogether separate method is to
%transfer the atoms from one region of the chamber to another by
%means of a far-off resonant laser beam that acts to push the atoms.
%This approach adds a considerable amount of heating to the cloud,
%usually necessitating a dual-MOT design, and is not very realistic
%for a chamber of our length.

The track-based moving coil system has its merits and is worth
comparing and contrasting to our system.  We have heard that the
moving coil system is easy to use, robust, and simple to set up. At
JILA, it worked on the first attempt \cite{lewandowski2003ssc}.  The
ease of use of the moving coil system makes it appear rather
advantageous compared to our machine.  However, the moving coil
system is not without its disadvantages.  It places a moving
mechanical system on the optical table, which can potentially have
repercussions for vibration-sensitive equipment such as laser locks.
It is also large, bulky, and restricts optical access to the MOT
cell and science cell.  Finally, because the same coil pair and
associated mount need to translate from the MOT cell to the science
cell, optical access to the science cell will be particularly
blocked in this design.

One of the main advantages of our design is the large amount of
optical access to the science cell from five directions, which will
be highlighted in the description of the TOP trap in
Section~\ref{subsubsection:fnum}. Setting up complex experiments
such as 3-D optical lattices while imaging from multiple directions
and using an optical trapping beam is possible with our open
geometry design. The biggest disadvantage to our current system is
the MOSFET lifetime. We continue to blow up MOSFETs, particularly
the DC TOP MOSFET and occasionally one of the switching MOSFETs.
This is something that can possibly be alleviated by increased or
more efficient cooling, waiting longer between runs, or, in the case
of the DC TOP MOSFET, distributing the power dissipation over two
MOSFETs.

%% file: Chap_Making.tex
\chapter{BEC FORMATION AND IMAGING}\label{chapter:making}

\section{Introduction}
Chapters~\ref{chapter:experimental} and~\ref{chapter:transfer}
describe the first two steps in our BEC formation process: (1)~laser
cooling and magnetic trapping in the MOT cell; and (2) magnetic
transfer of atoms from the MOT cell to the science cell.  This
chapter describes the final step in the BEC formation process:
evaporative cooling of atoms in the science cell. After transfer
into the science cell, the atoms are held in a Time-averaged
Orbiting Potential (TOP) trap, described in
Section~\ref{section:top}.  The atoms are then evaporatively cooled
in the TOP trap until a BEC is formed; this process is described in
Section~\ref{section:evap}. The methods used for imaging the
condensate and techniques used for image processing are described in
Section~\ref{section:imaging}. The final characteristics of our BEC
are presented in Section~\ref{section:final}; and
Section~\ref{section:complete} summarizes our complete BEC formation
sequence.

\section{TOP Trap} \label{section:top}
In order to evaporatively cool atoms to the BEC transition in a
magnetic trap, the atoms cannot be held in a quadrupole magnetic
field. This is because as atoms get colder during the evaporation
process, more atoms will occupy the region of the potential minimum
(the $|\vec{B}|=0$ point) and be ejected from the trap by
\mbox{Majorana} losses~\cite{migdall1985fom}. In order to prevent
this problem, two main approaches are most commonly used to create a
BEC in a magnetic trap, the Time-averaged \mbox{Orbiting}
\mbox{Potential} (TOP) trap~\cite{petrich1995stc, anderson1995obe,
ensher1999feb} or the Ioffe-Pritchard (IP)
trap~\cite{pritchard1983cna, davis1995bec, mewes1996bec}.

We use a TOP trap consisting of a pair of anti-Helmholtz coils with
an axis of symmetry along the vertical direction, referred to as the
DC TOP coils, and two orthogonal pairs of Helmholtz coils with an
axis of symmetry in the horizontal plane, referred to as the AC TOP
coils. The TOP trap provides a simple method for creating an axially
symmetric trapping potential.  Also, it is relatively
straightforward to elongate the TOP trap in one of the radial
directions and rotate this elongated trap. This is one way of
nucleating a vortex lattice in the BEC, a procedure that will be
described in Section~\ref{subsection:lattice}.  We have chosen to
use a TOP trap because of its vertical axial symmetry, and the fact
that the most tightly confining axis is in the vertical direction.
This makes it easier to make `pancake-shaped' condensates that can
easily be manipulated with light in a 2-dimensional plane.

%and the relative insensitivity to electrical noise in a
%weakly-confining trap.

\subsection{Physical layout of TOP trap}
\begin{figure}
\begin{center}
\leavevmode
\includegraphics[angle=-90,width=1\linewidth]{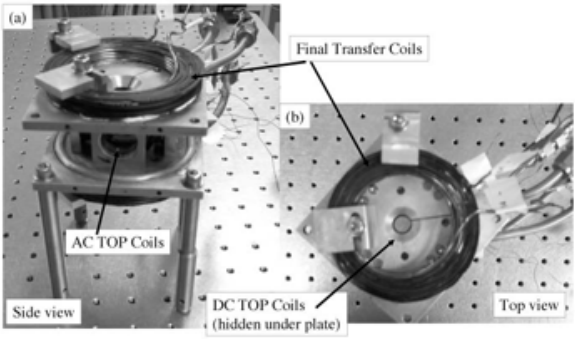}
\end{center}
\caption[Photograph of TOP trap]{Photograph of TOP trap.  (a) Side
view of the TOP trap and mount. (b) Top view of the TOP trap. The
final transfer coils are mounted above and below the water-cooled
copper plates, and are indicated by arrows.  Four aluminum plates
with holes for horizontal optical access to the science cell join
the upper and lower copper plates. On the inside of each of these
four aluminum plates sits one of the AC TOP coils, one of these is
indicated with an arrow. Two of these four aluminum plates are
electrically isolated from the upper copper plate, and the other two
are electrically isolated from the lower copper plate; this provides
us with four (rather than eight) contact points between the four
aluminum plates and the two copper plates, allowing for cooling of
the aluminum plates but minimizing eddy currents. The slot through
the upper copper plate that prevents eddy current flow around the
central hole is visible in (b). The DC TOP coils, obscured from
view, are epoxied to the inside of the upper and lower copper plates
around a hole used for vertical optical access to the science cell.
The electrical connections for the final transfer coils, the DC TOP
coils, the AC TOP coils, the RF coils, and the copper tubing
connections for the water cooling all point off in the same
direction. The entire mount fits around the science cell.}
\label{fig:TOP-trap}
\end{figure}

The TOP trap is an axially symmetric trap created by the addition of
the fields from the two DC TOP coils and the four AC TOP coils. The
DC TOP coils are concentric with the final transfer coils and held
on the same copper mount, but lie closer to the science cell.  The
AC TOP coils are mounted to aluminum plates on all four sides of the
science cell. A photograph of the TOP trap showing a side view and
top view is shown in Figure~\ref{fig:TOP-trap}. A table showing the
locations and other relevant information about the coils surrounding
the science cell is shown in Table~\ref{table:TOP-coils}.

\begin{table}
\begin{center}
\leavevmode
\includegraphics[angle=-90,width=1\linewidth]{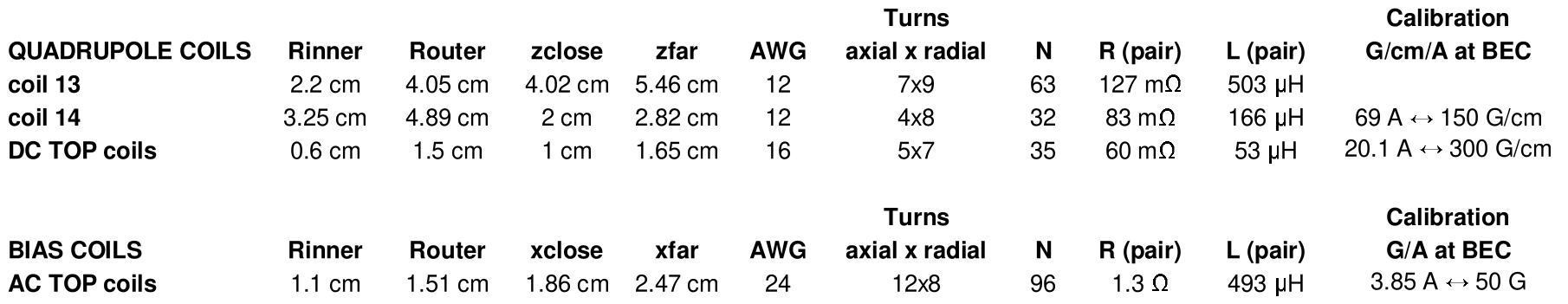}
\end{center}
\caption[Parameters for DC and AC TOP coils]{Information about the
coils surrounding the science cell is tabulated in two sections. The
`QUADRUPOLE COILS' section describes transfer coil pair 13, transfer
coil pair 14 (the final transfer coil pair), and the DC TOP coils.
The `BIAS COILS' section describes the AC TOP coils.  The inner
radius $Rinner$ and outer radius $Router$ of each coil are
tabulated, as  are $zclose$ and $zfar$, defined as the $z$ distance
from the center of the trap to the closest and farthest part of the
coil. The wire gauge AWG, axial and radial number of turns, total
number of turns $N$, resistance $R$, inductance $L$, and coil
calibrations are given. Similar information is provided for the AC
TOP coils, where the coil axis of symmetry is now along the $x$
axis, and the calibration is in G/A for these Helmholtz coils.}
\label{table:TOP-coils}
\end{table}

\newpage

Because of the long evaporation times, typically around 70 seconds,
the DC TOP coils heat up to a temperature that is hotter than any
other coils used in the transfer process.  This could pose a problem
with overheating of the BEC cell, since the coils lie only a few mm
from the glass cell. However, forced air cooling from a directed fan
alleviates the overheating.

\subsubsection{Optical access to the science cell} \label{subsubsection:fnum}
\begin{figure}
\begin{center}
\leavevmode
\includegraphics[angle=-90,width=1\linewidth]{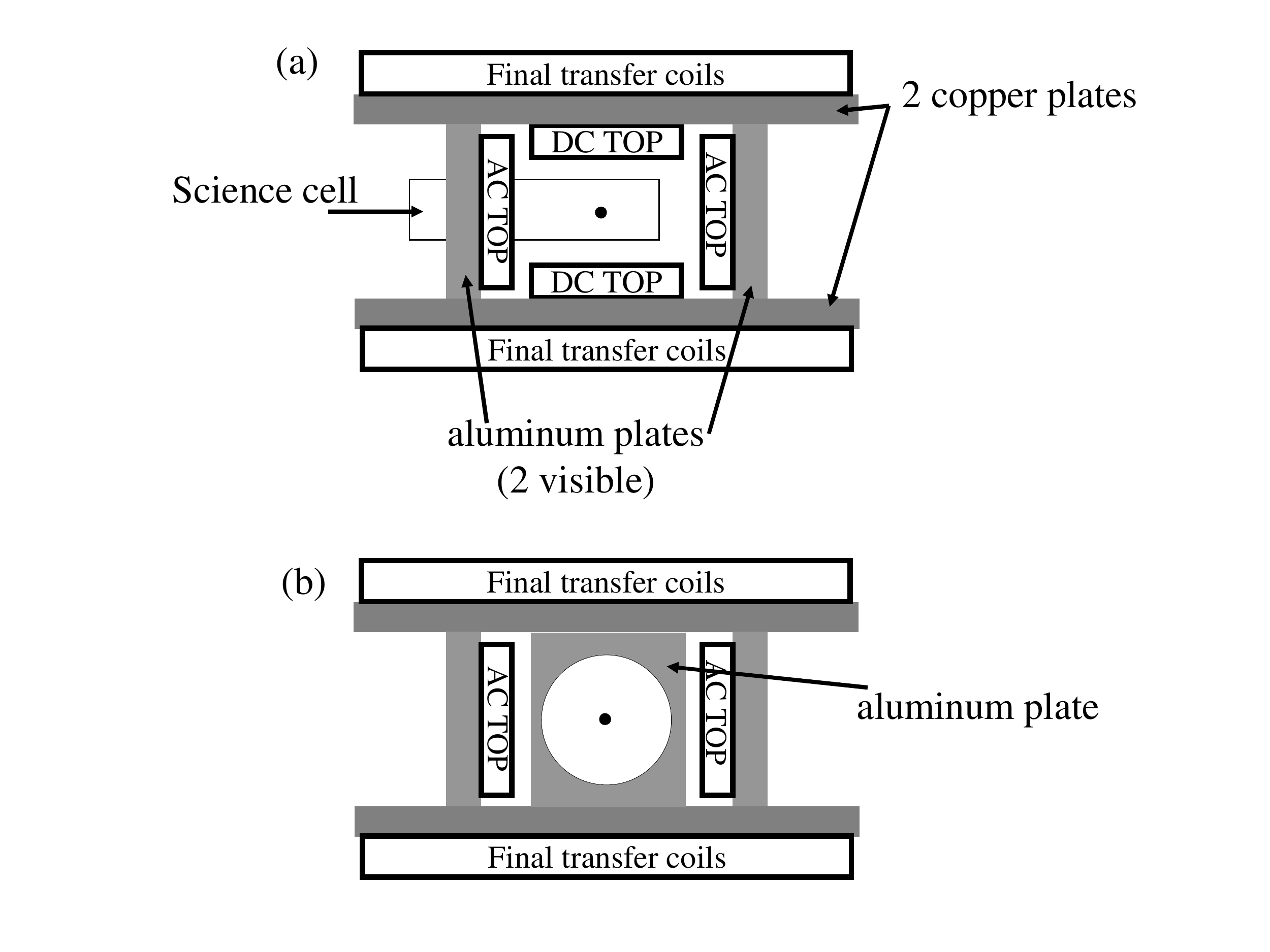}
\end{center}
\caption[Schematic of the TOP trap mount and nearby coils]{Schematic
of the TOP trap mount and nearby coils, highlighting optical access
to the science cell.  (a) Side view of the final transfer coils, DC
TOP coils, AC TOP coils, science cell, and BEC location.  The two
copper plates and two of the four aluminum plates are shown in gray.
(b) A side view of the aluminum plate, with a hole cut for optical
access, is shown.} \label{fig:TOPmount}
\end{figure}

Because we believe that the large amount of optical access to the
science cell is one of the principal advantages of our experimental
system, it will be described in detail in this section.  To
highlight this, a to-scale rendering of the TOP trap mount and
nearby coils is shown in Figure~\ref{fig:TOPmount}.

\newpage

The DC TOP coils have an inner radius of 6~mm, an axial thickness of
7.9~mm, and the closest part of the coils lie 4~mm from the top and
bottom of the science cell. The DC TOP coils are epoxied to a copper
plate around a 12.7~mm diameter hole that allows for vertical
optical access to the science cell. The distance from the BEC to the
copper plate is 18~mm. The AC TOP coils have an inner radius of
11~mm, an outer radius of 16~mm, and a thickness of 6.9~mm.  They
are epoxied to an aluminum plate around a 22~mm diameter hole that
lies 18.6~mm from the BEC.

A useful figure of merit that characterizes the amount of light
collected by an optical system is the $f/\#$, conventionally defined
as $f/\# = f/D$ where $f$ is the focal length of a collecting lens
that is placed a distance $f$ from an object and $D$ is the
\mbox{diameter} of the lens \cite{grievenkamp2004fgt}. This is the
traditional definition given for infinite-conjugate imaging systems,
and is a useful measure for determining the light-collection
efficiency of an optical system. We will modify this definition
somewhat for our finite-conjugate system in order to quantify it. We
will use $f/\# = z/D$, where $z$ is the distance from the object
(the BEC) to the limiting aperture in object space (in theory we
could place a lens there but in practice it is usually a few mm
behind this aperture) and $D$ is the diameter of that aperture.
Using this expression, the $f/\#$ from the top and side are:
\begin{equation}
f/\#_{top} = \frac{18~\mbox{mm}}{12.7~\mbox{mm}} = 1.417
\end{equation}
\begin{equation}
f/\#_{side} = \frac{18.6~\mbox{mm}}{22~\mbox{mm}} = 0.85
\end{equation}

With our open geometry system we achieve $f/0.85$ imaging from three
horizontal directions and $f/1.417$ imaging from two vertical
directions. Despite the fact that the sixth direction faces the
transfer tube, we still make use of it by sending the horizontal
absorption imaging beam to the BEC from that direction. There is
also the possibility of using laser beams to access the science cell
that are oriented in the horizontal plane at 45$^\circ$ to the TOP
trap mount. Although the $f/\#$ here is too poor to use these axes
for imaging, beams can access the science cell through one of the
$\sim$3~mm wide gaps in between the four aluminum plates.

\subsection{Electronics used in TOP trap}
\subsubsection{DC TOP coils}
\begin{figure}
\begin{center}
\leavevmode
\includegraphics[angle=-90,width=0.7\linewidth]{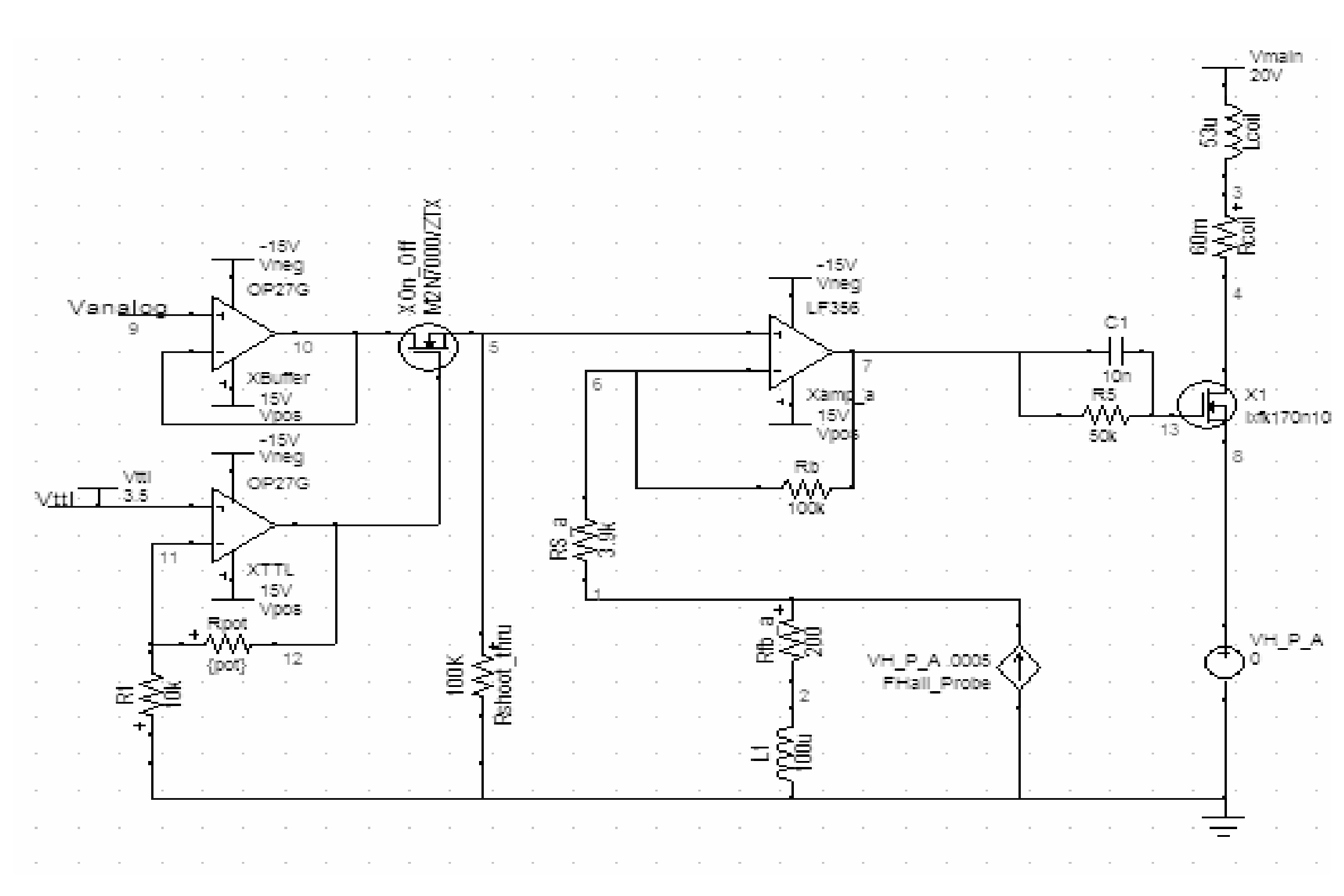}
\end{center}
\caption[Schematic of DC TOP circuit]{Schematic of the DC TOP
circuit.} \label{fig:TOP-circuit}
\end{figure}

The DC TOP coils operate on a circuit conceptually similar to the
transfer circuit, described in Section~\ref{subsection:transfer}.
The coils use the same power supply, there is one IXYS 180N10
MOSFET, a fly-back diode, and Hall Probe in the DC TOP MOSFET box. A
schematic of the circuit is shown in Figure~\ref{fig:TOP-circuit}.
This circuit differs from the transfer circuit of
Figure~\ref{fig:Transfer-circuit} in three ways: (1) this circuit
controls only one pair of coils (the DC TOP coils), rather than
five; (2) only one power MOSFET is used, rather than using one
switching MOSFET and two ramping MOSFETs to operate each coil pair;
and (3) there is an additional inductor in series with a feedback
resistor, as shown in Figure~\ref{fig:TOP-circuit}. Other than these
differences, the principles of operation of the circuit are similar
to those of the transfer circuit, described in
Section~\ref{subsection:transfer}.

Using only one MOSFET in this circuit is the principal drawback in
this design, as the DC TOP MOSFET is the most frequently blown-up
MOSFET in our experiment, despite the fact that it is the only
MOSFET that receives forced air cooling by a directed fan.  We are
currently installing a new circuit that includes two power MOSFETs,
two Hall Probes, and two separate feedback loops to operate the DC
TOP coils. There are numerous safety interlocks on the DC TOP coil
temperature, integrated current, and instantaneous current, as
described in Section~\ref{subsection:interlocks}.

\subsubsection{AC TOP coils}
The AC TOP coils provide a rotating radial bias field that acts to
shift the $|\vec{B}|=0$ point away from the origin to a circle in
the horizontal plane referred to as the circle of death $R_{cod}$
\cite{ensher1999feb}.
\begin{equation}
R_{cod}=\frac{B_{bias}}{\partial B/\partial r}
\end{equation}
The time average of this effect is the creation of an harmonic
potential with a minimum at the origin.  Atoms that have enough
total energy to access the outer reaches of this time-averaged
potential will experience a $|\vec{B}|=0$ point and be ejected from
the trap due to Majorana losses.

\begin{figure}
\begin{center}
\leavevmode
\includegraphics[angle=-90,width=0.9\linewidth]{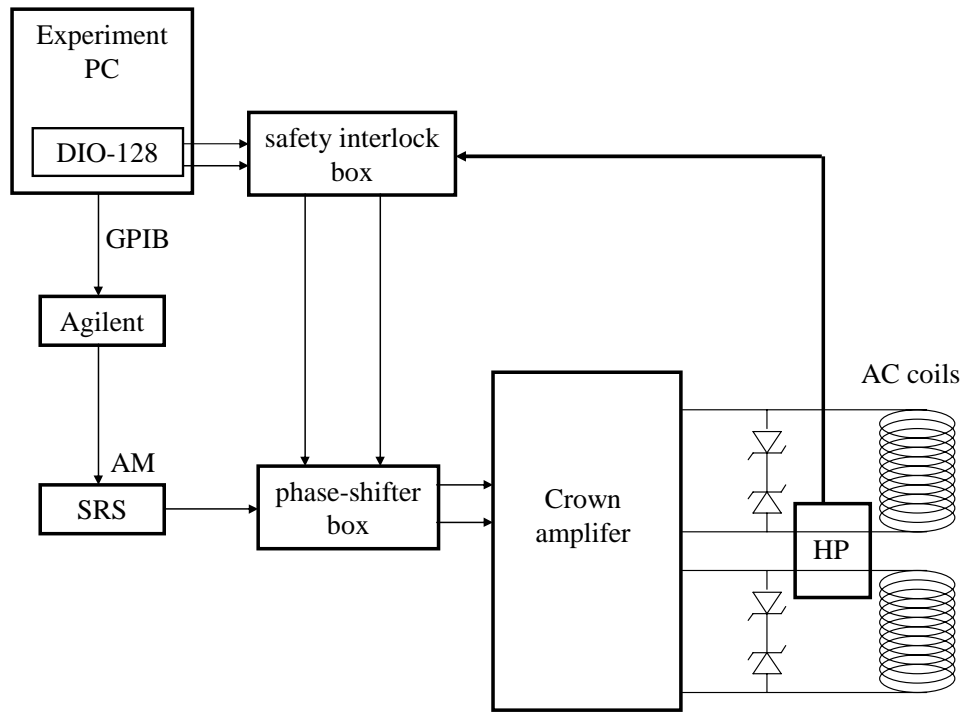}
\end{center}
\caption[AC TOP coils driving circuit]{Schematic of the AC TOP coils
driving circuit.  Two orthogonal pairs of coils are connected to the
output of the Crown amplifier.  There are two pairs of
oppositely-oriented Zener diodes placed in parallel with each of the
two pairs of coils. A hall probe (HP) reads the amount of AC current
flowing simultaneously through both pairs of coils, and sends this
signal to the safety interlock box.  If the amount of current is too
high, the safety interlock box will switch off the two TTL signals
going to the phase-shifter box, shutting off the current flow from
the Crown.} \label{fig:AC-coils}
\end{figure}

A schematic showing the hardware and electronics used to operate the
AC TOP coils is shown in Figure~\ref{fig:AC-coils}.  A Crown XLS402
power amplifier is used to drive the AC TOP coils, and an SRS
function generator is used to create the 2~kHz sine wave that is
sent to the Crown amplifier.

\newpage

Over the $\sim$70~second evaporation process, the amplitude of the
rotating radial bias field decreases from 41~G to 5~G.  The
amplitude of this waveform is created in software on the experiment
PC and loaded onto an Agilent 33120A arbitrary waveform generator
through a GPIB connection.  The Agilent output signal is applied to
the Amplitude-Modulation (AM) input of an SRS DS345 function
generator to produce an AM-modulated 2~kHz sine wave.

\begin{figure}
\begin{center}
\leavevmode
\includegraphics[angle=-90,width=1\linewidth]{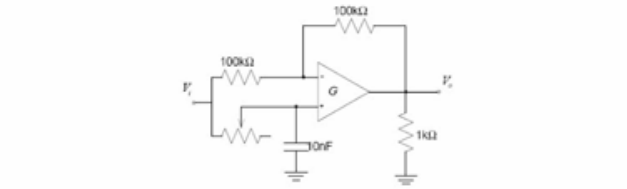}
\end{center}
\caption[Phase-shifter circuit]{Schematic of the phase-shifter
circuit. The circuit uses an OP27 op amp and produces an output
$V_o$ that is a phase-shifted version of the input $V_i$, based on
the resistance of the 20~k$\Omega$ potentiometer.}
\label{fig:Phase-shifter}
\end{figure}

The role of the phase-shifter box is to convert its sine wave input
into two outputs: an identical version of the sine wave input, and a
90$^\circ$ phase-shifted version of its input.  The phase-shifter
circuit operates on a circuit similar to that shown in
Figure~\ref{fig:Phase-shifter}.

%Th two outputs of the phase-shifter circuit drive the orthogonal
%pairs of AC TOP coils, and cause the amplitude of the resultant bias
%field to trace out a circle in the radial plane.

The phase-shifter box has one analog input (from the SRS) and two
logic inputs, one for each channel of the Crown.  When both logic
inputs are high, the phase-shifter box outputs a sine wave identical
to its input on one channel, and a 90$^\circ$ phase-shifted version
of this signal on the other channel.  These signals are sent to the
two inputs on the Crown amplifier.  The two connections on each
channel of the {\newpage}

\noindent Crown's output are connected to either end
of each of the two pairs of AC TOP coils.

There are two kinds of safety interlocks on the Crown amplifier, but
neither are as robust as the safety interlocks on the Agilent power
supply, where an external TTL signal can instantly disable the
supply.  Because the Crown does not have this function, we had to
invent ways of preventing unwanted current from flowing in case of a
computer or electronics error.  The first thing we did was to apply
two back-to-back $-68$~V zener diodes across the output of each
channel of the Crown, in parallel with the AC TOP coils.  These act
to protect the amplifier from a voltage spike, such as what would
occur if the power were suddenly lost.  If a voltage spike of more
than 68~V is generated, the zener diodes turn on and shunt current
through the diodes instead of back into the amplifier.

The second safety interlock monitors how much current is flowing
through the AC TOP coils with a Hall Probe and disables the two
logic inputs to the phase-shifter box if the current is too high. AC
current through both pairs of coils passes through the same Hall
Probe, which sends a signal to the safety interlock box.  The safety
interlock box rectifies this signal, integrates it, and if the
integrated value is too high, turns off the two logic signals sent
to the phase-shifter box, stopping current flow. There is a relay in
the safety interlock box that prevents further current flow until a
front-panel switch is depressed.

\subsection{Transfer into TOP trap}
Immediately after transferring atoms into the science cell, we apply
a short pulse of light near-resonant with the $|F=2 \rightarrow
F'=3\rangle$ transition to drive away any wayward atoms that are
still magnetically trapped in the $|F=2\rangle$ state in the science
cell due to imperfect optical pumping into $|F=1\rangle$ in the MOT
cell. This pulse, referred to as the `kill pulse', is 5~ms long and
the process is not very dependent on laser intensity or detuning.
Using the kill pulse will increase the number of atoms in the final
$|F=1,m_F=-1\rangle$ BEC by removing any $|F=2\rangle$ atoms, which
only provide collisions that are deleterious to the evaporative
cooling process.

After the kill pulse, the $|F=1\rangle$ atoms are still trapped in a
quadrupole field created by the final transfer coil pair, coil pair
14, at an axial field gradient of 180~G/cm. The atoms are then
transferred into a quadrupole field created by the DC TOP coils,
concentric with coil pair 14. This is done by ramping the current in
coil pair 14 down to zero while ramping up the DC TOP coils from
zero to 180~G/cm over 100~ms. The atoms are then transferred into
the TOP trap by switching the axial field gradient in the DC TOP
coils from 180~G/cm to 300~G/cm and simultaneously turning on the
2~kHz rotating radial bias field to 41~G.

We have held atoms in the 180~G/cm quadrupole trap in the science
cell for as long as 8~minutes, and were limited in a measurement of
the TOP trap lifetime due to heating of the DC TOP coils.

\section{RF Coils}
We use a radio-frequency (RF) magnetic field in the region from
$4-70$~MHz for the primary evaporative cooling of the atomic cloud;
this occurs in addition to the TOP-induced evaporation caused by
decreasing the radius of the circle of death. This RF field causes
transitions from the magnetically trapped $|F=1,m_f=-1\rangle$ level
to the untrappable $|F=1,m_f=0\rangle$ and $|F=1,m_f=1\rangle$
levels, thereby ejecting atoms from the TOP trap. The two RF coils
are each made up of one single loop of magnet wire, above and below
the BEC cell, wired in Helmholtz configuration.  The RF coils, which
are wrapped around the outside of the DC TOP coils, have a diameter
of 3~cm and lie $\sim$1~cm from the plane of the BEC; only the glass
surface of the science cell lies between the RF coils and the BEC.

The desired RF frequencies are created with an Agilent E4400B RF
synthesizer.  This signal is then sent through a Mini-Circuits
ZYSW-2-50DR RF switch before being amplified by an OPHIR 5303055 RF
amplifier, then sent to the RF coils.
% where are the RF coils in the picture of the TOP trap?

\section{Evaporative Cooling} \label{section:evap}
Evaporative cooling is the process by which the hottest atoms are
ejected from the trap and the remaining atoms are allowed to
rethermalize at a lower temperature \cite{davis1995ame,
sackett1997oec, ketterle1996ect}. This process occurs while the
atoms are held in the TOP trap in the science cell. The evaporative
cooling process will eventually eject $\sim99\%$ of the atoms from
the TOP trap, but is the process by which the remaining atoms are
cooled to the critical temperature necessary to reach the BEC
transition.

%Because only the hottest atoms in the trap are energetically able to
%oscillate out to the circle of death, they will be the atoms that
%are ejected (evaporated) from the trap.
%Evaporative cooling is accomplished by two processes.  First, the
%amplitude of the radial bias field is reduced, bringing the
%$|\vec{B}|=0$ point (the circle of death) closer to the origin.
%Second, the frequency of the RF magnetic field is reduced. Because
%the atoms are in a spatially varying magnetic field, the cloud
%experiences spatially varying Zeeman shifts.  The RF frequency is
%tuned to eject the hottest atoms (those that happen to sit at the
%periphery of the cloud) from the trap by driving them from the
%trapped $|F=1,m_f=-1\rangle$ state to the untrappable
%$|F=1,m_f=0\rangle$ state.

The evaporation process occurs in two phases, described in the
following two sections.  In brief, Phase I occurs in the tight TOP
trap and cools the atoms to a point just above the BEC critical
temperature. Then, the axial field gradient is reduced, placing the
atoms in a weakly confining potential that allows us to generate
spatially larger BECs. The Phase II evaporation, which cools the
atoms through the BEC critical temperature, occurs in this
weakly-confining TOP trap.

\subsection{Initial evaporation in the tight TOP trap}
Phase I evaporation occurs over a period of 72~seconds in the tight
TOP trap with a magnetic field gradient of 300~G/cm.  In the first
2~seconds the bias field is held constant at 41~G and the RF
frequency is reduced from 70~MHz to 57~MHz.  This fast sweep acts to
eject the hottest atoms from the trap.  Over the next 58~seconds the
RF frequency ramps down from 57~MHz to 7~MHz, and the amplitude of
the rotating bias field ramps down from 41~G to 5~G. During this
part of the evaporation sequence, the RF frequency changes along
with the bias field in such a way that the RF frequency is always
ejecting atoms that lie directly at the circle of death.  All of the
ramps used in our evaporation scheme are exponentially decreasing
ramps with an adjustable time constant. The 58~second ramps
described above follow an exponential decay with a 100~second time
constant.

For the rest of the evaporation sequence the bias field is left at
5~G and all additional evaporation is done solely with the RF
frequency. In the next 12~seconds the RF frequency is brought from
7~MHz to 4.5~MHz, this cools the atoms to a point just above the BEC
critical temperature. This 72~second sequence concludes the Phase I
evaporation sequence in the tight TOP trap.

\subsection{Evaporative cooling in a weakly-confining TOP trap} \label{subsection:weak}
The atoms are then moved to the gravity-sagged TOP trap by reducing
the axial magnetic field gradient from 300~G/cm to 55~G/cm over
2~seconds, decreasing the radial trapping frequency from 40~Hz to
7.4~Hz. This reduction in field strength uses a nonlinear current
ramp that approximates a linear shift in the $|\vec{B}|=0$ point
over the 2~second sag time. The atoms sag 0.6~mm due to the larger
role played by the gravitational potential in the weak TOP trap
compared to the tight TOP trap \cite{ensher1999feb}. During the sag
process the RF frequency is increased to 5~MHz to prevent unwanted
additional evaporative cooling.

Phase II evaporation occurs in the gravity-sagged TOP trap with an
axial field gradient of 55~G/cm and radial bias field of 5~G.
Because the atoms no longer lie at the center of the TOP trap coils,
the Zeeman shifts are different at the top and bottom of the cloud,
and evaporation occurs primarily closer to the bottom (the south
pole) of the atomic cloud \cite{thomas2004ucc}. This
reduced-dimension evaporation is less efficient than the \mbox{2-D}
(around the equator) evaporation afforded by evaporation in a tight
TOP trap. Phase II evaporation reduces the RF frequency from 5~MHz
to 4.7~MHz over 10~seconds, using a 2~second exponential decay time
constant, after which a pure BEC is observed.

%The phase II evaporation time can be decreased to as short as
%5~seconds, but there will be a reduction in the number of atoms in
%the eventual BEC.

\begin{figure}
\begin{center}
\leavevmode
\includegraphics[angle=0,width=1\linewidth]{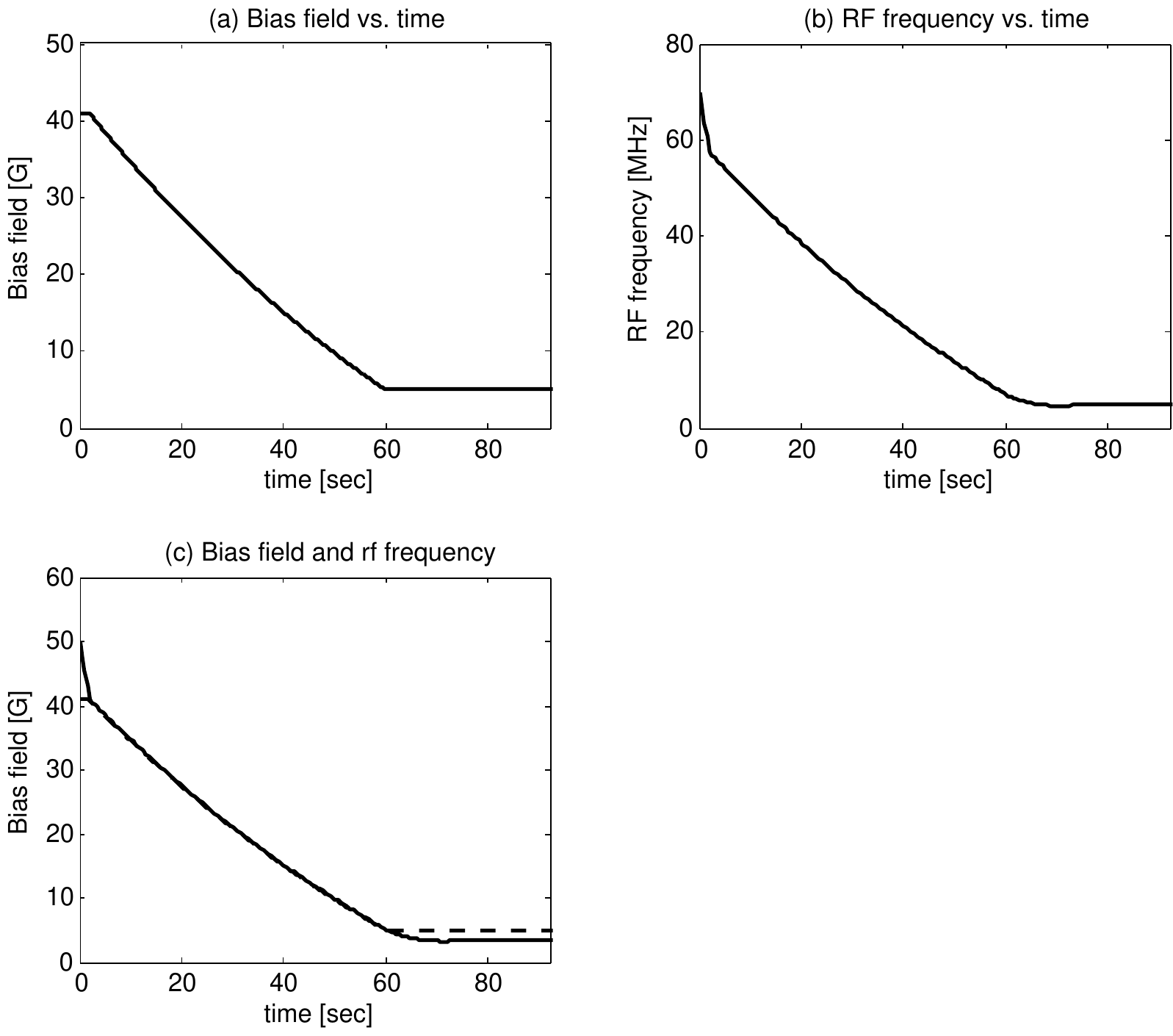}
\end{center}
\caption[Evaporative cooling sequence]{Evaporative cooling sequence.
(a) Shown is the bias field [G] vs.\ time~[s].  (b) Shown is the RF
frequency [MHz] vs.\ time [s].  (c) Shown is the bias field and RF
frequency plotted together, with the RF frequency shown as a solid
line and the bias field shown as a dashed line. The gradient field
from the DC TOP coils is 300~G/cm until time $t=72$~s, after which
it ramps down to 55~G/cm over 2~seconds.} \label{fig:Evap-scheme}
\end{figure}

Plots of the bias field strength and RF frequency vs.\ time used in
the above evaporation sequence are shown in
Figure~\ref{fig:Evap-scheme}. Figure~\ref{fig:Evap-scheme}~(a) shows
the bias field ramp vs.\ time over the course of the Phase I and
Phase II evaporation steps. Figure~\ref{fig:Evap-scheme}~(b) shows
the RF frequency ramp vs.\ the same time.
Figure~\ref{fig:Evap-scheme}~(c) shows the bias field and RF
frequency plotted together, with the RF evaporation point converted
to G to show both the RF evaporation point and the TOP-induced
evaporation point on the same scale.

To understand how to plot both quantities on the same scale,
consider the following example\footnote{For addition information,
see the lab notebook entry on 3/25/2005.}. If the instantaneous
$|\vec{B}|=0$ point happens to be at $x=R_{cod}$, the point in the
cloud that experiences the largest magnetic field will be at
$x=-R_{cod}$, where the magnitude of the field is $B_{max}=2 \cdot
\partial B/\partial r \cdot R_{cod} = 2 B_{bias}$.  Knowing that the Zeeman
splitting between the $|F=1,m_f=-1\rangle$ and $|F=1,m_f=0\rangle$
states is 0.7~MHz/G for $^{87}$Rb \cite{metcalf1999lca}, the
appropriate RF frequency $f$ needed to evaporate away atoms based on
the bias field $B_{bias}$ is $f = 2 \cdot B_{bias} \cdot 0.7~$MHz/G.
In Figure~\ref{fig:Evap-scheme}~(c) it is clear that the RF
frequency keeps pace with the TOP-induced evaporation point for most
of the evaporation sequence, until the end of the sequence when it
alone continues to decrease and acts to evaporate atoms from the
trap.
% for a plot showing the COD and evaporation point, see 3/25/05

%\paragraph{Unwanted consequences of RF}
%The propagation of RF frequencies through an antenna has other
%serious consequences in our lab, such as driving lasers and
%temperature controllers out of lock.  For frequencies in the range
%of 11 to 16~MHz, we need to decrease the RF power from its nominal
%value of -5~dBm to -15~dBm, and between 16 and 23~MHz we decrease
%the power to -23~dBm. The RF coil also has different responses to
%different frequencies, and we boost the power in the ranges from 31
%to 36~MHz as well as 43 to 52~MHz up to -3~dBm.
%%(!!CHECK ATTENUATOR!!)

\section{Imaging} \label{section:imaging}
100$\%$ of the data that comes from our experiment results from the
analysis of images of atomic clouds.  Because of this, the ability
to take images of these atomic clouds and extract useful information
from them is a major component of our experiment.  We have one
computer dedicated to image acquisition and analysis.  This imaging
PC (shown in Figure~\ref{fig:PC-and-boards}) communicates with a
Princeton Instruments/Acton PIXIS 1024BR CCD camera via a USB
connection, and the camera receives a TTL trigger from the
experiment PC via a digital line the DIO-128 board's Port A. The
following sections describe the design and layout of our imaging
systems, a description of absorption and phase-contrast imaging, and
a summary of our image processing routine.

\subsection{Overview, physical layout of imaging systems}
Atoms are trapped in the science cell in a TOP trap that is axially
symmetric with the axis of symmetry aligned along the direction of
gravity.  The vortices created in our experiment are a signature of
orbital angular momentum about this axis, and will thus be visible
only by imaging along the vertical direction. Our experiment
includes a vertical and horizontal imaging system, for reasons
described below.

\subsubsection{Imaging along multiple axes}
\begin{figure}
\begin{center}
\leavevmode
\includegraphics[angle=-90,width=1\linewidth]{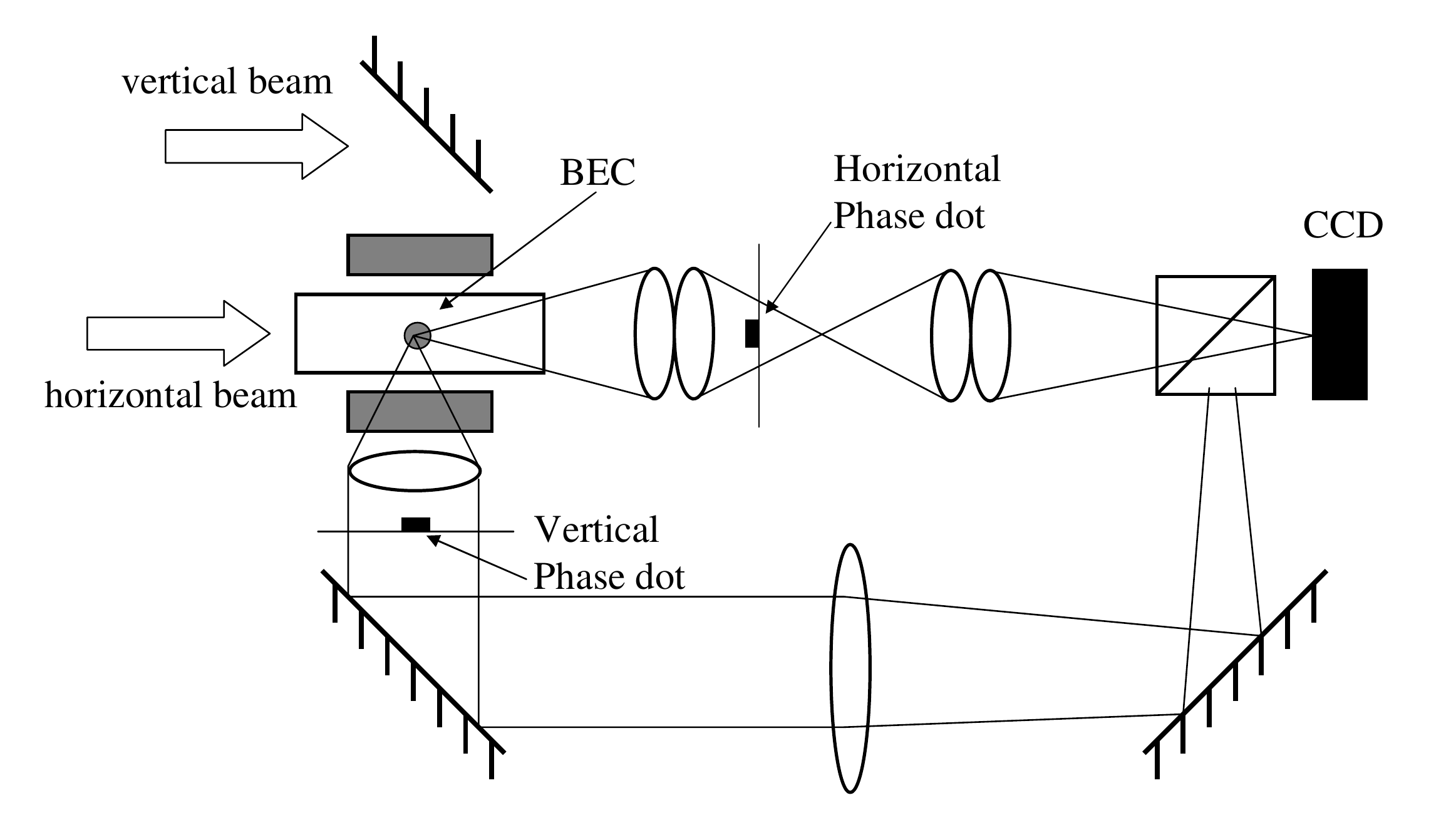}
\end{center}
\caption[Layout of horizontal and vertical imaging systems]{Layout
of horizontal and vertical imaging systems.  The horizontal imaging
system contains two sets of paired achromats, and includes an
intermediate focal plane between the achromat pairs.  The horizontal
probe beam focuses onto the plane of the horizontal phase dot.  The
vertical imaging system contains two imaging achromats, and has a
collimated intermediate image space.  The vertical probe beam
focuses onto the plane of the vertical phase dot.  Both imaging
systems have an optical magnification of 5.36.}
\label{fig:Imaging-layout}
\end{figure}

A schematic showing the layout of the vertical and horizontal
imaging systems is shown in Figure~\ref{fig:Imaging-layout}.  The
BEC is imaged onto the CCD camera via the four lenses that comprise
the horizontal imaging system, and is also imaged onto the CCD
camera via the two lenses of the vertical imaging system.

The horizontal system uses an imaging beam that passes through the
MOT cell, travels 80~cm down the transfer tube, and exits through
the end port of the science cell, passing through two pairs of
paired achromats on its way to the CCD camera.  The vertical imaging
beam passes through the two 1~mm thick glass surfaces of the science
cell, which are separated by 10~mm.  Because these glass surfaces
are close to parallel, the beam undergoes multiple reflections
within the science cell that put visible interference fringes on the
beam. When imaging a small BEC using phase-contrast imaging, these
fringes can significantly degrade image quality.  The horizontal
axis does not suffer from any of these multiple reflections due to
the glass science cell. Additionally, the use of only two achromats
on the vertical imaging system rather than the two pairs of paired
achromats (four total) on the horizontal axis produces a slightly
larger diffraction-limited spot size on the vertical system.  The
sum total of these effects has meant that the images on the vertical
system have never been as high in image quality as those achievable
on the horizontal system.  We typically use the horizontal system
for in-trap phase-contrast imaging of the cloud to characterize the
BEC, and the vertical imaging system for absorption imaging of the
cloud after ballistic expansion.

%Another drawback to the vertical imaging system as compared to the
%horizontal is the longer overall beam path, requiring a 2" diameter
%lens as the second lens.  This will mean that there is slightly more
%vignetting on the vertical axis.

For phase-contrast imaging on either axis, a phase dot is slipped
into or out of the focus of the probe laser beam in order to shift
the phase of the laser light by 3$\pi$/2 relative to the BEC image
light, which is spatially much larger than the 100~$\mu$m diameter
phase dot.  The 100~$\mu$m diameter phase dot was made using the
Maskless Lithography Tool at the College of Optical Sciences; this
machine will be described in Section~\ref{subsection:making} on
Page~\pageref{subsection:making}.

Both imaging systems have an optical magnification of 5.36, which
was measured experimentally on the horizontal axis by calibrating
the distance a cloud of atoms fell vertically due to gravity when
released from the trap. The vertical axis was then cross-calibrated
against the horizontal axis by ensuring that the cloud sizes on both
axes were the same when both systems were in focus.  Each system was
focussed by using on-resonance absorption imaging to image as small
a cloud as possible immediately after release from a tight TOP trap.
On both axes the first lens in the imaging system (lens pair on the
horizontal system) was placed on a translation stage.  The lens
position was optimized to produce the smallest size atom cloud in
the image plane.  By tuning a few linewidths on either side of
resonance, lensing of the probe laser beam by the BEC is visible
(see Ref.~\cite{matthews1999tcb} Section~4.1.1). The high optical
depth of a small BEC can cause an off-resonant probe beam to refract
through the BEC, which acts like a lens. The sign of the lensing
caused by the BEC will change on either side of the focus position,
and analysis of the defocussed clouds can help to set the focus
position correctly.

\subsubsection{Optical design}
\begin{figure}
\begin{center}
\leavevmode
\includegraphics[angle=90,width=1\linewidth]{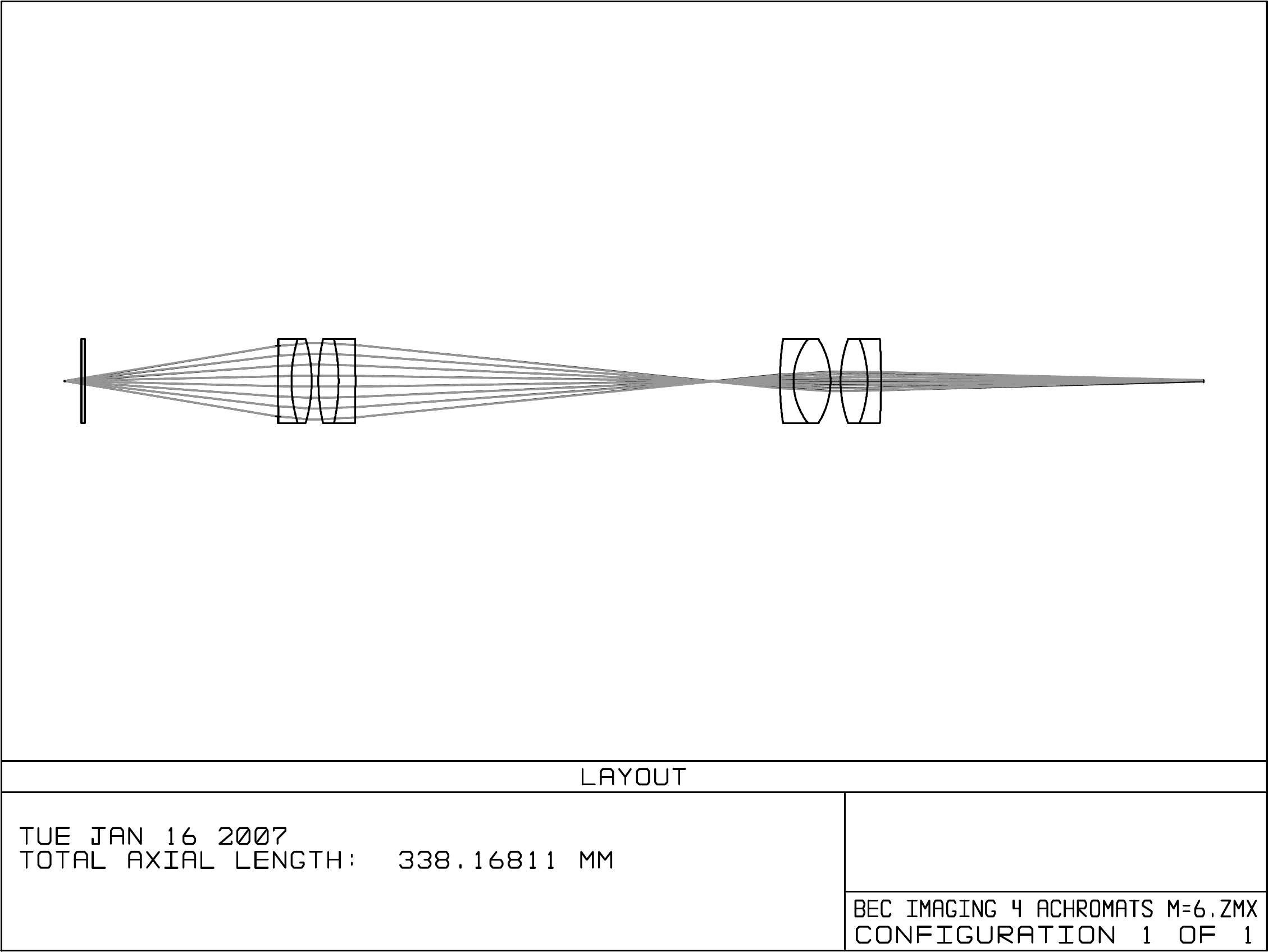}
\end{center}
\caption[ZEMAX layout of horizontal imaging system]{Layout of
horizontal imaging system, generated using ZEMAX.  The BEC is imaged
through a 1~mm thick plate of glass (a simulation of the glass
science cell), through 4 achromats, and onto the CCD camera.  The
lenses used are all Edmund Optics TECH-SPEC Near-Infrared Achromats
with focal lengths of $f=75$~mm, $f=100$~mm, $f=35$~mm, and
$f=60$~mm, from left to right.} \label{fig:layout}
\end{figure}

\begin{figure}
\begin{center}
\leavevmode
\includegraphics[angle=90,width=1\linewidth]{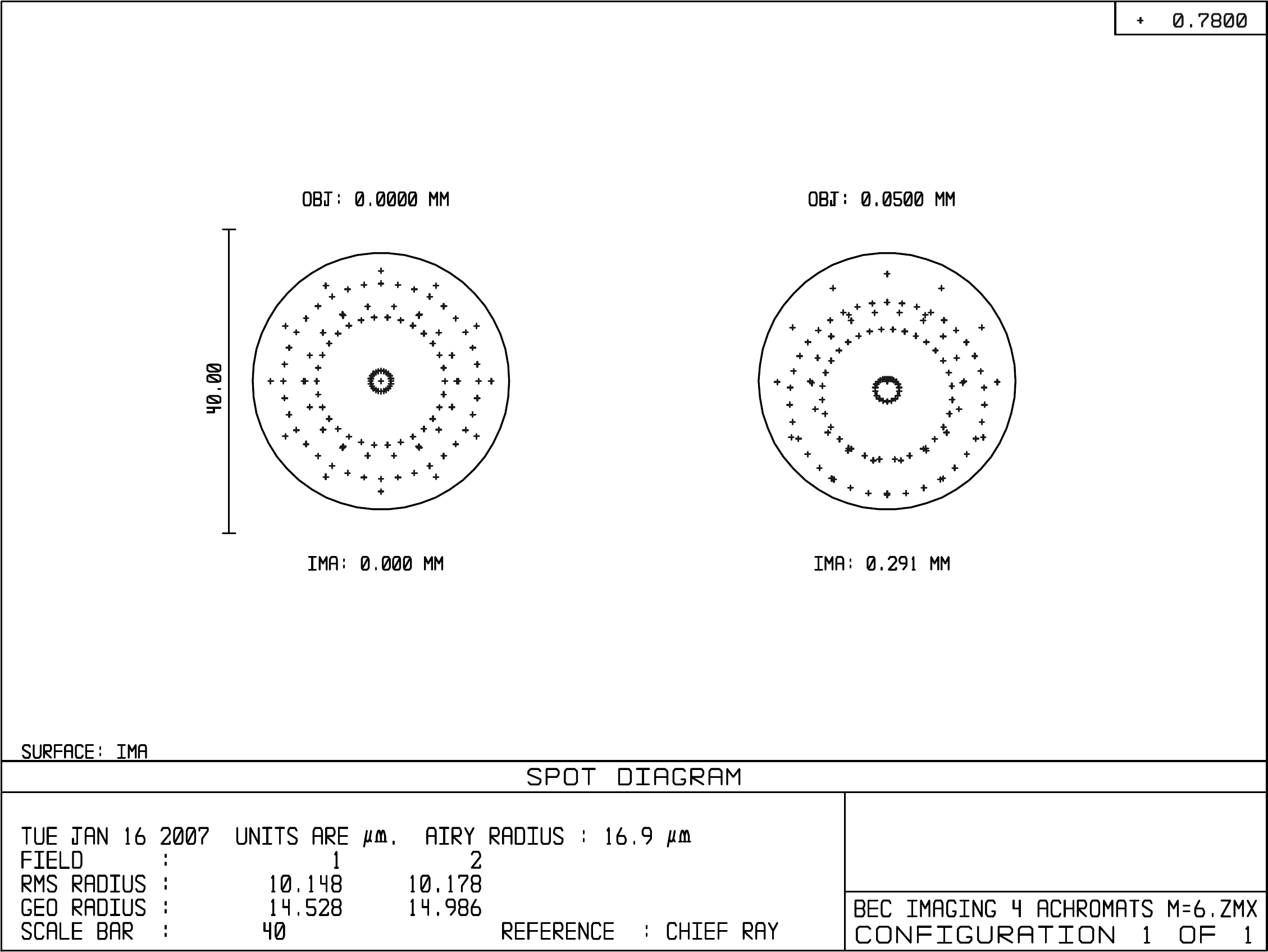}
\end{center}
\caption[ZEMAX spot diagram for horizontal imaging system]{Spot
diagram of the horizontal imaging system, generated using
\mbox{ZEMAX}.  The spots from the marginal and chief rays all fall
clearly within the Airy disk of radius 16.9~$\mu$m, indicating a
diffraction-limited optical system.} \label{fig:spotdiagram}
\end{figure}

Both the horizontal and vertical imaging systems were designed using
the optical design software ZEMAX, which allowed us to design
diffraction-limited optical systems.  The layout for the horizontal
imaging system is shown in Figure~\ref{fig:layout}. A spot diagram
showing the rays traced through the horizontal system for a
100~$\mu$m diameter BEC is shown in Figure~\ref{fig:spotdiagram}.

\newpage

\subsection{Absorption imaging} \label{subsection:abs}
We use absorption images taken on the vertical axis to image the
cloud after a period of expansion, as described  below.  It is
necessary to allow the cloud to expand in order to optically resolve
vortex cores, which would not be visible using in-trap
phase-contrast images because they are smaller than the optical
diffraction limit.

In order to image the vortex cores in our experiment we need to
allow the cloud to expand in size by a factor of $\sim$5. The atomic
cloud can only fall in free space for about 30~ms before it hits the
bottom of the science cell 5~mm away; this is not enough time for
the cloud to expand enough for vortex cores to be observable. In
addition to the fact that using a short ballistic expansion time
would not allow the cloud to expand enough, it would also place the
cloud at a different vertical position, requiring a change in lens
position to focus the expanded cloud on the CCD camera.  We needed
to find a way of allowing for ballistic expansion times of up to
$\sim$70~ms while keeping the cloud in place. We solved this problem
by using a levitation coil during the expansion phase.

\subsubsection{Levitation coil} \label{subsubsection:levitation}
% see 6/8/06 in lab notebook
Our absorption imaging expansion scheme incorporates a levitation
coil\footnote{For additional information, see the lab notebook entry
on 6/5/2006.} that operates in conjunction with the DC TOP coils to
provide a field gradient that is similar to that obtained in a QUIC
trap \cite{esslinger1998bec}. The difference is that in our case the
single levitation coil is oriented along the axis of symmetry of the
DC TOP coils rather than orthogonal to it. The basic idea of
operation is that the combined field gradient provides a very weak
harmonic trapping potential in the axial (gravity) direction
centered at the position of the trapped BEC. In our case, the
trapped BEC is initially held in a gravity-sagged weakly-confining
TOP trap. The radial field can be very weakly trapping or
anti-trapping. This is achieved when the DC TOP coils operate at a
field gradient near 55~G/cm, the levitation coil produces a field
near 32.4~G, and the AC TOP coils are off. After 56~ms of hold time
in the combined magnetic fields of the DC TOP coils and the
levitation coil, the center of the atomic cloud will not have
changed in vertical position, making it still in-focus in the
vertical imaging system, whereas the cloud will expand by a factor
of $\sim$5 radially.

The levitation coil has an inner diameter of 39~mm and is made from
18~AWG magnet wire, with 8, 8, 7, 7, 8, and 7 turns, for a total of
45 turns.  We have measured experimentally that the coil calibration
is 2.7~G/cm/A when the closest part of the coil is placed 2~cm from
the BEC.  The levitation coil operates on the same power supply and
circuit topology as the MOT, transfer, and DC TOP coils.  A Hall
Probe monitors how much current is flowing through the levitation
coil, and will inhibit the power supply if the instantaneous amount
of current is greater than 15~A.

\subsubsection{Absorption imaging sequence} \label{subsubsection:absimaging}
The formation of a BEC occurs in a weak TOP trap in the science cell
as previously described. After BEC formation, a period of 56~ms of
expansion occurs with the AC TOP coils off, the levitation coil at
32.4~G, and the DC TOP coils at 55~G/cm.  We create a condensate in
the $|F=1,m_F=-1\rangle$ state, but the near-resonant light used for
absorption imaging drives the $|F=2 \rightarrow F'=3\rangle$
transition, so we optically pump the atoms from $|F=1\rangle$ into
$|F=2\rangle$ by using a short pulse of light, referred to as the
`repump flash', in resonance with the $|F=1 \rightarrow F'=2\rangle$
transition. The repump flash occurs in free space, using a pulse of
10~$\mu$s duration and $\sim$~1~mW/cm$^2$ intensity.  We have found
that longer or more intense repump flash pulses can significantly
degrade the quality of absorption images, and can cause an increase
in size of the atomic cloud by a factor of 2 to 3. Immediately after
the repump flash pulse, we take an image by using a 20~$\mu$s
exposure with an intensity on the order of $I_{sat}$.

\subsubsection{Absorption image processing}
\begin{figure}
\begin{center}
\leavevmode
\includegraphics[width=1\linewidth]{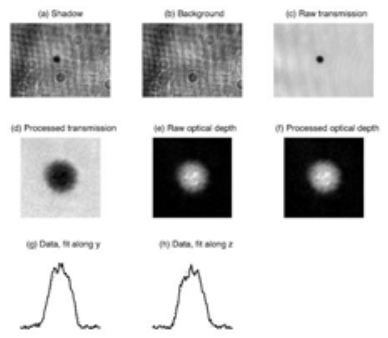}
\end{center}
\caption[Vertical absorption image]{Vertical absorption image. (a)
Shown is the $shadow$ frame, taken with the atoms present. (b) Shown
is the $background$ frame, taken without atoms present.  (c) The raw
transmission, where the black dot represents atoms.  (d) The
processed transmission, where the black hole represents atoms. (e)
The raw optical depth, where the white hole now represents atoms.
(f) The processed on-resonance optical depth.  Images (g) and (h)
show the data (black line) and a corresponding fit to a Thomas-Fermi
distribution (dashed line) along the $y$ (horizontal) and $z$
(vertical) directions, respectively.  The scale on images (d) - (h)
is $121 \times 121~\mu$m.} \label{fig:vert-abs}
\end{figure}

% from Fit2D_v9.m with 2006-08-27 file bec11 bec fit, 50 x 50 pixels
% don't have that file on my PC =(
A typical vertical absorption image taken after 56~ms of expansion
using the levitation coil expansion scheme is shown in
Figure~\ref{fig:vert-abs}.

Three frames on the CCD camera are taken per data run, one with the
atoms present, one without atoms, and a dark frame with no light
present. Figure~\ref{fig:vert-abs}~(a) shows the $shadow$ and
Figure~\ref{fig:vert-abs}~(b) shows the $background$, where $shadow$
and $background$ both have the dark frames already subtracted off.
Figure~\ref{fig:vert-abs}~(c) shows the raw transmission $T$,
defined as $T=shadow/background$. Figure~\ref{fig:vert-abs}~(d)
shows the processed transmission, which is just a centered and
zoomed-in version of the raw transmission.
Figure~\ref{fig:vert-abs}~(e) shows the raw optical depth $OD_r$,
defined as
\begin{equation}
OD_r = -ln(T)
\end{equation}
for near-resonant absorption imaging. Figure~\ref{fig:vert-abs}~(f)
shows the processed optical depth, which differs from the raw
optical depth due to two nonlinear corrections.

\newpage

We first apply a correction to account for the maximum optical depth
measurable in our system\footnote{This is discussed in
Ref.~\cite{matthews1999tcb} section 4.1.}, which we have determined
to be $OD_{max} \cong 4$. Defining $\alpha=e^{-OD_{max}}$, the new
value for the $OD$ is
\begin{equation}
OD_a = \frac{shadow - \alpha \cdot background}{background - \alpha
\cdot background}
\end{equation}
The second correction accounts for saturation effects and finite
detuning\footnote{This is discussed in Ref.~\cite{ensher1999feb} eq.
5.6.}.
\begin{equation}
\beta = \frac{I/I_{sat}}{1 + 4 \cdot (\Delta / \Gamma)^2}
\end{equation}
\begin{equation}
OD_b = [OD_a + (1-e^{-OD_a}) \cdot \beta] \cdot [1 + 4 \cdot (\Delta
/ \Gamma)^2]
\end{equation}

We then use an image processing routine that spatially filters the
value for $OD_b$ to remove the effects of interference fringes at
specific spatial frequencies.  Figures~\ref{fig:vert-abs}~(g) and
(h) show the data (from the processed optical depth) in solid lines
and a Thomas-Fermi fit to the data in dashed lines along the $y$ and
$z$ axes.  %The scale of the images is $121 \times 121~\mu$m.

\subsection{Phase-contrast imaging}
We use phase-contrast images of the BEC taken along the horizontal
axis to determine BEC characteristics such as number, temperature,
and size. Because of the better image quality on this axis, we
almost exclusively use the horizontal system for characterization of
the atomic cloud in the science cell.

\subsubsection{Phase-contrast imaging sequence}
We typically image the atoms in the weak TOP trap or immediately
after release from the trap in order to characterize the number,
size, and temperature of the cloud. For phase-contrast images, we
use light from the probe laser that is 800~MHz red-detuned from the
$|F=1 \rightarrow F'=2\rangle$ transition, and optical pumping is
not necessary. We use a 20~$\mu$s pulse of light with an intensity
of $\sim$1~mW/cm$^2$ for imaging.

\subsubsection{Phase-contrast image processing}
\begin{figure}
\begin{center}
\leavevmode
\includegraphics[width=1\linewidth]{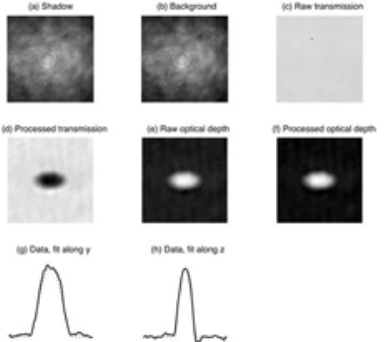}
\end{center}
\caption[Horizontal phase-contrast image]{Horizontal phase-contrast
image.  (a) Shown is the $shadow$ frame, taken with the atoms
present. (b) Shown is the $background$ frame, taken without atoms
present.  (c) The raw transmission, where the black dot represents
atoms. (d) The processed transmission, where the black hole
represents atoms. (e) The raw optical depth, where the white hole
now represents atoms.  (f) The processed optical depth.  Images (g)
and (h) show the data (black line) and a corresponding fit to a
Thomas-Fermi distribution (dashed line) along the $y$ (horizontal)
and $z$ (vertical) directions, respectively.  The scale on images
(d) - (h) is $73 \times 73~\mu$m.} \label{fig:Horiz-pc}
\end{figure}

A typical horizontal phase-contrast image taken immediately after
release from the weak TOP trap is shown in
Figure~\ref{fig:Horiz-pc}.

Figure~\ref{fig:Horiz-pc}~(a) shows the $shadow$ frame and
Figure~\ref{fig:Horiz-pc}~(b) shows the $background$.
Figure~\ref{fig:Horiz-pc}~(c) shows the raw transmission $T$,
defined as $T=shadow/background$. Figure~\ref{fig:Horiz-pc}~(d)
shows the processed transmission, which is just a centered and
zoomed-in version of the raw transmission.
Figure~\ref{fig:Horiz-pc}~(e) shows the raw optical depth $OD_r$,
defined as
\begin{equation}
OD_r = -\left[ \frac{\Gamma}{2\Delta} + \frac{I}{I_{sat}}
\frac{2\Delta}{\Gamma} \right](T-1)
\end{equation}
for far-off-resonant phase-contrast imaging.
% need to reference this eq.?
Figure~\ref{fig:Horiz-pc}~(f) shows the processed optical depth, in
this case this is only a filtered version of $OD_r$.  We use an
image processing routine that spatially filters the image to remove
the effects of interference fringes. Figures~\ref{fig:Horiz-pc}~(g)
and (h) show the data (from the processed optical depth) in solid
lines and a Thomas-Fermi fit to the data in dashed lines along the
$y$ and $z$ axes. %The scale of the images is $73 \times 73~\mu$m.

\subsection{Image processing and analysis}
Images are taken by triggering a Princeton Instruments/Acton PIXIS
1024BR CCD camera.  The camera has a BNC input that accepts a TTL
signal to trigger the camera.  The software program that controls
the camera is set to capture three frames, and the LabVIEW program
on the experiment PC sends three triggers to the CCD camera. Three
frames are taken for each data run: a shadow, background, and dark
frame. The shadow frame consists of the probe laser beam passing
through the BEC and onto the camera, the background consists of the
same probe laser beam shining directly onto the CCD camera, and the
dark frame consists of an image exposure with no light present. The
dark frame is subtracted from the shadow and background frames to
get rid of the effects of room lights, stray light, and bad pixels.

We have constructed an image processing program using Matlab that
will process images of thermal, bimodal, or BEC clouds and provide
relevant parameters. The program is adaptable for either absorption
or phase-contrast imaging, and is variable for use with either the
horizontal or vertical imaging systems.  The basic algorithm is as
follows: a 2D array labeled `optical depth' is constructed, based on
either the absorption or phase-contrast imaging scenarios,
corrections for detuning or saturation effects are applied to the
optical depth array, the array gets smoothed and/or filtered, then
the array is fit to either a Bose-enhanced Gaussian distribution,
Thomas-Fermi distribution, or a combination of the two to extract
relevant parameters~\cite{ketterle1999mpa}.

In everyday operation, we normally take one image at the start of
the day at a baseline RF evaporation frequency to ensure that BEC
formation is normal, comparing the number of atoms in the BEC and
the size of the cloud to the baseline values.  It is also often
useful to take a series of images through the BEC transition to
determine at what RF frequency a pure BEC is formed, as we have
noticed that this value can change slightly from day-to-day.

\section{Final BEC Characteristics} \label{section:final}
We have measured the trap frequencies in the tight TOP trap to be
$\omega_r = 2\pi \cdot$~39.8~Hz radially and $\omega_z = 2\pi
\cdot$~110~Hz axially, and the trap frequencies in the
gravity-sagged weak TOP trap to be $\omega_r = 2\pi \cdot$~7.4~Hz
radially and $\omega_z = 2\pi \cdot$~14.1~Hz
axially\footnote{For additional information, see the lab notebook entry on 4/11/06.}. %footnote here!

The final RF evaporative cooling stage in the weak TOP trap produces
condensates of up to $2 \cdot 10^6$ atoms, with condensate fractions
above 65$\%$ and thermal cloud transition temperatures of $\sim$
35~nK. The chemical potential of the BEC is $\sim k_B \cdot 8$~nK. A
typical BEC has an axial Thomas-Fermi radius of a$_z \approx
18~\mu$m and a radial Thomas-Fermi radius of a$_r \approx 34 ~\mu$m.

%We start out with $\sim 3 \cdot 10^9$ atoms in the MOT and transfer
%$\sim 3 \cdot 10^8$ atoms into the quadrupole trap in the science
%cell.  We estimate that we transfer most of these atoms from the
%quadrupole trap into the TOP trap, and achieve a BEC of $\sim 4
%\cdot 10^5$ atoms after evaporative cooling.

\begin{figure}
\centering \subfigure[t=0 seconds] {
    \label{fig:transition:a}
    \includegraphics[width=3.3cm]{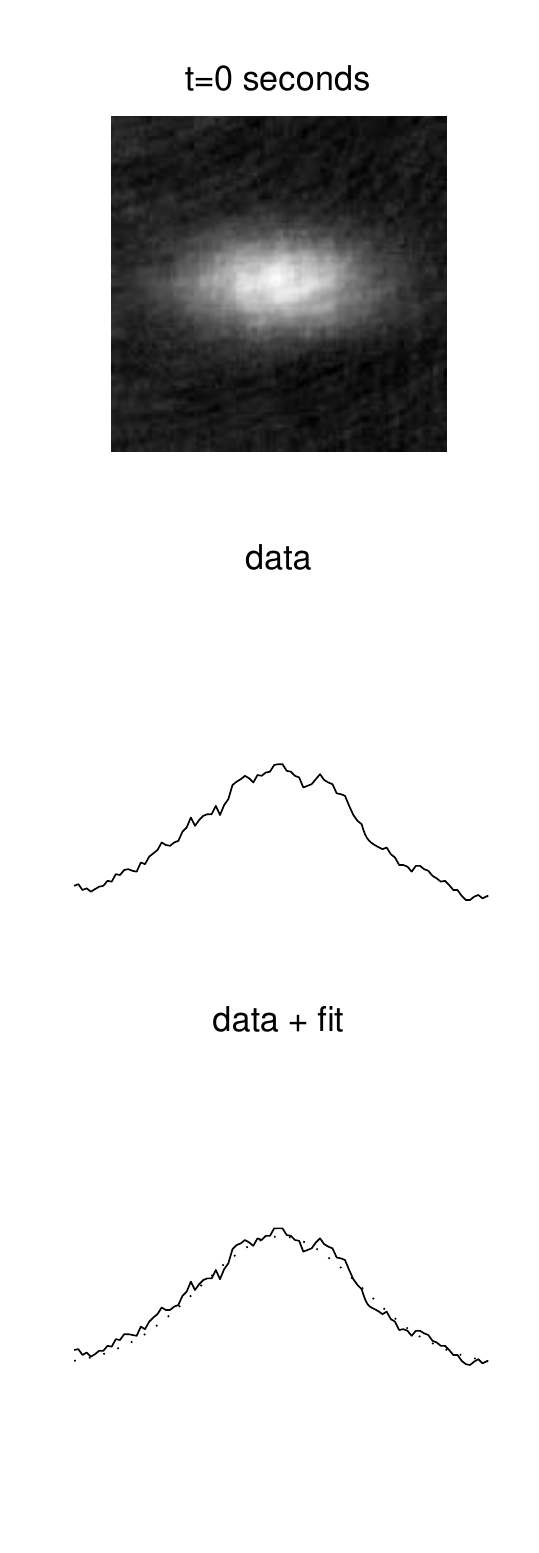}
} \subfigure[t=5 seconds] {
    \label{fig:transition:b}
    \includegraphics[width=3.3cm]{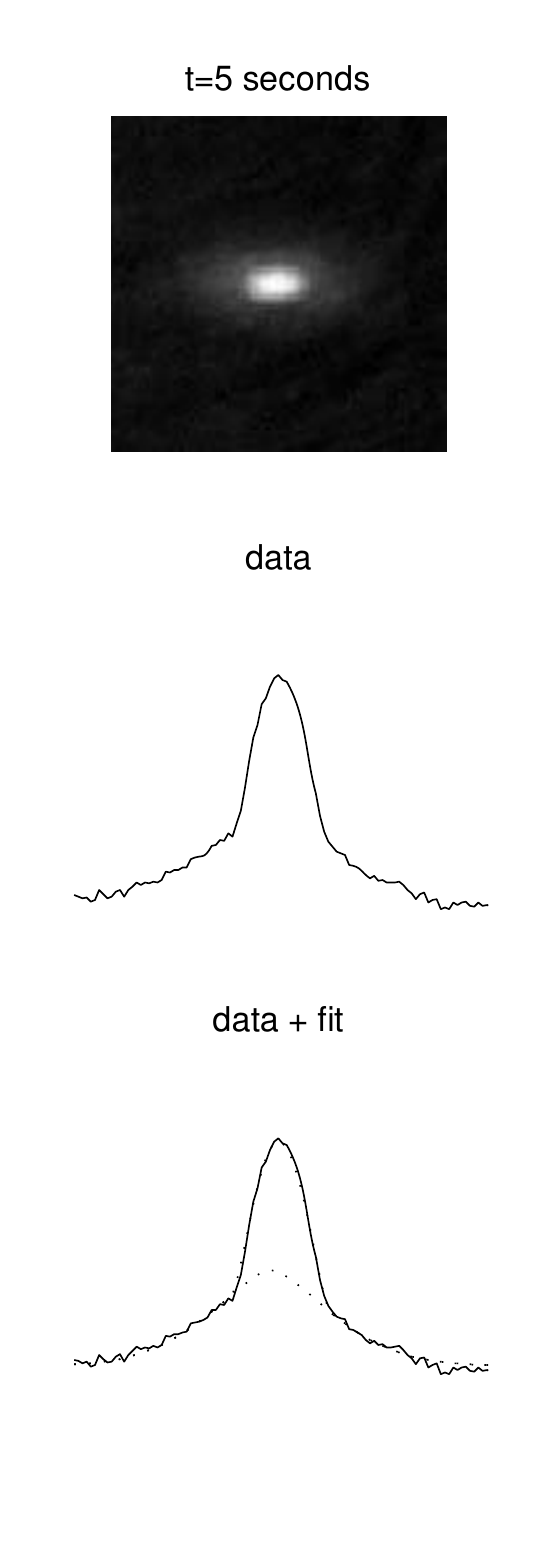}
} \subfigure[t=10 seconds] {
    \label{fig:transition:c}
    \includegraphics[width=3.3cm]{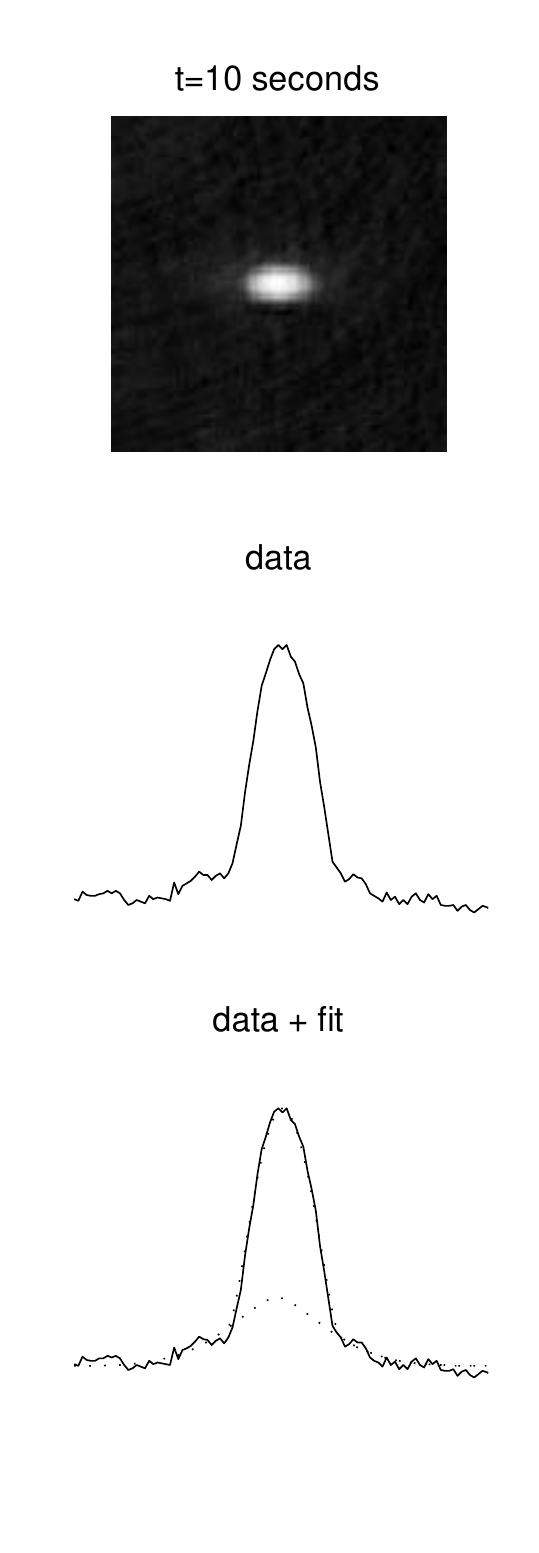}
} \subfigure[t=20 seconds] {
    \label{fig:transition:c}
    \includegraphics[width=3.3cm]{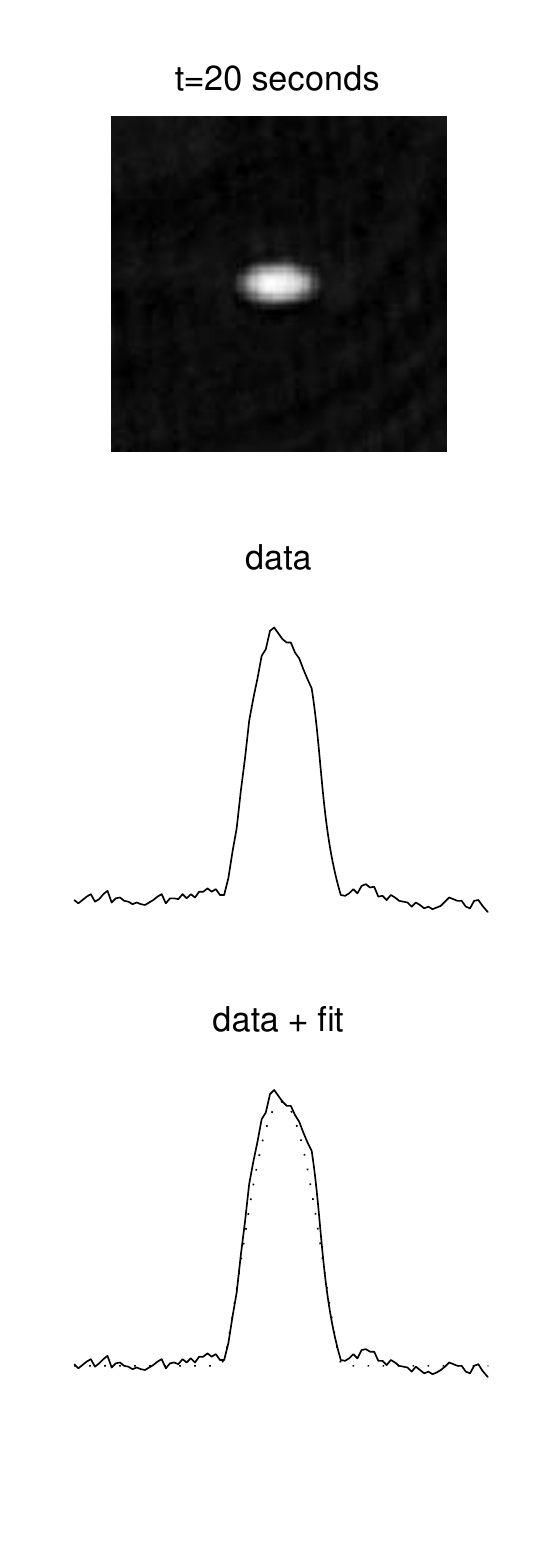}
} \caption[BEC transition data]{A series of four in-trap,
phase-contrast images of the atomic cloud at progressive times in
the RF evaporative cooling process. For each of the four times, the
processed optical depth, a horizontal slice of the data, and a
horizontal slice of the data alongside the appropriate fit are shown
in columns. Image (a) is taken at $t=0$~s and the data is fit to a
Gaussian distribution.  Image (b) is taken at $t=5$~s and the data
is fit to a bimodal (Thomas-Fermi + Bose-enhanced Gaussian)
distribution. Image (c) is taken at $t=10$~s and the data is fit to
a bimodal distribution.  Image (d) is taken at $t=20$~s and the data
is fit to a Thomas-Fermi distribution.  The Thomas-Fermi fits are
difficult to see, as they line up so well with the data. The scale
on the images is $121 \times 121~\mu$m.}
\label{fig:transition} % caption for the whole figure
\end{figure}

% using 09/06/2006 transition data 50 pixels x 50 pixels
% 50 pixels = 13 um*50/5.36 = 121 um
%\footnote{Data is from 09/07/2006.} files 7,14,10,6 are 0,5,10,20 seconds
A series of four images showing the atomic cloud undergo the
Bose-Einstein condensation transition is shown in
Figure~\ref{fig:transition}. Images were taken at four different
times in the 20~second Phase II evaporation process, using the
horizontal phase-contrast imaging system with a probe detuning of
-900~$\Gamma$. For each of the four images, the optical depth is
shown on a $121 \times 121~\mu$m scale. A horizontal slice of the
data as well as the data overlaid with the fit are also shown.
Figure~\ref{fig:transition}~(a), taken at the start of the Phase II
evaporation scheme, shows a thermal cloud that fits well to a
Gaussian distribution.  In Figure~\ref{fig:transition}~(b), taken
after 5~seconds, evidence of the nascent BEC starts to appear as the
strong density peak in the middle of the cloud.  This image was fit
to a bimodal (Thomas-Fermi + Bose-enhanced Gaussian) distribution,
the condensate fraction is 42$\%$. Figure~\ref{fig:transition}~(c),
taken after 10~seconds, shows a bimodal distribution with even less
of a thermal component visible; the condensate fraction here is
74$\%$. Figure~\ref{fig:transition}~(d), taken at the end of a
longer-than-usual 20~second evaporation process, shows a nearly pure
BEC with an undetectable thermal component. This image is fit to a
Thomas-Fermi distribution.
% transition data on 11/08/06 looks like crap.  Do we have better?

\section{Complete Timing Sequence} \label{section:complete}
\begin{figure}
\begin{center}
\leavevmode
\includegraphics[angle=-90,width=1\linewidth]{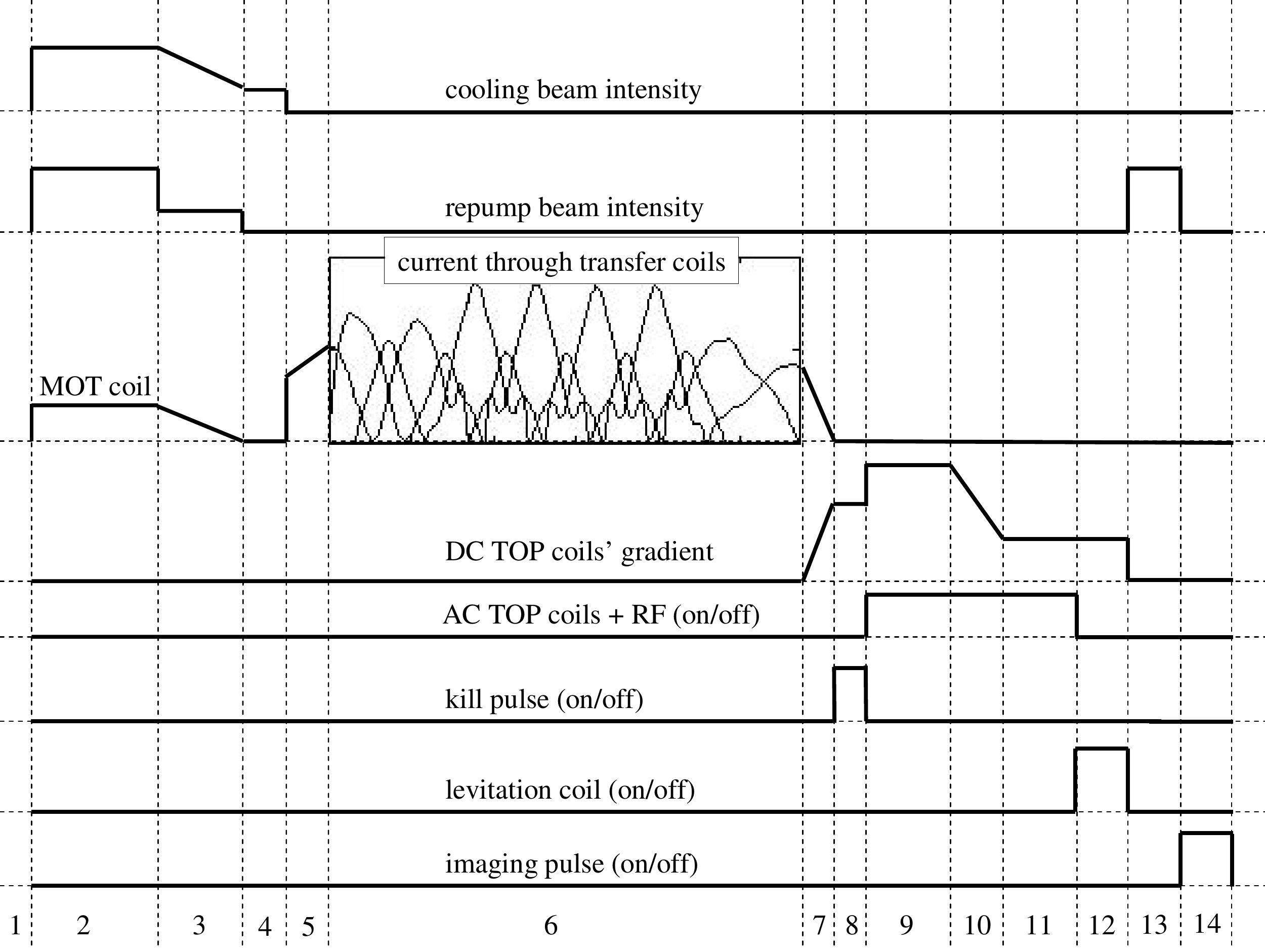}
\end{center}
\caption[Complete BEC formation timing sequence]{Complete BEC
formation sequence vs.\ time.  Not to scale on the time or amplitude
axes. The different segments of the BEC formation process, described
in the text, are: (1) all beams are turned off; (2) MOT loading; (3)
CMOT; (4) optical pumping; (5) initial magnetic trap; (6) magnetic
transfer to science cell; (7) transfer into quadrupole trap formed
by DC TOP coils only; (8) kill pulse; (9) transfer into TOP trap and
Phase I evaporation; (10) transfer into gravity-sagged weak TOP
trap; (11) Phase II evaporation; (12) \emph{(absorption imaging
only)} expansion; (13) \emph{(absorption imaging only)} repump flash
pulse; and (14) imaging pulse.} \label{fig:Complete-timing}
\end{figure}

The complete timing sequence used to create a BEC is shown in
Figure~\ref{fig:Complete-timing}.  The numbers at the bottom of the
figure correspond to the following stages:
\begin{enumerate}
\item All the lasers and magnetic fields are turned off momentarily
prior to MOT loading.
\item The cooling beam, repump beam, and MOT magnetic field are
turned on to load a new MOT for about 20~seconds.  We load the MOT
until a MOT of $\sim 3 \cdot 10^9$ atoms is obtained.
\item The CMOT phase occurs for 40~ms by ramping the cooling beam
detuning from -4~$\Gamma$ to -7~$\Gamma$, switching the repump power
from 30~mW to 260~$\mu$W, and ramping the MOT magnetic field
gradient from 8~G/cm to 0~G/cm.
\item The optical pumping phase occurs for 1~ms with
only the cooling beam on.
\item The magnetic field `catch' occurs by turning on a 40~G/cm
magnetic field using the MOT coils.  We then ramp up the magnetic
field to 180~G/cm over 100~ms.
\item Magnetic transfer from the MOT cell into the science cell
occurs by employing the series of current ramps through the 14
transfer coil pairs, taking a total of 5.9~seconds.
\newpage
\item Atoms are transferred from a quadrupole trap created by the final
transfer coil pair into a quadrupole trap created by the DC TOP
coils by simultaneously ramping the final transfer coils off and the
DC TOP coils on over 100~ms.
\item A 5~ms `kill pulse' eliminates any stray $|F=2\rangle$ atoms that
have made it into the science cell.
\item Atoms are transferred into the TOP trap by switching the DC TOP
coils' field strength from 180~G/cm to 300~G/cm and turning the
amplitude of the AC TOP coils on to 41~G at 2~kHz.  Phase I
evaporation occurs in the tight TOP trap over a period of 72~seconds
by ramping the bias field from 41~G to 5~G and ramping the RF
frequency from 70~MHz to 4.5~MHz.
\item The atoms are transferred to a gravity-sagged TOP trap
by ramping the DC TOP coils' field strength from 300~G/cm to 55~G/cm
over 2~seconds, while increasing the RF field.
\item Phase II evaporation occurs over 10~seconds by ramping the RF
frequency from 5.0~MHz to 4.7~MHz, forming a BEC.
\item \emph{(Absorption imaging only)} The atomic cloud is allowed to
expand for 56~ms with the DC TOP coils on at 55~G/cm, the levitation
coil on at 32.4~G, and the AC TOP coils off.
\item \emph{(Absorption imaging only)} A 10~$\mu$s `repump flash' pulse
pumps the atoms into the $|F=2\rangle$ state before imaging.
\item The imaging pulse (using either phase-contrast or absorption
light) occurs, followed by a background and dark exposure.
\end{enumerate}

%% file: Chap_Vortices.tex
\chapter{VORTEX FORMATION\label{chapter:vortices}}

\section{Introduction}
Questions such as \emph{``How do vortices form in superfluids? Why
is it important to study new mechanisms for vortex generation in
superfluids?"} have motivated our experimental search for a new
mechanism of vortex formation that relies of the interference
between BECs as they are merged.  The results of our investigation,
published in ``Vortex Formation by Merging of Multiple Trapped
Bose-Einstein \mbox{Condensates}''~\cite{scherer2007vfb}, are
presented in this chapter along with a more detailed description of
the background information and concepts relevant to our experiment.

The outline of this chapter is as follows: the background
information and concepts relevant to the experiment will be
presented, paying particular attention to the notion of phase in a
BEC. Then, the specific \emph{research question} will be posed and
an explanation of why this research project is novel, important, and
unanswered will be given. Finally, the experimental details,
procedure, results, and discussion will be presented.

\section{The Concept of Phase in a BEC}
An atomic-gas BEC is a macroscopic occupation of a single quantum
state by an ensemble of atoms, and can be described by a macroscopic
wavefunction, or order parameter, $\Psi (\vec{r})$, given by
\begin{equation} \label{eq:nephi}
\Psi (\vec{r}) = \sqrt{n(\vec{r})} e^{i \phi (\vec{r})}
\end{equation}
where $n(\vec{r})$ and $\phi (\vec{r})$ are a spatially varying
density and phase,
respectively~\mbox{\cite{dalfovo1999tbe,leggett2001bec}\footnote{Other
mathematical descriptions of BECs exist, but this description is
adequate for the experiments discussed in this dissertation.}.}
This usage of $\Psi(\vec{r})$ is derived from a quantum field theory
approach, where formally $\Psi(\vec{r})$ is the expectation value of
an atomic field annihilation operator~\cite{leggett1991csb}.  In the
interpretation of $\Psi(\vec{r})$ as a complex wavefunction, one can
think of $\sqrt{n(\vec{r})}$ as the amplitude and $\phi(\vec{r})$ as
the phase of a complex function within the bounds of mean-field
theory. This $\Psi(\vec{r})$ is a solution to the Gross-Pitaevskii
equation, which governs the dynamics of the wavefunction
$\Psi(\vec{r})$ in an external trapping
potential~\cite{dalfovo1999tbe}. One fundamental consequence of this
description is that the addition of a constant, overall phase $a$ to
the wavefunction to make
\begin{equation}
\Psi_2 (\vec{r}) = \sqrt{n(\vec{r})} e^{i (\phi (\vec{r})+a)}
\end{equation}
does not alter the physical properties of the
wavefunction~\cite{cohentannoudji1986qmv}.

%This should not be very surprising, since the very \emph{concept} of
%phase is a relative idea; the overall, or global phase of any valid
%Schrodinger wavefunction can be changed without altering the
%physical properties of the wavefunction.

\subsection{The superfluid velocity and quantized vortices} \label{subsection:superfluid}
One consequence of the above description is the derivation of the
superfluid velocity $\vec{v}(\vec{r})$ as
\begin{equation} \label{eq:vgradphi}
\vec{v}(\vec{r}) = \frac{\hbar}{m} \nabla \phi (\vec{r}),
\end{equation}
where $\hbar$ is Planck's constant and $m$ is the atomic mass. The
superfluid velocity $\vec{v}(\vec{r})$ satisfies the following two
constraints.  First, because it can be written as the gradient of
the scalar field $\phi (\vec{r})$, $\vec{v}(\vec{r})$ is an
irrotational vector field that satisfies
\begin{equation}
\nabla \times \vec{v}(\vec{r}) = 0.
\end{equation}
Second, by taking the integral of Equation~\ref{eq:vgradphi} around
a closed loop, it can be seen that there is a quantization condition
on the flux around a closed loop, referred to as the Onsager-Feynman
quantization condition:
%\cite{onsager1949ncs,feynman1955plt}
\begin{equation} \label{eq:quantization}
\oint \vec{v}(\vec{r}) \cdot dl = n \frac{h}{m}
\end{equation}
where $n$ is an integer \cite{leggett2001bec}.

\begin{figure}
\begin{center}
\leavevmode
\includegraphics[angle=-90,width=0.65\linewidth]{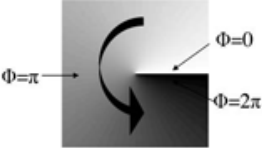}
\end{center}
\caption[$2\pi$ phase winding of a singly-charged vortex]{0 to
$2\pi$ phase winding, corresponding to the phase profile of a
singly-charged vortex in a superfluid.} \label{fig:Phase-winding}
\end{figure}

The consequences of the above superfluid formulation are myriad for
the case of BEC.  First, although the addition of a constant overall
phase to the wavefunction carries no physical significance, it is
clear from Equation~\ref{eq:vgradphi} that the phase gradient
$\nabla \phi(\vec{r})$ determines the rate and direction of
superfluid flow given by $\vec{v}(\vec{r})$.  Second, the
quantization condition of Equation~\ref{eq:quantization} has
important implications for the existence of quantized vortices
within BECs. The phase of the wavefunction $\Psi(\vec{r})$ can be
spatially varying, but must be single-valued at any point in space.
A situation where the phase increases azimuthally is shown in
Figure~\ref{fig:Phase-winding}. If the phase at a particular point
is 0 and increases azimuthally, the phase must be 0, $2\pi$, or some
integer multiple thereof upon returning to the original point.  This
$2\pi$ phase winding corresponds to a circulation of $n=1$ in the
quantization condition, and corresponds to the phase winding of a
singly-charged vortex.  The vorticity or circulation $n$ of
Equation~\ref{eq:vgradphi} is the number of $2\pi$ phase windings
around a central point, referred to as the vortex core.  This phase
profile results in circular atomic flux around the hollow vortex
core, where the density drops to zero and there is a phase
singularity.

\newpage

\subsection{On the relative phase of independent superfluids}
A famous \emph{gedanken} experiment of P.\ W.\ Anderson proposed the
following scenario~\cite{anderson1986}: imagine that two separate,
independent buckets of superfluid He are brought together. When the
two superfluids are brought together, will there be superfluid flow?
And in which direction?  If one had \emph{a priori} knowledge of the
relative phase between the superfluid in bucket 1 and the superfluid
in bucket 2, one could calculate the direction of fluid flow
$\vec{v}(\vec{r})$.

%But this produces a logical inconsistency, since
%we know that the overall phase of any one of the superfluids is
%physically meaningless.
%It turns out that \emph{`What is the relative phase of independent
%superfluids?'} is an ill-posed question.

However, a relative phase does not exist until a phase measurement
has been made~\cite{anderson1986, javanainen1997pap, mullin2006opi}.
Upon each realization of the experiment, a direction of fluid flow
will be observed, from which a relative phase can be calculated.
Repeated trials of the experiment will produce different results. It
is the quantum measurement process of interfering the two
superfluids that produces a relative phase.  It should be emphasized
that the \emph{relative} phase is the only physically meaningful
variable here, individual overall phases are not physical.  We will
discuss this \emph{gedanken} experiment in the context of BECs in
the following sections.

\subsection{Example case: interference between two BECs}
In this section, we consider the case of two interfering BECs as an
example to show how the direction of fluid flow depends on their
relative phase. Consider a wavefunction that is a superposition of
two Gaussian density distributions, centered at $x=x_0$ and
$x=-x_0$, having phases $\phi_1$ and $\phi_2$
\begin{equation} \label{eq:super}
\Psi(\vec{r}) = A \cdot [e^{-(x-x_0)^2 + i \phi_1} + e^{-(x+x_0)^2 +
i \phi_2} ]
\end{equation}
where $A$ is a normalization coefficient and $x$ is a dimensionless
spatial variable. An elementary consequence of the formalism of
quantum mechanics shows that the probability current
$\vec{J}(\vec{r})$ can be written as
\begin{equation} \label{eq:current}
\vec{J}(\vec{r}) = \frac{\hbar}{2 m i} [\Psi^{*}(\vec{r}) \nabla
\Psi(\vec{r}) - \Psi(\vec{r}) \nabla \Psi^{*}(\vec{r})]
\end{equation}
for a wavefunction $\Psi(\vec{r})$~\cite{cohentannoudji1986qmv}. By
inserting Equation~\ref{eq:super} into Equation~\ref{eq:current}, it
can be shown that the current at $x=0$ is proportional to
sin($\phi_1 - \phi_2$). In this case, with a wavefunction that can
be written as the superposition of two overlapping but otherwise
uncorrelated BECs, the direction of fluid flow depends on the sine
of the phase difference between the two states.  This idea will be
expanded upon in Section~\ref{subsection:direction} to explain the
direction of fluid flow in our experiment, which involves three
interfering BECs.

\subsection{Matter-wave interference between two expanding BECs} \label{subsection:matter-wave}
A experiment similar to the proposed \emph{gedanken} experiment
described above was performed in the group of Wolfgang Ketterle at
MIT in 1997~\cite{andrews1997oib}. In this experiment, interference
between two independent BECs was observed by looking at interference
fringes between two spatially overlapped condensates after a period
of ballistic expansion. This experiment was the first to
experimentally investigate matter-wave interference between BECs, an
idea that will be expanded upon in the experiments reported in this
dissertation. In the MIT experiment, two independent BECs were
created in a double-well potential and then the two BECs were
allowed to expand (and hence overlap) during a period of ballistic
expansion. High-contrast interference fringes were observed,
providing the initial experimental evidence that BECs are coherent
\cite{kasevich2002ca} and capable of displaying `laser-like'
matter-wave interference.  In this experiment, the phase of the
interference fringe pattern, a physical quantity that is the result
of the relative phase of the two BECs upon interference, would
theoretically vary with each trial of the experiment based on the
preceding description.  However, mechanical instabilities in the
experimental setup would have made even a fixed relative phase
appear as random in this experiment.

% atom interferometry with BECs in a double-well potential
In a later publication, the MIT group continued to explore atom
interferometry in a double-well potential \cite{shin2004aib}.  A BEC
was formed in a single-well potential that was coherently split into
a double-well potential, separating one BEC into two spatially
distinct components. The two BECs were then held in the double-well
potential for a variable amount of time before being released. As
before, in the ballistic expansion process the separate clouds will
overlap and interfere, producing visible interference fringes.  In
this experiment, however, the interference fringe pattern was stable
and reproducible from shot-to-shot.  A deliberate phase shift could
then be applied to one of the condensates, producing a detectable
shift in the spatial location of the center of the interference
fringes. Reproducible interference patterns between condensates
separated for up to 5~ms were measured; this measurement of the
coherence time of the separated condensates was limited to 5~ms due
to technical difficulties.

In a third experiment, the MIT group observed interference of BECs
split with an atom chip \cite{shin2005ibe}.  A BEC was formed in a
magnetic potential using a microfabricated atom chip, and the BEC
was split into two parts by deforming the trapping potential into a
double-well potential.  As in the experiment above, the trap was
turned off and the separated condensates were allowed to expand and
overlap. Interference fringes were again observed, and,
surprisingly, in some experimental runs, a fork dislocation appeared
in the interference fringe pattern, this fork shape represents the
phase winding around a vortex core \cite{inouye2001ovp}. The vortex
generation mechanism was not fully understood, but it was
hypothesized that it was due to the BEC splitting procedure.

%could have been due to topological imprinting when the zero point of
%the magnetic field crossed through the condensates.

These experiments set the stage for an understanding of the basic
concepts of matter-wave interference between BECs.  It should be
pointed out that in these experiments the spatial overlap (and hence
interference) of separate BECs was facilitated by a period of
ballistic expansion, and that the interference was always between
two initially isolated BECs.

\section{Research Question: \emph{Can vortices be made by merging and \mbox{interfering} \mbox{independent} \mbox{superfluids}?}}
In the preceding sections we reviewed two concepts that are
fundamental to the understanding of our experiment: (1) the
relationship between a $2\pi$ phase winding and the existence of
quantized vortices in superfluids; and (2) the ability of
independent BECs to interfere, with a direction of fluid flow that
depends on their relative phase, a quantity that is indeterminate
prior to BEC interference. These two concepts will be combined in an
original way in order to answer the research question posed in this
dissertation. The purpose of this section is to provide some
motivation for our interest in this research project.

The existence of quantized vortices is one of the signature features
of superfluid systems \cite{tilley1986sas, donnelly1991qvh}, and
atomic-gas BECs have been a fruitful playground in which to explore
superfluid vortex properties since their original observation in BEC
in 1999 \cite{matthews1999vbe}.  Due to the relative ease of
microscopic manipulation and detection techniques, BECs are
well-suited to address open questions regarding superfluid mixing
and associated vortex generation~\cite{aftalion2004pvb,
kevrekidis2004vbe, srinivasan2006vbe}. Vortices were originally
created in BEC using quantum phase manipulation
\cite{matthews1999vbe,leanhardt2002ivb}, and have also been created
using methods more analogous to those in classical fluid dynamics
\cite{batchelor1980ifm}, namely through rotating traps
\cite{madison2000vfs, aboshaeer2001ovl, hodby2001vnb,
haljan2001dbe}, turbulence \cite{inouye2001ovp}, and dynamical
instabilities \cite{anderson2001wds, dutton2001oqs}. Yet in contrast
with classical fluid dynamics, vortex generation via the mixing of
initially isolated superfluids has been an unexplored research area.

In this chapter we describe our experiments demonstrating that
merging together three condensates in a trap can lead to the
formation of quantized vortices in the merged BEC. We ascribe the
vortex generation mechanism to matter-wave interference between the
initially isolated BECs, and show that vortices may be induced for
both slow and fast merging rates. While it is now well-known that
matter-wave interference may occur between BECs
\cite{andrews1997oib}, and that condensates can be gradually merged
together into one larger BEC \cite{chikkatur2002csb}, our experiment
demonstrates a physical link between condensate merging,
interference, and vortex generation, providing a new paradigm for
vortex formation in superfluids.  We emphasize that no stirring or
quantum phase engineering steps are involved in our work; the vortex
formation process is stochastic and uncontrollable, and partially
depends on relative quantum phases that are indeterminate prior to
condensate merging.  This vortex formation mechanism may be
particularly relevant for developing further understanding of the
roles of potential-well defects, roughness, and disorder on
establishing a superfluid state.  Furthermore, this work may be
viewed as a model for studies of spontaneous symmetry breaking and
topological defect formation during phase transitions
\cite{kibble1976tcd, zurek1985ces, drummond1999qde, anglin1999vwr,
marshall1999eca, sadler2006ssb}.
%
%This chapter describes experiments demonstrating that the merging
%together of multiple BECs in a trap can lead to the formation of
%potentially long-lived quantized vortices in the resulting BEC. We
%ascribe the vortex generation mechanism to matter-wave interference
%between the initially spatially isolated but otherwise identical
%BECs, and show that vortex formation may be induced even for slow
%mixing time scales.  Our experiment demonstrates a physical link
%between matter-wave interference and vortex generation, providing a
%new paradigm for vortex formation in superfluids.  We emphasize that
%no stirring or quantum phase engineering steps are involved in our
%work, nor are any other means for controllably nucleating vortices
%in the trapped atomic gas. The vortex formation process itself is
%stochastic and uncontrollable, and depends on relative quantum
%phases that are indeterminate prior to condensate mixing.  The
%vortex formation mechanism identified here may be particularly
%relevant when defects or roughness are present in a trapping
%potential, or when multiple condensates are otherwise joined
%together. Our experiment may also illuminate aspects of vortex
%formation at site defects in other superfluids, for which
%microscopic studies may be exceedingly difficult and questions
%regarding vortex formation mechanisms are unresolved.

\section{Experimental Details}
This section describes some of the technical details of our
experimental procedure.  Our BEC formation process is reviewed in
Section~\ref{subsection:becformation}.  Use of our three-well atom
trap is described in Section~\ref{subsection:three-well}, and our
method for creation of the three-well trap is described in
Section~\ref{subsection:making}.  When the three condensates created
in our three-well trap are merged, there will be an atomic fluid
flow between the initially independent BECs.  The direction of fluid
flow is discussed in Section~\ref{subsection:direction} and the
timescale for merging is discussed in
Section~\ref{subsection:timescale}.  As a separate, test experiment,
a deliberate method for creating vortices in our BEC is discussed in
Section~\ref{subsection:lattice}.

\subsection{BEC formation} \label{subsection:becformation}
Our basic single BEC creation technique was described in detail in
Section~\ref{section:complete}.  In brief, we first cool a thermal
gas of $|F$=1, $m_F$=$-1\rangle$ $^{87}$Rb atoms to just above the
BEC critical temperature in an axially symmetric TOP trap with
radial and axial trapping frequencies of 39.8~Hz and 110~Hz,
respectively. We then ramp the TOP trap magnetic fields such that
the final trap oscillation frequencies are $7.4$~Hz (radially) and
$14.1$~Hz (axially). A final 10~second stage of RF evaporative
cooling produces a BEC of $\sim$4$\times$10$^5$ atoms, with a
condensate fraction near 65\% and a thermal cloud temperature of
$\sim$22~nK. The BEC chemical potential is $k_B \times$8~nK, where
$k_B$ is Boltzmann's constant.

\subsection{Three-well atom trap} \label{subsection:three-well}
\begin{figure}
\begin{center}
\leavevmode
\includegraphics{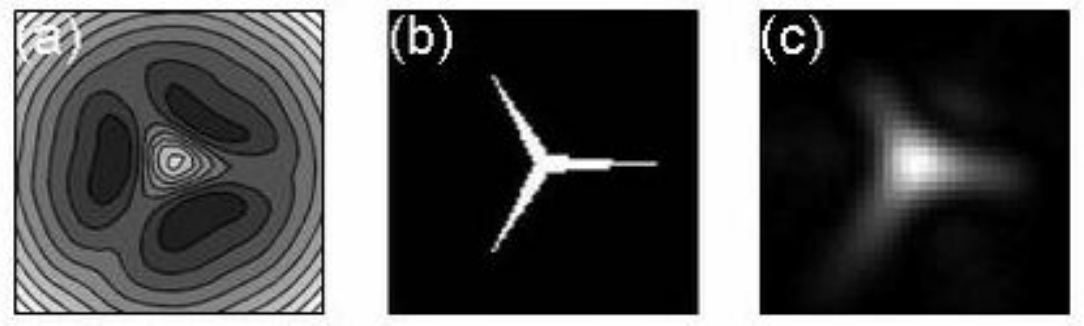}
\end{center}
\caption[Three-well potential]{(a)  Potential energy contour showing
a horizontal slice through the center of our three-well trap,
representing the addition of the harmonic TOP trap with the measured
intensity profile of the optical barrier beam, scaled to a potential
energy. (b) The binary mask profile used to create the optical
barrier, where white represents the transmitting area. (c) An image
of the actual optical barrier profile in the plane of the BEC. The
size of images (a) and (c) are 85~$\mu$m~$\times$~85~$\mu$m.}
\label{fig:Potential}
\end{figure}

To illustrate the basic concept underlying our experiment, we first
describe our atom trap, which is formed by the addition of a TOP
trap and a central repulsive barrier of axially (vertically)
propagating blue-detuned laser light shaped to segment the harmonic
oscillator potential well into three local potential minima.
Figure~\ref{fig:Potential}~(a) shows an example of potential energy
contours of our three-well potential.

We will assume throughout the ensuing descriptions that the energy
of the central barrier is low enough that it has negligible effect
on the thermal atom cloud; such is the case in our experiment.
However, the central barrier does provide enough potential energy
for an independent condensate to begin forming in each of the three
local potential minima from the original thermal cloud. If the
central barrier is weak enough, condensates with repulsive
interatomic interactions will grow together during evaporative
cooling; this is because the individual chemical potentials of each
condensate will grow to be greater than the barrier height, thus
enabling direct above-barrier transport.  If the barrier is strong
enough, the condensates will remain independent. In this latter
case, the central barrier height may be lowered while keeping the
condensed atoms held in the TOP trap. Overlap and interference
between the heretofore independent condensates would then be
established as the condensates merge together into one.

% Brian's second version
% We will assume throughout the ensuing
%discussions that the energy of the central barrier is low enough
%that it has negligible effect on the thermal atom cloud, as in our
%experiments, but high enough for independent condensates to begin
%forming in the three local potential minima from the single thermal
%cloud. There are two important regimes in this range of barrier
%energies: if the central barrier is weak, condensates with repulsive
%interatomic interactions will grow and merge together during
%evaporative cooling; if the barrier is strong, the condensates will
%remain independent, but may be merged together by lowering the
%barrier while keeping the atoms trapped \cite{TUNNELING}. We have
%examined both scenarios.

\subsection{Making a custom-shaped optical potential} \label{subsection:making}
An experiment like this would not have been possible had it not been
for our ability to make custom-shaped optical potentials such as
that shown in Figure~\ref{fig:Potential}~(c). We have been able to
create such light fields by using the Maskless Lithography Tool
(MLT), part of the Laboratory for Diffractive and Micro-Optics run
by Dr.\ Tom Milster at the College of Optical Sciences, Tucson, AZ.
The MLT uses a micro-positioned optical beam to etch off chrome from
a glass slide.  By removing the chrome from specific points on the
slide, we can create an arbitrary binary pattern on the mask, such
as that shown in Figure~\ref{fig:Potential}~(b).  By focussing a
beam through the binary transmission mask, we can create a light
beam that looks very similar to the transmission profile of the
mask, differing in shape from the mask due to beam quality and the
effects of diffraction. The versatility of this tool for experiments
in ultracold atomic physics cannot be underestimated, as one can use
this tool to create binary masks, and hence light fields, in
arbitrary shapes for myriad uses in atom interferometry and quantum
state engineering.

\begin{figure}
\begin{center}
\leavevmode
\includegraphics[angle=-90,width=1\linewidth]{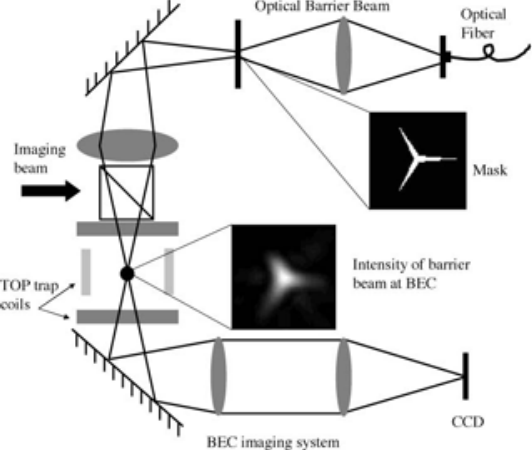}
\end{center}
\caption[Experimental setup, including optical barrier
beam]{Schematic of the experimental setup.  The optical barrier beam
is coupled from an optical fiber and focussed onto the binary
transmission mask.  The beam that passes through the binary
transmission mask then passes through a polarizing beam-splitter
cube and gets imaged by one focussing lens onto the plane of the
BEC; the intensity of the optical barrier beam at the plane of the
BEC is shown. The vertical imaging beam passes through the other
port of the polarizing beam-splitter cube and passes through the BEC
vertically, then gets imaged onto the CCD camera. The horizontal
imaging system is not shown.} \label{fig:Experiment-setup}
\end{figure}

The barrier that partitions our harmonic trap into three wells is
formed by illuminating the binary mask shown in
Figure~\ref{fig:Potential}~(b) with a focused blue-detuned Gaussian
laser beam of wavelength 660~nm (Mitsubishi model 101J27-01 laser
diode). After passing through the mask, the beam enters our vacuum
chamber along the vertical trap axis. The mask is imaged onto the
center of the atom cloud with a single lens. Due to diffraction, the
beam has an intensity profile as shown in
Figure~\ref{fig:Potential}~(c), with a maximum intensity and thus
barrier energy aligned with the center of the TOP trap. The
barrier's potential energy decreases to zero over $\sim$35~$\mu$m
along three arms separated by azimuthal angles of approximately
$120^{\circ}$.  A schematic showing the imaging system used for the
optical barrier beam is shown in Figure~\ref{fig:Experiment-setup}.

\subsection{Direction of fluid flow for given relative phases} \label{subsection:direction}
Our experiment involves BEC formation in a three-well potential
formed by the addition of the central repulsive optical barrier beam
and the TOP trap potential.  In the case of a weak barrier, the
condensates will merge together during the final stages of RF
evaporative cooling, or, in the case of a strong barrier, will only
merge together when the intervening barrier's strength is ramped
down to zero while the atoms are held in the TOP trap. Depending on
the relative phases of the three condensates and the rate at which
the condensates merge together, the final merged BEC may have
nonzero net orbital angular momentum about the vertical trap axis.

To demonstrate this, we first envision two condensates in two
potential minima merged slowly enough that although interference
occurs between the condensate pair, interference \emph{fringes} do
not. As merging begins, a directional mass current between the pair
is established, with the initial direction of above-barrier fluid
flow depending on the sine of the phase difference between the
overlapping states (as also occurs in the Josephson Effect
\cite{josephson1962pne, feynman1965flp} for the case of tunneling).
Recalling that the relative phase between two independent
condensates is indeterminate until it is measured via interference,
the relative phase and hence fluid flow direction will vary randomly
upon repeated realizations of the experiment.

When the three condensates of our experiment are merged together
while remaining trapped, a net fluid flow over the barrier arms may
occur that is simultaneously either clockwise, counter-clockwise, or
neither, relative to the trap center.  Recalling that the direction
of fluid flow depends on the phase differences between the
individual condensates, one can label the condensates formed in the
three local minima of Figure~\ref{fig:Potential}~(a) as $j$=1, 2,
and 3 in clockwise order. Upon merging of the three condensates, a
value of the relative phases $\Delta \phi_{ij}$ will be established,
which we will write as $\Delta \phi_{12}$, $\Delta \phi_{23}$, and
$\Delta \phi_{31}$, where $\Delta \phi_{ij} = \phi_i - \phi_j$.

Upon merging of the three condensates, if it turns out that for
example, $\Delta \phi_{12} = 0.7\pi$ and $\Delta \phi_{23} = 0.8\pi$
(thus necessarily $\Delta \phi_{31} = 0.5\pi$ since the phase
everywhere must be single valued), then some finite amount of
clockwise-directed fluid flow will be established between each pair,
hence also for the entire fluid. More generally, if the three
merging condensates have relative phases $\Delta \phi_{12}$, $\Delta
\phi_{23}$, and $\Delta \phi_{31}$ that are each simultaneously
between 0 and $\pi$, or each between $\pi$ and 2$\pi$, the resulting
BEC will have acquired nonzero net orbital angular momentum after
the merger, which will be manifest as a vortex within the BEC.

\begin{figure}
\begin{center}
\leavevmode
\includegraphics[angle=-90,width=0.7\linewidth]{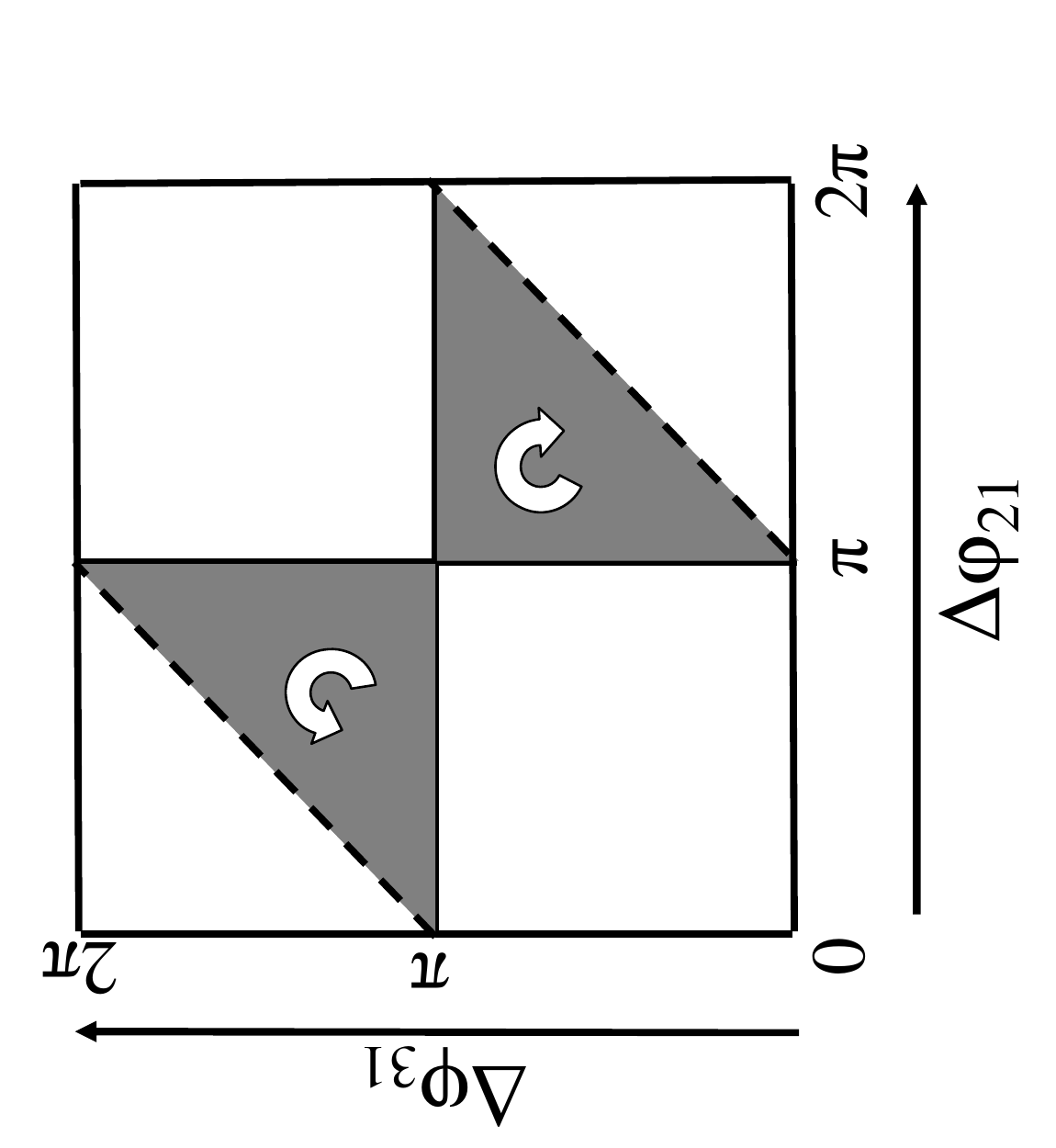}
\end{center}
\caption[Illustration showing which relative phases can lead to a
vortex]{Illustration showing which relative phases can lead to a
vortex, $\Delta \phi_{21}$ is plotted on the horizontal axis from 0
to $2\pi$, and $\Delta \phi_{31}$ is plotted on the vertical axis
over the same range. The phase conditions necessary for clockwise
fluid flow over all three barrier arms are contained in the shaded
gray region in the lower right-hand corner. A similar region
corresponding to counter-clockwise fluid flow is shown as the shaded
region in the upper left-hand corner. Given random initial phases
between the three condensates, the necessary phase conditions for
vortex formation will occur 25\% of the time.}
\label{fig:Phase-plot}
\end{figure}

To understand which conditions in the relative phases can produce
vorticity, consider Figure~\ref{fig:Phase-plot}, which examines the
full range of phase difference possibilities, plotted as a function
of $\Delta \phi_{21}$ and $\Delta \phi_{31}$ over their full range
of values from 0 to $2\pi$. The condition for a clockwise fluid flow
over all three intervening barriers is $\phi_2 - \phi_1 \geq \pi$,
$\phi_3 \geq \phi_2 + \pi$, and $\phi_3 - \phi_1 \leq \pi$, which
corresponds to a region occupying $1/8$ of all the possible values
of $\Delta \phi_{21}$ and $\Delta \phi_{31}$, shown as a shaded gray
region in the lower right-hand corner of
Figure~\ref{fig:Phase-plot}. Another $1/8$ of the chart is occupied
by the region allowing for three counter-clockwise fluid flows,
therefore there will be a net probability of $P_v=0.25$ for either
direction of fluid flow, given uniformly random relative phases of
the condensates upon merging. Because our experiment is not set up
to detect the difference between a clockwise and counter-clockwise
fluid circulation, $P_v$ is the probability for a singly-charged
vortex to form as the three condensates merge together.

This relationship between vortex trapping and relative phases is an
application of the so-called geodesic rule \cite{berry1987apa,
samuel1988gsb}. Related work includes a theoretical investigation of
three Josephson-coupled BECs \cite{kasamatsu2005mia}, yet the
geodesic rule also applies to a wider range of topics including
spontaneous defect formation in liquid crystals \cite{chuang1991cld,
bowick1994ckm}.

\subsection{Timescale for merging} \label{subsection:timescale}
For faster merging rates and correspondingly steeper phase
gradients, interference \emph{fringes} may develop as the
condensates merge. To estimate the longest time scale $\tau_f$ over
which two merging condensates can support a single dark interference
fringe, we envision two condensates that are initially atomic point
sources separated by a distance $d$, and that each expands to a
radius of $d$ over a time $\tau_{f}$ such that the condensates
overlap in the intervening region.  The condensate expansion speed
$v \sim d/\tau_{f}$ corresponds to a phase gradient at the side of
each condensate of
\begin{equation}
\nabla \phi = \frac{v \cdot m}{\hbar} \sim \frac{d \cdot m}{\tau_{f}
\cdot \hbar}
\end{equation}
where $m$ is the atomic mass. To create a single interference fringe
in the overlap region, $\nabla \phi \sim \pi/d$. With $d \sim
35$~$\mu$m, appropriate for our experiment, $\tau_{f} \sim 550$~ms;
shorter merging times would produce more interference fringes, while
longer times correspond to slow merging and no fringes. Each dark
fringe will be subject to the same dynamical instabilities as dark
solitons and decay to vortices, antivortices, and possibly vortex
rings over times on the order of 50~ms \cite{feder2000dss,
anderson2001wds, dutton2001oqs}. Similar decay has been seen in
recent numerical simulations \cite{whitaker2006fvv}. For condensates
merged together over times of $\tau_f$ or shorter, we may thus
expect to find multiple vortex cores in a BEC, or to find a value of
$P_v$ exceeding 0.25.

\subsection{Making a vortex lattice by rotating an asymmetric TOP trap} \label{subsection:lattice}

 Before attempting to produce vortices by merging and
interfering multiple BECs, which would only produce vortices with a
finite probability, we performed a basic system check to determine
if vortex cores using our novel method of generation would be
visible. We performed this system check by creating a vortex lattice
by placing a BEC in a rotating asymmetric TOP trap; this method was
first described by Arlt \emph{et al}.\ in 1999~\cite{arlt1999bec}
and used to create a vortex lattice shortly
thereafter~\cite{hodby2001vnb}. Successful generation of vortex
lattices using this method ensured that we were allowing enough time
for expansion using the levitation coil scheme and that the vertical
imaging system was in focus.

This method `squeezes' the TOP trap along one axis of the $x-y$
(radial) plane, and rotates the axis of trap asymmetry at a
frequency that is resonant with the trap's quadrupolar mode, adding
orbital angular momentum to the condensate.  The asymmetric trap is
characterized by three harmonic oscillator frequencies $\omega_x <
\omega_y < \omega_z$.  The amount of `squeezing' of the trap in the
$x-y$ plane is characterized by the deformation parameter
$\epsilon$, given by
\begin{equation}
\epsilon = \frac{\omega_y^2 - \omega_x^2}{\omega_y^2 + \omega_x^2},
\end{equation}
where the situation $\epsilon =0$ corresponds to an axially
symmetric TOP trap.

We create this asymmetry in the two horizontal trapping frequencies
by combining the signals from two function generators operating at
frequencies $\omega_1 = \omega_{TOP}+\omega_{spin}$ and
$\omega_2=\omega_{TOP}-\omega_{spin}$.  By letting
$\omega_2=\omega_1$, we can turn off the rotating asymmetry.  These
signals are both phase-shifted and summed to produce a signal that
consists of a slow rotation (the spinning asymmetry at
$\omega_{spin}$) superimposed on a fast rotation (the TOP trap
rotation at $\omega_{TOP}$ = 2~kHz), yielding radial components of
the magnetic field that oscillate according to
\begin{equation}
B_x(t)=B_0 cos(\omega_1 t) + \epsilon \cdot cos(\omega_2 t)
\end{equation}
\begin{equation}
B_y(t)=B_0 sin(\omega_1 t) - \epsilon \cdot sin(\omega_2 t).
\end{equation}
The amount of spinning is characterized by $\epsilon$ and
$\omega_{spin}$, the amplitude and frequency of the rotating
asymmetry, respectively. The average radial trapping frequency
$\omega_{\perp}$ is defined as
\begin{equation}
\omega_{\perp} = \sqrt{\frac{\omega_x^2 + \omega_y^2}{2}}.
\end{equation}
Resonant excitation of the quadrupolar mode of the asymmetric trap
will occur when the spinning frequency
$\omega_{spin}=\frac{1}{\sqrt{2}} \cdot \omega_{\perp}$
\cite{parker2006rab}.

The procedure for creating a vortex lattice using this method
involves creating a condensate, allowing for a few seconds of
spinning in the asymmetric trap at the \mbox{resonant} spinning
frequency, turning off the spinning and letting the cloud
equilibrate into a stable triangular vortex lattice in a symmetric
TOP trap, and then allowing for a period of expansion before
imaging.  We have created vortex lattices using this method in the
weak gravity-sagged TOP trap with radial trapping frequency of
7.4~Hz, as well as in the tight TOP trap with radial trapping
frequency 40~Hz, albeit at a different spinning frequency.  In order
to generate a high-contrast lattice, it was necessary for us to
begin with as pure of a BEC as possible, as even a small amount of
thermal component greatly reduced the vortex lattice visibility.
However, a small thermal component was still present, enabling
damping of the vortex array into an ordered lattice.

% example shown if 8/24/06 file bec40 40x40 pixels
Conditions to produce a regular lattice are the following: atoms are
transferred into the weak TOP trap with radial trapping frequency of
7.4~Hz. A pure condensate is produced in a 10~second Phase II
evaporation step by decreasing the RF evaporation frequency to
4.7~MHz. By instantaneously turning on the rotating trap asymmetry,
spinning was induced in the asymmetric trap for 4~seconds, with an
amplitude of 0.026~V (peak-to-peak) on the function generator that
controls spinning.  The frequency of the function generator used for
spinning is 11.5~Hz, which corresponds to an actual spinning
frequency of 5.75~Hz, or $\sim0.78~\omega_{\perp}$. During this spin
time, the RF frequency was increased to 8~MHz.

\begin{figure}
\begin{center}
\leavevmode
\includegraphics[width=0.5\linewidth]{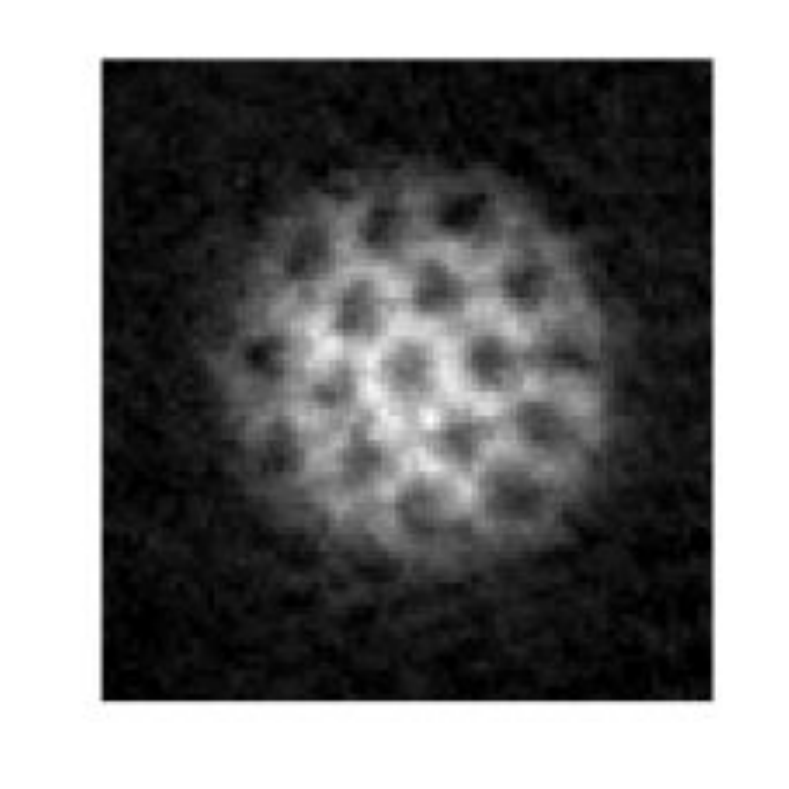}
\end{center}
\caption[Vortex lattice]{Image of a vortex lattice created by
rotating an asymmetric TOP trap.  The size of the image is $97~\mu$m
$\times$ 97~$\mu$m.} \label{fig:Vortex-lattice}
\end{figure}

After this spinning time, the rotating trap asymmetry was turned off
and 6~seconds of equilibration time occurred in the symmetric TOP
trap with the RF frequency at 6.5~MHz. It is essential that the RF
shield remain on and at a higher than normal value during the
spinning and equilibration periods, otherwise the atoms will heat up
during these long hold times. We have found that it is necessary to
let the BEC equilibrate in the symmetric TOP trap for at least
4~seconds to allow enough time for the lattice to form. After the
equilibration period, 56~ms of expansion using the levitation coil
scheme occurs before absorption imaging on the vertical axis. An
image of a vortex lattice generated using the above recipe is shown
in Figure~\ref{fig:Vortex-lattice}.

\begin{figure}
\begin{center}
\leavevmode
\includegraphics[width=0.95\linewidth]{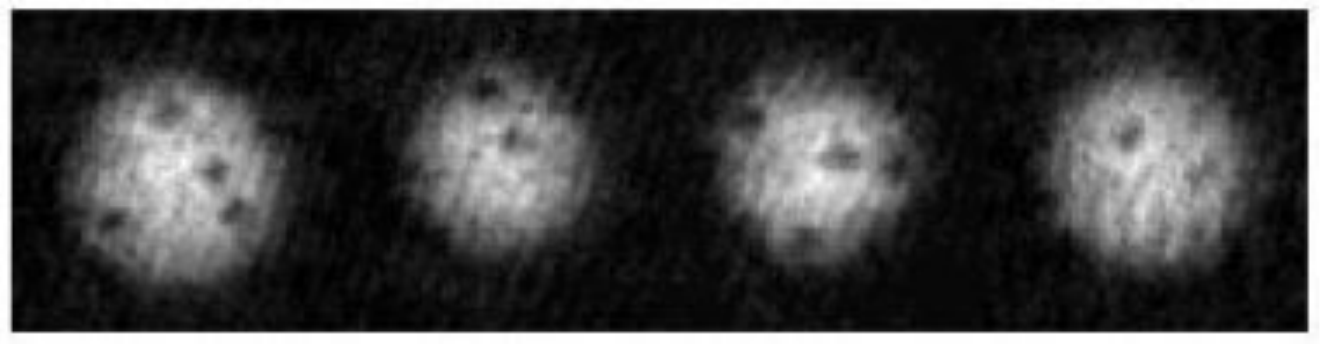}
\end{center}
\caption[Images showing only a few vortices created by rotating an
asymmetric TOP trap]{Images showing only a few vortices created by
rotating an asymmetric TOP trap.  The size of each image is
$83~\mu$m $\times$ 83~$\mu$m.} \label{fig:Few-vortices}
\end{figure}

Using the above procedure, we were able to create a vortex lattice
of up to $\sim$15 vortices by spinning in the weak TOP trap, and
were able to see 2 or 3 times more vortices by spinning in a tighter
TOP trap. We also investigated the regime of very weak spinning, to
see what only a few vortex cores would look like, as our matter-wave
interference experiment was designed to look for only a few
vortices.  Example images of small numbers of vortices generated by
weak spinning are shown in Figure~\ref{fig:Few-vortices}.  These
images were all generated by using weak spinning amplitudes and
short spinning times.
% see 8/24/06 for files, few-vortices uses 42, 44, 46, 48, 34x34 pixels
%see raman bragg spectroscopy xxx
%hodby vortex nucleation 2002

\section{Vortex Formation by Merging Independent BECs} \label{sec:independent}
%The remainder of this chapter will be devoted to the scientific
%results obtained from matter-wave interference of independent BECs,
%in an attempt to answer the research question posed above.
The concept behind our experiment and the experimental details have
been presented in the previous sections,
sections~\ref{sec:independent} through~\ref{section:splitting} will
describe the results of the three major scientific projects
presented in this dissertation.

\subsection{Formation of independent BECs}
\begin{figure}
\begin{center}
\leavevmode
\includegraphics[width=0.3\linewidth,height=0.3\linewidth]{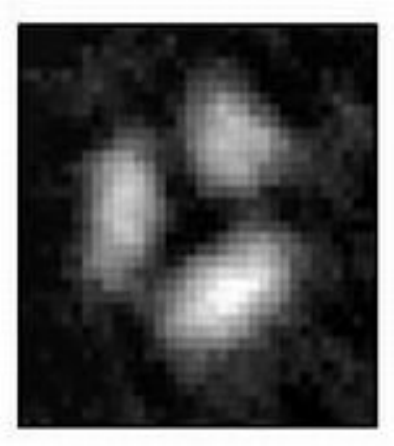}
\end{center}
\caption[Image of trapped condensates formed in the presence of a
strong barrier]{In-trap, phase-contrast image of a condensate formed
in the presence of a strong barrier.  The maximum strength of the
optical barrier beam is $170~\mu$W, strong enough of a barrier to
allow for three independent condensates to form in the bottom of
each of the wells in the three-well potential.  The size of the
image is 85~$\mu$m $\times$ 85~$\mu$m.} \label{fig:Strong-Y}
\end{figure}

In the first of three investigations, three spatially isolated
condensates were created in the presence of a strong barrier of
maximum potential energy $k_B \times$26~nK. To create three isolated
condensates in a segmented trap, we modify the BEC formation
procedure described in Section~\ref{subsection:becformation} by
ramping on the optical barrier beam over 500~ms immediately before
the final 10~second stage of evaporative cooling in the weak TOP
trap. With 170~$\mu$W in the beam, which corresponds to a maximum
barrier height of $k_B \times$ 26~nK for our beam, three condensates
are created and do not merge together during their growth.  With a
strong barrier, tunneling plays no role due to the large barrier
width, and the BECs can be considered independent until merged. A
set of three BECs created under these conditions is shown in
Figure~\ref{fig:Strong-Y}.

\subsection{Vortex observation fraction vs.\ merging time}
% data taken on 09/13/06
After BEC formation, the three independent condensates were then
merged by linearly ramping down the strength of the barrier to zero
over a variable time $\tau$. After the barrier ramp-down, the cloud
expands for 56~ms using the levitation coil scheme, and the expanded
cloud is imaged using absorption imaging along the vertical axis.
This entire process was repeated between 5 and 11 times for each of
6 different values of the barrier ramp-down time $\tau$ between
50~ms and 5~s.

\begin{figure}
\begin{center}
\leavevmode
\includegraphics[width=0.8\linewidth]{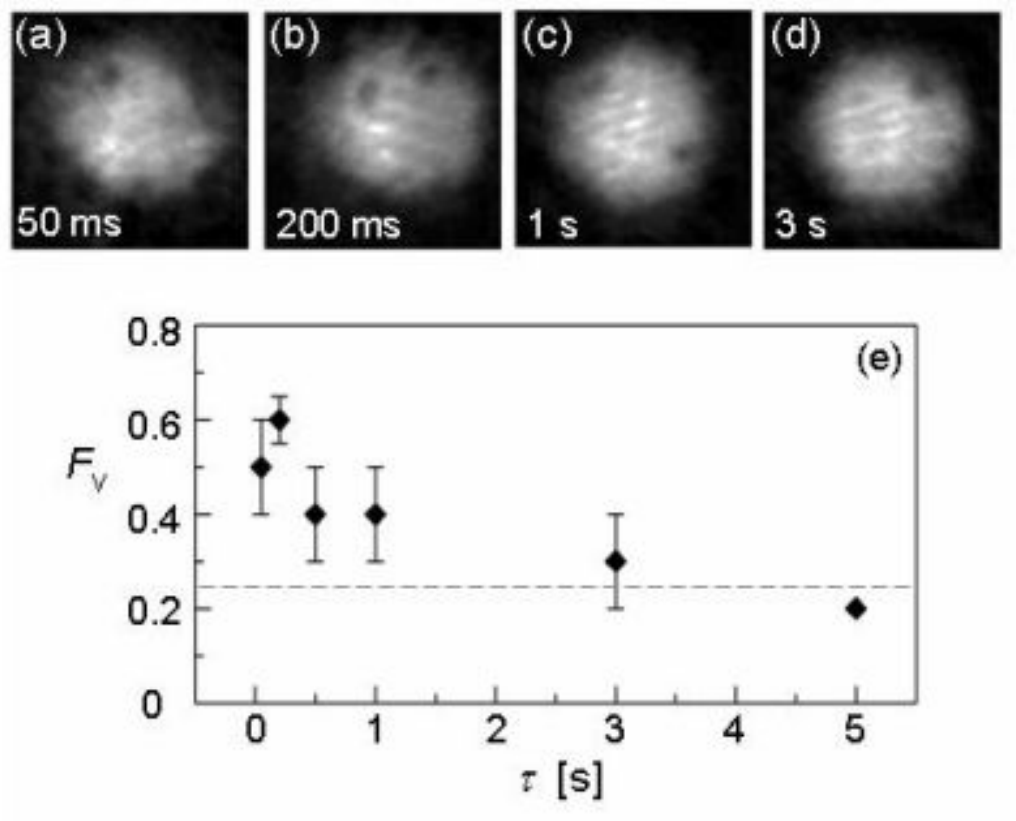}
\end{center}
\caption[Example images of vortices created by merging independent
BECs]{Example images of vortices created by merging independent
BECs.  Images (a)--(d) show vortices in condensates created in the
presence of a strong ($k_B \times$26 nK) barrier. Prior to release
from the trap, the barrier was ramped off over the time $\tau$
indicated.  The size of each image is 170~$\mu$m $\times$
170~$\mu$m.  (e) Vortex observation fraction $F_{v}$ vs.\ barrier
ramp-down time $\tau$. The data for $\tau$ values of 50~ms, 200~ms,
500~ms, 1~s, 3~s, and 5~s consisted of 5, 11, 10, 10, 5, and 5
images, respectively.  For clarity, statistical uncertainties due to
finite sample sizes are not shown, but they generally exceed our
counting uncertainties.} \label{fig:Merging-times}
\end{figure}

In a significant fraction of our images, one or more vortex cores
were visible, a clear indication that condensate merging can indeed
induce vortex formation. The spatial density distributions varied
from shot to shot, as would be expected with indeterminate phase
differences between the initial condensates, and many images were
absent of vortices. Example images of expanded BECs in
Figure~\ref{fig:Merging-times}~(a)-(d) show the presence of vortex
cores after various barrier ramp-down times.

\newpage

An analysis of vortex observation statistics is given in
Figure~\ref{fig:Merging-times}~(e) for the different values of
$\tau$ examined. We define the vortex observation fraction $F_{v}$
as the fraction of images, for each value of $\tau$, that show at
least one vortex core. The error bars reflect our uncertainty in
determining whether or not an image shows at least one vortex. For
example, core-like features at the edge of the BEC or core-like
features obscured by imaging noise may lead to uncertainty in our
counting statistics and determination of $F_{v}$. As the plot shows,
$F_{v}$ displays a maximum of $\sim$0.6 for the smaller $\tau$
values, and decreases to $\sim$0.25 for long ramp-down times. We
expect that with larger sample sizes, $F_{v}$ should approximate
$P_v$ for each $\tau$. Thus our results are consistent with our
conceptual analysis, where $P_v > 0.25$ for fast merging times, and
$P_v=0.25$ for slow merging times according to the geodesic rule for
random initial phase differences. This expected lower limit is
represented by the dashed line in
Figure~\ref{fig:Merging-times}~(e).

\subsubsection{Multiple-vortex lifetime vs.\ single-vortex lifetime}
For $\tau \leq 1$~s, multiple vortices were often observed in our
images, as those of Figure~\ref{fig:Merging-times} show, perhaps
signifying the creation of both vortices and antivortices. Although
we are unable to determine the direction of fluid circulation around
our observed vortex cores, we performed an additional test in which
the barrier was ramped off in 200~ms, forming multiple vortex cores
with a high probability such as the ones shown in
Figure~\ref{fig:Merging-times}~(b). We then inserted additional time
in order to hold the final BEC in the unperturbed harmonic trap
before our expansion imaging step. After such a sequence, the
probability of observing multiple vortices dropped dramatically: for
no extra hold time, we observed an average of 2.1 vortex cores per
image, whereas this number dropped to 0.7 for an extra 100~ms of
hold time, suggestive of either vortex-antivortex recombination on
the 100~ms timescale, or some other dynamical processes by which
vortices leave the BEC. However, single vortices were observed even
after 5~s of extra hold time in our trap following the 200~ms
barrier ramp, indicating relatively long vortex lifetimes in our
trap.

\section{Vortex Formation During BEC Creation}
\subsection{Experimental sequence}
\begin{figure}
\begin{center}
\leavevmode
\includegraphics[width=0.3\linewidth,height=0.3\linewidth]{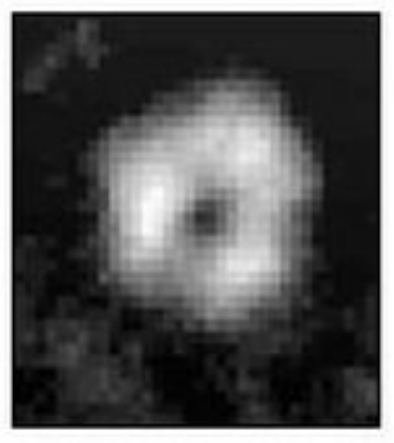}
\end{center}
\caption[Image of a BEC formed in the presence of a weak
barrier]{In-trap, phase-contrast image of a BEC formed in the
presence of a weak barrier.  The maximum strength of the optical
barrier beam is $45~\mu$W, weak enough of a barrier to allow the
initial condensates to merge together during evaporative cooling.
The observable hole in the BEC is due to the beam displacing atoms
from the center of the trap. The size of the image is 85~$\mu$m
$\times$ 85~$\mu$m.} \label{fig:Weak-Y}
\end{figure}

% data taken on 09/15/06
In our second investigation, we used only 45~$\mu$W in the optical
barrier beam, corresponding to a maximum barrier energy of $k_B
\times$7~nK.  Three independent condensates again \emph{initially}
form, but as the condensates grow in atom number, their repulsive
interatomic interactions eventually provide enough energy for the
three condensates to flow over the barrier arms.  The three
condensates thus naturally merge together into one BEC during
evaporative cooling, as shown in the in-trap, phase contrast image
of Figure~\ref{fig:Weak-Y}, taken at the end of the evaporative
cooling stage.

\begin{figure}
\begin{center}
\leavevmode
\includegraphics[width=1\linewidth]{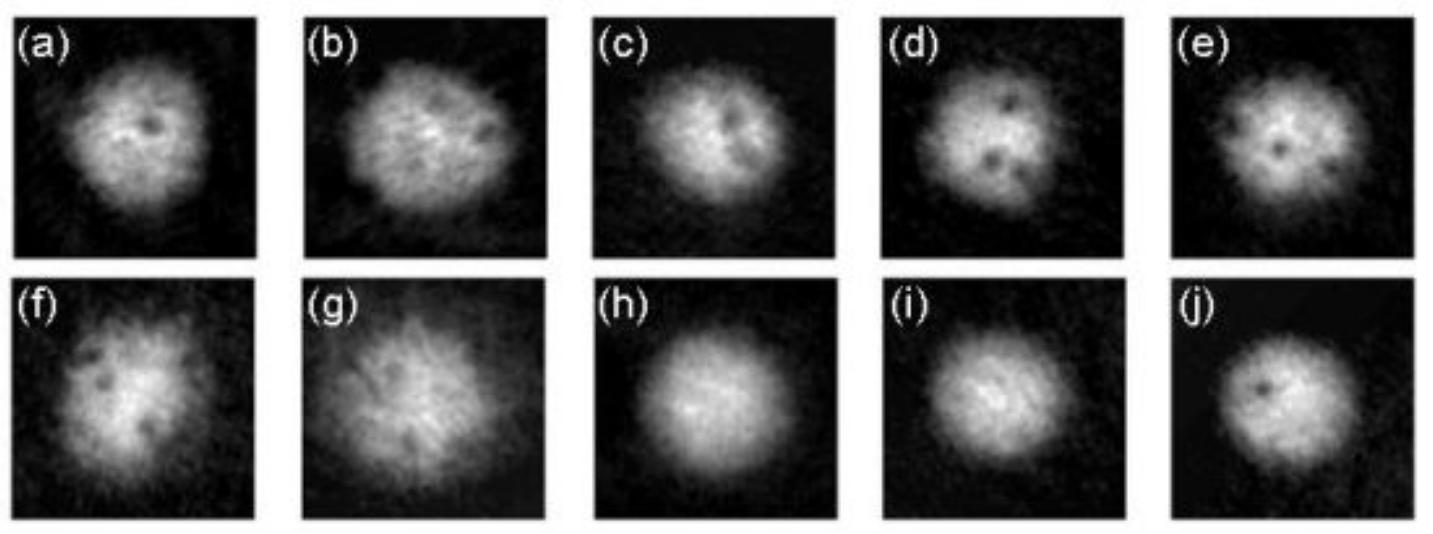}
\end{center}
\caption[Gallery of images from weak, intermediate and zero barrier
strengths]{Gallery of images from weak, intermediate, and zero
barrier strengths.  Images (a) and (b) show vortices occurring in
condensates created in a potential well with a $k_B \times$7~nK
segmenting barrier.  Images (c)--(g) show an array of vortices
obtained using intermediate barrier heights.  For comparison, image
(h) is a typical example of an expansion image without vortex cores
present.  Images (i) and (j) show BECs created without any central
optical barrier.  Note the presence of a spontaneously formed vortex
in (j).  All images are 170~$\mu$m $\times$ 170~$\mu$m wide.}
\label{fig:Gallery}
\end{figure}

We emphasize that this merging process is due solely to the
increasing chemical potential of each condensate exceeding the
potential energy of the barrier arms between the wells; the barrier
strength remained constant throughout the growth and merging of the
condensates.  After evaporative cooling, we ramped off the optical
barrier over 100~ms and released the atoms from the trap to observe
the BEC after a period of expansion. Under these conditions, our
vortex observation fraction was $F_{v}$=0.56$\pm$0.06 in a set of 16
images, with example images shown in Figure~\ref{fig:Gallery}.

\subsection{Discussion}
By adding an extra 500~ms of hold time after the final 10~second
stage of BEC formation but \emph{before} the start of the 100~ms
barrier ramp-down and ballistic expansion, the vortex observation
fraction decreased to $F_{v}$=0.28$\pm$0.14. Again, this drop in
probability may be due to vortex-antivortex recombination during
extra hold time in the weakly perturbed trap. We thus conclude that
with a maximum barrier energy of $k_B \times$7~nK, vortices are
formed during the BEC creation process rather than during the ramp
down of the weak barrier, consistent with our phase-contrast images
of trapped BECs that show a doughnut-like rather than segmented
final density distribution.

\subsection{Intermediate potential barriers}
Barrier strengths between the two limits so far described can also
lead to vortex formation, either during BEC growth or during barrier
ramp-down.  With a barrier strength in this range, up to at least
four clearly defined vortex cores have been observed in single
images, as the examples in Figure~\ref{fig:Gallery}~(c)--(g) show.
Density defects other than clear vortex cores have also appeared in
our images, as in the upper left of Figure~\ref{fig:Gallery}~(g),
where a ``gash''-like feature may be a possible indicator of
vortex-antivortex recombination; similar features have been seen in
related numerical simulations. Often, however, no vortices are
observed; an example with no vortex cores is shown in
Figure~\ref{fig:Gallery}~(h). For comparison, a BEC created in a
trap without a barrier is shown in Figure~\ref{fig:Gallery}~(i).

\section{Vortex Formation by Splitting and Recombining a BEC} \label{section:splitting}
% see data on 09/04/06
In our third investigation, a single BEC was formed in the tight TOP
trap by evaporating to a lower than normal RF frequency. The BEC was
then transferred to the weak TOP trap, and a strong optical barrier
potential was then turned on.  The maximum barrier strength was even
larger than in our first investigation, ensuring that we coherently
split the pre-formed BEC into three segments. The 10~second Phase~II
evaporation phase proceeded as usual in the weak TOP trap by
decreasing the RF frequency while holding the cloud in the
three-well potential\footnote{The BEC had already been formed at the
beginning of this evaporation phase, which only served as an RF
shield, continuing to evaporate atoms away from the already-formed
BEC.}. It should be noted that 10~seconds is longer than the phase
diffusion time of our BEC, an estimate of which can be made based on
the number-phase uncertainty principle.  For Poissonian number
fluctuations about a mean number of atoms $N$, the phase diffusion
time $\tau$ is
\begin{equation}
\tau \approx \frac{5 h \sqrt{N}}{2 \mu}
\end{equation}
where $\mu$ is the chemical potential~\cite{shin2004aib}. Using
$N=1\cdot 10^5$ atoms and $\mu = k_{B} \cdot 8$~nK, we expect the
phase diffusion time to be $\tau \approx 5$~seconds.

\begin{figure}
\begin{center}
\leavevmode
\includegraphics[angle=-90,width=0.9\linewidth]{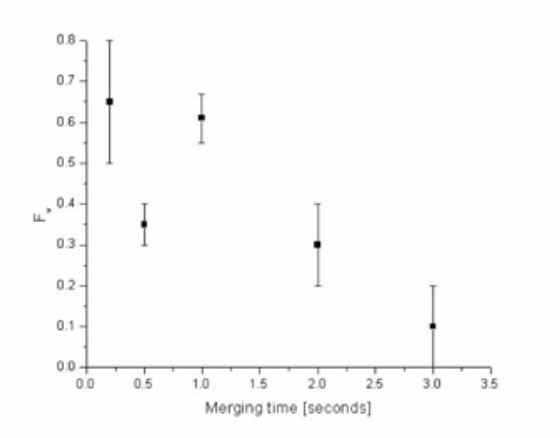}
\end{center}
\caption[Vortex observation fraction for splitting and then
recombining three condensates]{Vortex observation fraction $F_v$
vs.\ merging time $\tau$ for splitting and then recombining
condensates 10~seconds later.  The data for $\tau$ values of 200~ms,
500~ms, 1~s, 2~s, and 3~s consisted of 10, 10, 9, 5, and 5 images,
respectively.} \label{fig:Splitting-recombining}
\end{figure}

%Because of this long hold time, there should be no form of phase
%coherence between the split condensates.
Because the three condensates do not maintain a phase relationship
after the 10~second separation time, they can be considered
uncorrelated, independent BECs.  Just like in our first
investigation, by ramping down the strength of the optical barrier
beam, the three independent condensates will interfere, and vortices
may form.  By looking at images of the cloud after a period of
expansion, we observed a strong agreement with the results of our
first investigation, with a high vortex observation fraction for
short merging times decaying to a smaller observation fraction for
longer merging times. A plot of the vortex observation fraction
$F_v$ vs.\ merging time $\tau$ for $\tau$ = 200~ms, 500~ms, 1~s,
2~s, and 3~s is shown in Figure~\ref{fig:Splitting-recombining}.

\section{Spontaneous Vortices} \label{sec:spontaneous}
During the course of the aforementioned experiments, we noticed that
single vortex cores appear in $\sim$10\% of our expansion images
taken in the absence of the optical barrier beam. In other words,
for our basic single BEC creation procedure as outlined above, and
\emph{without a segmenting barrier ever turned on}, vortices
occasionally form spontaneously and are observable in expansion
images.  Figure~\ref{fig:Gallery}~(i)--(j) show two example
expansion images of condensates taken without any optical barrier in
place. Figure~\ref{fig:Gallery}~(j) shows a clear spontaneous vortex
core visible in an expansion image taken without the presence of any
optical barrier. These observations may be related to predictions of
spontaneous vortex formation due to cooling a gas through the BEC
transition \cite{kibble1976tcd, zurek1985ces, drummond1999qde,
anglin1999vwr, marshall1999eca, sadler2006ssb}. We are currently
investigating these intriguing observations further.

\section{Numerical Simulations}
As a check on our results and analysis, we used the split-step
numerical method~\cite{agrawal2001nfo} to solve the Gross-Pitaevskii
equation in simulations of three merging two-dimensional
condensates. Details of the simulations will not be given here;
however, we mention that the simulations display features
qualitatively similar to those seen in our experiment, namely: (1)
arbitrarily slow merging gives a 25\% probability for vortex
formation, given random initial phases, and without formation of any
interference fringes (dark solitons); (2) rapid merging leads to
interference fringes that decay to multiple vortices and
antivortices, which may annihilate each other in the BEC; and (3) as
merging times decrease, $P_v$ increases.  Our simulations have shown
two additional features: (1) slightly asymmetric or off-center
barriers, or unequal numbers of atoms in the three wells, can also
lead to vortex formation upon merging; and (2) a single vortex core
may migrate to and be pinned at the center of the barrier where the
energy cost of displacing fluid is low; this may help explain why a
weak barrier does not appear to readily destroy all BEC angular
momentum.

\section{Where Does the Angular Momentum Come From?}
To generate vortices by the mechanisms described above it is
important for three reasons that the condensates merge and interfere
while held in a trap. First, in a trapped BEC, the nonlinear
dynamics due to interatomic interactions play a key role in the
structural decay of interference fringes, which may be responsible
for generation of multiple vortices and antivortices seen with fast
merging times. Second, by keeping condensates trapped during their
mixing, arbitrarily slow merging times and related vortex generation
can be studied.  Finally, a gas confined in an asymmetric potential
can acquire orbital angular momentum from the trap. The barrier in
this experiment breaks the trap's cylindrical symmetry, and thus
allows for the exchange of orbital angular momentum between the
atoms and the trap. The law of conservation of angular momentum is
satisfied because the increase in angular momentum of the atomic gas
comes from the trap itself.

%In summary, we have demonstrated vortex generation by merging
%isolated and initially uncorrelated condensates into one BEC. Our
%main results are: (1) subsequent vortex observations are consistent
%with a conceptual analysis regarding merging rates and indeterminate
%phase differences between the initial condensates; and (2) BECs
%created in the presence of weak trapping potential defects or
%perturbations, such as our weak optical barrier, may naturally
%acquire vorticity during BEC creation. This second result challenges
%the notion that a BEC \emph{necessarily} forms with no orbital
%angular momentum in the lowest energy state of a trapping potential;
%rather, the shape of a static confining potential may be sufficient
%to induce vortex formation during BEC growth, a concept important to
%current and future BEC experiments and perhaps to experiments with
%other superfluids as well.

%% file: Chap_Conclusion.tex
\chapter{CONCLUSIONS\label{chapter:conclusion}}

This dissertation includes a description of two major
accomplishments: (1) the complete construction of an experimental
apparatus used to create $^{87}$Rb Bose-Einstein condensates; and
(2) an experiment that resulted in the formation of vortices by
merging and interfering multiple trapped BECs. The following
sections summarize the main ideas presented in the description of
these two achievements.

%; these steps are presented in Chapters~\ref{chapter:experimental},
%\ref{chapter:transfer}, and \ref{chapter:making}, respectively.

\section{A Review of the Experimental Apparatus}
A description of the components, techniques, and procedure used in
our method of BEC formation is given in
Chapters~\ref{chapter:experimental}, \ref{chapter:transfer}, and
\ref{chapter:making}. In summary, the three major steps in our BEC
formation process are: (1) loading atoms in a MOT and transferring
them into the initial magnetic trap in the MOT cell; (2) magnetic
transfer of atoms from the MOT cell to the science cell; and (3)
evaporatively cooling atoms in the TOP trap to form a BEC.

This dissertation includes a description of several novel
implementations of previously demonstrated experimental techniques.
For example, the advantages attained in using a diverging beam MOT
are presented in Section~\ref{subsection:diverging}.  In addition, a
description of a novel system for long-distance transfer of
magnetically trapped atoms is presented in
Chapter~\ref{chapter:transfer}.  Also, a description of the
evaporative cooling sequence used for formation of a $^{87}$Rb BEC
in a weak ($f$ = 7.4~Hz radially) TOP trap is presented in
Section~\ref{subsection:weak}.  Finally, the use of the Maskless
Lithography Tool to create arbitrary optical potentials is presented
in Section~\ref{subsection:making}.

\section{A Summary of the Experimental Results}
The experiments presented in Chapter~\ref{chapter:vortices}
demonstrate the formation of vortices by merging initially isolated
condensates into one BEC. These experiments link together the ideas
of vortex formation, merging of independent condensates, and
matter-wave interference between independent condensates in a new
way, providing a novel paradigm for vortex formation in superfluids.

Our main results are: (1) vortex observations are consistent with a
simple conceptual model regarding the indeterminate phase
differences between the initial condensates and their merging rates;
and (2) BECs created in the presence of weak trapping potential
defects or perturbations, such as our weak optical barrier, may
naturally acquire vorticity during BEC formation.

In a broader context, these experiments fit within the field of atom
optics, but our observations may be relevant to other areas as well.
For example, the formation of vortices in a condensate created in
the presence of a weak optical barrier, which may be considered a
defect, challenges the notion that a condensate \emph{necessarily}
forms with no orbital angular momentum in the lowest energy state of
a trapping potential. We have shown that the shape of a static
confining potential may be sufficient to induce vortex formation
during condensate growth, a concept that may be important in the
closely related field of superconductivity~\cite{tung2006ovp,
baert1995cfl}.

%
%
%important to current and future BEC experiments and perhaps to
%experiments with other superfluids as well.
%
%
%  For example, our results show that
%formation of vortices  may naturally acquire vorticity during BEC
%creation, a result which may be relevant to superfluids and
%superconductors.
%
%
%%
%
%\begin{enumerate}
%\item Demonstration of vortex formation by merging of independent
%superfluids.
%\item Characterization of the vortex observation fraction vs.\
%merging time for independent condensates; results in agreement with
%a simple conceptual theory.
%\item Bose-Einstein condensation of an atomic gas in a state that
%includes orbital angular momentum; this is due to the shape of the
%trapping potential.
%\item Observation of spontaneous vortex formation in a BEC.
%\end{enumerate}

\section{Directions for Future Research}
The work presented in this dissertation provides many avenues for
further research.  First, it is very scientifically interesting to
be able to form a condensate in a smooth harmonic trap in a state
that contains vorticity. This is something that we have seen in our
experiment, the observation of single vortices in $\sim 10\%$ of our
experimental runs.  Because these spontaneous vortices occur with
such infrequency, we do not currently have enough information on
their occurrence to make quantitative predictions.  Additional data,
such as spontaneous vortex observation probability vs.\ condensation
rate, RF evaporation time, and TOP trap trapping frequency need to
be gathered in order to understand this new phenomenon and be able
to compare it to theoretical predictions. This is the subject of
current work in our research group.

Second, a measure of the phase coherence time of a split BEC could
be made using our experimental setup.  By forming one BEC and then
coherently splitting it into three segments by turning on a strong
optical barrier beam, we can partition a condensate into three
segments. Then, one can hold the atoms in the three-well potential
for a variable hold time before ramping down the barrier and looking
at the vortex observation fraction.  This is similar to the
experiment described in \mbox{Section~\ref{section:splitting}}, but
with varying hold times.  In the experiment described in
Section~\ref{section:splitting}, vortices have been observed to
occur $\sim 25 \%$ of the time when merging the condensates slowly,
an observation that is consistent with our conceptual model of three
interfering superfluids with indeterminate relative phases. For hold
times less than the phase diffusion time of the split condensates,
the condensates should still maintain a phase coherence, and the
vortex observation probability may differ from the above $\sim 25
\%$.  For hold times larger than the phase diffusion time, we expect
the vortex observation probability to approach the $\sim 25 \%$ seen
for independent, uncorrelated BECs. We could therefore use our
experimental apparatus for a measurement of the phase diffusion time
of a split condensate, a measurement that has been previously
performed by looking at interference fringes between condensates in
expansion \cite{shin2004aib}.  Our method has distinct advantages
compared to looking at interference fringes in expansion: (1) by
looking at interference between trapped condensates, we can explore
arbitrarily long merging times; and (2) vortices are long-lived
topological states in superfluids, whereas interference fringes are
subject to decay and instability.

%% file: appendix_A.tex
\chapter{COHERENT MBR-110 AND VERDI V-10 LASER \mbox{OPERATION} TIPS\label{apndxA}}
Over the years our research group has obtained considerable
experience in using the Coherent MBR-110 titanium:sapphire laser,
pumped by a Verdi V-10. This appendix is meant to serve as a
resource for other users of the MBR laser and provide information on
working with the optics and electronics of the MBR and Verdi.

\subsubsection{Alignment of the MBR-110}
As a safeguard in case misalignment should occur, two irises should
be placed in the beam path between the Verdi and the MBR.  These
will be helpful in case the alignment of either of the two steering
mirrors is lost.  If lasing of the MBR is lost because of the two
steering mirrors, lasing should be re-gained using only those two
steering mirrors. In the event that this is not possible, or if
something in the cavity gets misaligned, a full alignment of the MBR
will need to be performed.

\paragraph{Preliminaries} Generally, intracavity mirror adjustments should
be performed with the tweeter disconnected, as the tweeter is always
trying to maintain lock.  The etalon tune knob should be adjusted so
that it is fully clockwise, then the etalon will be at the maximum
angle with respect to the intracavity beam, and back-reflections off
of the etalon will diverge quickly and not be mistaken for something
else.

\paragraph{Pump Beam Initial Setup} The pump beam coming into the MBR should
be level with the optical table and at a height of 4".  This can be
checked by placing an iris at a height of 4" directly in front of
the MBR input window. The floor of the MBR cavity is 1" above the
optical table, and an iris at a height of 3" will just fit in
between the second focusing lens and mirror M1. The height of the
pump beam can be set using these two irises. The horizontal
adjustment of the pump beam can be set by placing the pump beam in
the center of the two focusing lenses and the ti:sapphire crystal.

\paragraph{Index Card}
As a position reference for the laser beam inside the cavity, make
an index card $\sim$1" wide and $\sim$2" tall with a horizontal line
28~mm from the bottom of the card and two vertical lines separated
by 2~mm running up and down the middle of the card.

\paragraph{Pump Beam Alignment into Cavity}
As this point it is helpful to turn the pump beam power all the way
down so you can see the beam comfortably.  The pump beam should be
centered on M2 and at the height of the horizontal line on the index
card (from now on, everything should be at the 28~mm height). Use M2
to align the pump beam onto the M3. The pump should be centered
vertically on M3 and approximately 1~mm to the left of center
(\emph{left, right} always mean from the viewpoint of an observer
traveling with the the beam we are aligning).  Use M3 to align the
pump beam onto M4, it should be at the correct height vertically and
horizontally centered on M4 and the output window. Use M4 to align
the pump beam back onto M1.  This is difficult because it is not
possible to put a card in front of M1, as this blocks the input pump
beam.  But with the card as close as possible, you can get the
height and horizontal position close enough to start.

\paragraph{Alignment of Fluorescence Spots}
The pump beam causes fluorescence out of the ti:sapphire crystal in
all directions.  The procedure to obtain lasing consists of getting
the fluorescence in the forwards direction (ti:sapphire
$\rightarrow$ M2 $\rightarrow$ M3 $\rightarrow$ M4 $\rightarrow$ M1)
and the fluorescence in the backwards direction (ti:sapphire
$\rightarrow$ M1 $\rightarrow$ M4 $\rightarrow$ M3 $\rightarrow$ M2)
to be coincident. It is helpful to block the green pump beam while
looking at the fluorescence spots, a convenient place to put the
orange filter is against the walls 1" in front of M3. Remember that
the orange filter will prevent lasing.

\paragraph{Fluorescence Spots on M3}
After placing the index card in front of M3, two fluorescence spots
should be visible to the human eye (we are working with 780~nm). One
is the forwards spot coming from M2, this should be a round spot
2~mm in diameter.  The other is the backwards spot coming from M4,
this should be a slightly rectangular spot 1~$\frac{1}{2}$~mm tall
and 1~mm wide.  Set the spacing between the spots to be 2~mm on the
index card.  If you could place the card directly on M3, the two
fluorescence spots would be coincident and centered on M3. The green
pump beam will be slightly to the left of center on M3 because of
dispersion.

\paragraph{Fluorescence Spots on M4}
With the index card placed in front of M4, look for fluorescence
spots on M4. It is necessary to use goggles that block the green
light because of the large amount of scatter on M4.  At least 2
fluorescence spots will be visible on M4. The backwards spot comes
from M1 through the BRF; looking at this spot before the BRF, it
looks like a round spot 6~mm in diameter. A small bright dot should
be visible right in the middle of this spot.  This round spot gets
clipped by the BRF, so by the time it gets to M4, it is only a few
mm wide. The bright dot should still be visible, however, and it
will probably be right-of-center with respect to the clipped beam.
Depending on the angle of the BRF the clipping may not be purely
from the left and the right, it may be angled. There may also be a
much larger and weaker fluorescence glow behind all this, this is
just scatter from the BRF.  The second spot we need to align on M4
is the forwards spot, from M3. This is a slightly rectangular spot,
1~$\frac{1}{2}$~mm tall and 1~mm wide. Align the center of the
forwards spot to be 3-4~mm separated from the small bright dot of
the backwards spot, and at the correct height.  If you could place
the index card on M4 directly, the beams would be coincident and
approximately centered on M4. They should also be approximately
centered on the output window.

\paragraph{Closing the cavity}
Now all that's left is closing the resonator cavity, which is done
by `connecting' mirror M4 to M1.  It may be possible to look at the
forwards spot after one round trip near M1, but you can't put a card
right on M1.  The forwards fluorescence spot should be visible
outside of the cavity.  When you are close to lasing, you can see a
small second spot (from the second round trip through the cavity)
appear outside of the cavity as well.  This second spot (there is
even a third when you are very close) is very difficult to see, even
with an IR viewer.  At this point the best thing to do is to walk
the beam around horizontally with mirrors M1 and M4 and look for
lasing outside of the cavity.  Place a lens and a fast,
high-responsivity photodiode directly outside of the laser, and
remove the orange filter from the cavity and place it in front of
the photodiode, which we'll use to monitor lasing.  Small walks on
the horizontal adjustments of M1 and M4 should produce lasing, which
will register as a sharp increase on the photodiode.  If you still
can't get the laser to lase, another thing to try is to make the
spots on M3 slightly closer than 2~mm apart and the spots on M4 4~mm
apart.  You can also remove the etalon from the cavity if it is in
danger of clipping the beam.

\paragraph{Peaking up the Power}
Once lasing is obtained, peaking up the power is probably going to
be easiest with a power meter placed right at the output of the
laser.  In general, the intracavity mirrors should be walked in
order to peak up the power.  The two external steering mirrors
should NOT be walked, as this is counter-productive to our strategy.
We want to align the cavity so that it is optimal for the existing
pump beam, not the other way around.  However, small tweaks on one
of the external steering mirrors may be useful, after the
intracavity alignment procedure is performed.

Do a 2-mirror walk between mirrors M1 and M4, and M2 and M3, and
keep repeating this for a while.  When this no longer produces
gains, you can try walking M3 and M4, or M2 and M4. At some point
you will probably want to clean all the optics in the cavity, as
this can get you up to $\sim$200~mW more power.  Close the Verdi
shutter and re-obtain lasing after cleaning each component, in case
alignment gets lost by bumping something.

\paragraph{Peaking up the power}
The following things can be adjusted to peak up the output power:
\begin{enumerate}
\item The angle of the input polarizer before the input window to the
MBR.
\item The position of the focusing lenses, although adjusting these
is not recommended unless you have poor mode quality.
\item The angle of the quartz crystal in the optical diode. Also make sure
that it is not clipping.
\item The horizontal position of the housing for the magnet assembly of
the optical diode.  Make sure it is not clipping.
\item The angle of the BRF.
\item The position of the etalon. Dropping the etalon out
of the cavity will not stop lasing, you should lose no more than
$\sim$200~mW by letting the etalon back in the cavity.  Changing the
location of where the beam hits the etalon can make a difference in
output power, but be careful:  this can also affect what the
appropriate etalon drive frequency needs to be.
\end{enumerate}

%\paragraph{Hints}
%At some point in the alignment peak-up, you may obtain a multiple
%spatial mode output.  The power might drop by $\sim$300~mW and the
%output lasing beam can consist of 2 vertically or horizontally
%separated modes, TEM~01 or TEM~10 modes.  To get out of this local
%maximum in the power, it may be necessary to adjust one of the
%mirrors out of a local maximum in power and into the global maximum.
%This means you may have to adjust one mirror till the power drops
%very low and then it will rise again to a higher value and be in
%TEM~00 mode.

\subsubsection{Electronics of the MBR-110}
Occasionally it is necessary to peak up the electronics of the MBR
because the laser does not stay in etalon lock or servo lock.  The
main indicator of an electronics problem with the MBR is the etalon
sawtooth signal.  This signal should have a peak-to-peak voltage of
at least 1~V, and should display a rapid rise and slow fall when the
etalon tune knob is turned clockwise. If the etalon sawtooth signal
displays a slow rise and rapid fall when turning the tune knob
clockwise, the MBR will lock to and lase at multiple frequencies
until the phase is corrected by adjusting DIL switch SW10 in the
controller box.

%\footnote{Referred to as the `Scherer effect' due to our assistance
%in diagnosing this problem; see the Coherent MBR-110 Service Manual
%SVC-MBR-3.2 Rev. C.}

If the amplitude of the etalon sawtooth error signal is less than
500~mV peak-to-peak, the laser will have trouble locking and an
electronics peak-up should be performed by adjusted some of the
potentiometers in the MBR controller box.  This peak-up may be as
simple as adjusting PR14 (this is the fine adjustment of the etalon
drive frequency) while rotating the etalon tune knob until a larger
sawtooth signal is obtained. If this does not work, a full
electronics peak-up is recommended.

This procedure is designed to adjust the amplitude, frequency, and
phase of the etalon drive signal.  If done correctly, it should
adjust the drive signal to be in resonance with the natural
resonance frequency of the etalon, and the etalon sawtooth error
signal will increase.

\paragraph{Preliminaries}
Two phases of the etalon drive voltage are available at TP5 and TP7.
Use TP5 to trigger a scope for the rest of the procedure.

\paragraph{Etalon drive voltage adjustment}
Using a probe on the etalon itself, monitor the peak-to-peak
amplitude of the etalon drive voltage.  If this is different from
the voltage given on the spec. sheet ($\sim$4~V peak-to-peak) adjust
PR16 until it is corrected.

\paragraph{Etalon frequency adjustment}
Monitor TP6 on a scope.  AC couple the scope and increase the
sensitivity. Adjust PR14 (etalon drive frequency fine adjustment)
and PR15 (etalon drive frequency course adjustment) until the amount
of `breathing' on TP6 is maximized.  This breathing is when
alternate peaks in the modulation signal differ by the maximum
amount possible upon rotation of the etalon tune knob.  Zero
breathing corresponds to a sine wave, and full breathing corresponds
to an oscillatory modulation signal that alternates between high and
low peaks.

\paragraph{Etalon frequency fine adjustment}
Monitor TP8 on a scope.  Turn off the MBR controller power,
disconnect the umbilical cable to the laser head, and turn the power
back on.  Connect a jumper between TP9 and TP6.  Maximize the
amplitude of the signal on TP8 by adjusting capacitor CT1 (a
non-metal tool should be used to adjust this capacitor).  When you
are done, turn the power off, disconnect the jumper, reconnect the
umbilical, and turn the power back on.

\paragraph{Etalon phase adjustment}
Monitor TP10 on a scope.  Adjust the signal on PR17 (0-90$^\circ$
etalon phase adjustment) until TP10 looks as symmetric as possible.
After these adjustments, check that the amplitude of the etalon
sawtooth error signal has increased.

\subsubsection{Verdi V-10 Controller}
The diodes in the controller that pump the Verdi via the umbilical
cord are placed on temperature and current interlocks. Over time,
the amount of current needed to operate these diodes so that they
output the appropriate power to pump the Verdi will increase; this
is due to a gradual degradation in the efficiency of the diodes. Our
diodes, after being in use for $\sim$4000~hours, will go into
overcurrent mode when we try to turn on the Verdi to its maximum
power of 10~W, because the current in the pump diodes is too high.
We are able to operate the Verdi at high powers, however, by first
turning it on to 9~W and gradually increasing the output power.

\subsubsection{Verdi V-10 Output Window} We have noticed that the
output window of the Verdi V-10 will occasionally obtain a water
stain on the \emph{inside} of the hermetically sealed glass output
window of the shutter assembly. This is visible by shining a
flashlight onto the Verdi output window and looking directly into
the laser.  In addition to degrading the Verdi output beam quality,
this causes a small amount of focussing of the Verdi output beam,
and will result in decreased output power of the MBR.  We remove the
Verdi shutter assembly and clean the inside of the glass output
window about once a year.